\documentclass[10pt,pra,a4paper,final,nofootinbib]{revtex4}
\usepackage{graphicx}
\usepackage{amsmath}
\usepackage{amsfonts}
\usepackage{color}
\usepackage[breaklinks=true,colorlinks=true,linkcolor=blue,urlcolor=blue,citecolor=blue]{hyperref}

\usepackage{centernot}
\usepackage[francais,english]{babel}
\newcommand{\RR}{\mathbf{R}}
\newcommand{\rr}{\mathbf{r}}
\newcommand{\kk}{\mathbf{k}}

\newcommand{\KK}{\mathbf{K}}
\newcommand{\PP}{\mathbf{P}}
\newcommand{\qq}{\mathbf{q}}
\newcommand{\OO}{\mathbf{0}}

\newcommand{\kf}{k_{\rm F}}
\newcommand{\Ef}{\epsilon_{\rm F}}

\newcommand{\be}{\begin{equation}}
\newcommand{\ee}{\end{equation}}
\newcommand{\bea}{\begin{eqnarray}}
\newcommand{\eea}{\end{eqnarray}}

\newcommand{\yvan}{}

\newcommand{\qed}{\nobreak \ifvmode \relax \else
      \ifdim\lastskip<1.5em \hskip-\lastskip
      \hskip1.5em plus0em minus0.5em \fi \nobreak
      \vrule height0.75em width0.5em depth0.25em\fi}

\begin{document}

\author{Christian Trefzger and Yvan Castin}
\affiliation{Laboratoire Kastler Brossel, \'Ecole normale sup\'erieure, CNRS, UPMC, 24 rue Lhomond, 75231 Paris Cedex 05, France}
\title{{\yvan The} self-energy of an impurity in an ideal Fermi gas to second order in the interaction strength}

\begin{abstract}
We study in three dimensions the problem of a spatially homogeneous zero-temperature ideal Fermi gas of spin-polarized particles 
of mass $m$ perturbed by the presence
of a single distinguishable impurity of mass $M$. The interaction
between the impurity and the fermions involves only the partial $s$-wave through the scattering length $a$, and has negligible range $b$
compared to the inverse Fermi wave number $1/\kf$ of the gas. Through the interactions with the Fermi gas the impurity 
gives birth to a quasi-particle, which will be here a Fermi polaron (or more precisely a {\sl monomeron}).
We consider the general case of an impurity moving with wave vector $\KK\neq\OO$: Then the quasi-particle acquires a finite lifetime in
its initial momentum channel because it can radiate particle-hole pairs in the Fermi sea. A description of the system using a variational approach, based on a finite number of particle-hole excitations 
of the Fermi sea, then becomes inappropriate around $\KK=\mathbf{0}$. We rely thus upon perturbation theory, where the small and negative parameter $\kf a\to0^-$ excludes 
any branches other than the monomeronic one in the ground state (as e.g.\ the dimeronic one), and allows us a systematic study of the system. 
We calculate the impurity self-energy $\Sigma^{(2)}(\KK,\omega)$ up to second order included in $a$. Remarkably, we obtain an analytical
explicit expression for $\Sigma^{(2)}(\KK,\omega)$ allowing us to study its derivatives in the plane $(K,\omega)$. These present interesting 
singularities, which in general appear in the third order derivatives $\partial^3 \Sigma^{(2)}(\KK,\omega)$. In the special case of equal
masses, $M=m$, singularities appear already in the physically more accessible second order derivatives $\partial^2 \Sigma^{(2)}(\KK,\omega)$;
using a self-consistent heuristic approach based on $\Sigma^{(2)}$ we then regularise the divergence 
of the second order derivative $\partial_K^2 \Delta E(\KK)$ of the complex energy of the quasi-particle found in reference
[C. Trefzger, Y. Castin, Europhys. Lett. {\bf 104}, 50005 (2013)] 
at $K=\kf$, and we predict an interesting scaling law in the neighborhood of $K=\kf$.
As a by product of our theory we have access to all moments of the momentum of the particle-hole pair emitted by the impurity while damping its 
motion in the Fermi sea, at the level of Fermi's golden rule.
\end{abstract}

\pacs{03.75.Ss - Degenerate Fermi gases}

\maketitle

\section{Introduction and motivations}

In this work we study, in a three dimensional space, the problem of a mobile distinguishable impurity of mass $M$ undergoing    
elastic scattering with an ideal Fermi gas of fermions of mass $m$, all in the same spin-state, at zero temperature and in the 
thermodynamic limit.
The solution of this problem is a fundamental step in the microscopic comprehension of the concept of quasi-particle, which lays 
at the heart of the Fermi liquid theory developed by Landau \cite{Landau}.     
It allows us to show the expected effects resulting from the coupling of the impurity with the Fermi sea reservoir: From
reactive effects shifting the real energy of the impurity and changing its effective mass, to dissipative effects, appearing when the
impurity is moving, inducing a finite lifetime of the impurity in its initial momentum channel $\hbar\KK$ due to emission of 
particle-hole pairs in the Fermi sea that lower the impurity momentum. These two effects may be summarized by the notion of 
complex energy $\Delta E(\KK)$, counted from the energy of the unperturbed Fermi sea.

From the formalism perspective this leads naturally to the introduction of the impurity self-energy $\Sigma(\KK,\omega)$, a function
of the wave vector $\KK$ and of an angular frequency $\omega$, that enters into the Dyson equation satisfied by the space-time Fourier transform
$\mathcal{G}(\KK,\omega)$ of the two-points Green's function, which constitutes a building block of the diagrammatic methods 
for the $N$-body problem \cite{Fetter}.
Indeed, the complex energy $\Delta E(\KK)$ mentioned above, after division by $\hbar$, must be a pole of the analytic continuation 
of the function $\omega\mapsto\mathcal{G}(\KK,\omega)$ to the lower complex half-plane.

Our single impurity-problem has of course a long history. It emerged in a nuclear physics context, in the case of a $\Lambda$-particle
interacting with a Fermi sea of nucleons \cite{Walecka,BishopNucl} {\sl via} a spherical hard-core potential of radius $a>0$.
The results, obtained by resummation of diagrams in the $T$-matrix formalism, are limited to $\KK=\mathbf{0}$ but the expansion 
in powers of $\kf a$ was remarkably pushed up to the order four \cite{Bishop}, $\kf$ being here the Fermi wave number.

Recently the problem shows a renewed interest thanks to cold atoms experiments, where the impurity is an atom of the same chemical 
species as the fermions but in a different spin state \cite{Hulet,Ketterle,Kohl}, or even an atom of a different chemical species \cite{Zaccanti}.
The experiments of reference \cite{Ketterle} can then be very well interpreted by the fact that the fraction of atoms in the minority spin-state,
when sufficiently small, constitutes a Fermi ``liquid", that is an almost ideal gas of fermionic quasi-particles whose internal energy and
effective mass have been modified by the interaction with the Fermi sea of the atoms in the majority spin-state, in agreement with the theory
of  \cite{Chevy,LoboStringari} and as confirmed by the precise measurement of the equation of state of a spin-polarized gas \cite{Nascimbene,Navon}.

From a theoretical point of view, the model interactions appropriate to cold atoms strongly differ from those based on hard-spheres 
of the first references \cite{Walecka,BishopNucl,Bishop}. Indeed, for cold atoms and for the previously cited experiments, the interaction 
between the impurity and a fermion is resonant in the $s$-wave and negligible in the other partial waves. This means that the $s$-wave
scattering length $a$ is much larger, in absolute value, than the range $b$ of the interaction and it can have an arbitrary sign, two features 
that are missing in the hard-spheres model.     
One can even experimentally tend to the unitary limit $1/a=0$ thanks to the amazing tool of Feshbach resonances \cite{revue_feshbach}.
In this resonant regime $|a|\gg b$, one expects the interacting potential to be characterized only by the scattering length to the exclusion of 
any other microscopic details (as e.g.\ its position dependence); in such a case we speak of {\sl one-parameter universality}
and theoretically we are led to make the range of the potential $b$ tend to zero with a fixed scattering length, for any convenient model.
In reality, this one-parameter universality has not to be taken for granted. It breaks down when the mass ratio $m/M$ is too large 
so that the effective attractive interaction induced by the impurity between the fermions leads to the three-body Efimov effect, at the critical
mass ratio $m/M \simeq 13.607$ \cite{Efimov,Petrov,BraatenEfim}, or also to the four-body Efimov effect, at the critical mass ratio $m/M\simeq 13.384$ \cite{Efim4corps}. 
In the presence of such Efimov effects, additional three-body and four-body parameters must be introduced to characterize the interaction,
and in the limit of zero true range and effective range the energy spectrum is not bounded from below, which constitutes the 
Thomas effect \cite{Thomas}. Up to now, a necessary and sufficient condition on the mass ratio excluding all possible
Efimov effects, even at the thermodynamic limit, is unknown \cite{Teta} but we will suppose it to be satisfied in this work.

In this cold-atoms context, an important conceptual progress was to realize that our single impurity problem belongs to the general 
class of polaronic systems \cite{Svistunov}. By analogy with solid-state physics, in which the polaron is an electron dressed by the 
(bosonic) phonons describing quantum-mechanically the deformation of the crystal induced by the electromagnetic interaction 
with the electron charge, the impurity constitutes a Fermi polaron because it is dressed by the particle-hole pairs induced by its now short-ranged interaction,
with the fermions.  
 The picture is indeed very rich since many classes of polarons may exist, depending on whether the quasi-particle is constructed
 by particle-hole pair dressing of the bare impurity \cite{Chevy,LoboStringari}, or by dressing a two-body bound state (dimer) between
 the impurity and a fermion, that preexists in free space \cite{Svistunov,Zwerger,MoraChevy,Combescot_bs}, or even by dressing a 
 three-body bound state (trimer) between the impurity and two fermionic particles \cite{Parish}.
 By systematically extend the terminology of reference \cite{TrefzgerCastin}, we can then refer to monomeron, dimeron
 or trimeron to underline the quasi-particle character of the considered object, {\yvan as done by Lobo in \cite{revue_polaron}}; we can also refer to dressed atom, dressed dimer
 or dressed trimer as in the review article \cite{revue_polaron}. 
The advantage of the former terminology appears in the more surprising case where the binding between the impurity and a small number
of fermions does not preexist in vacuum but is itself induced by the presence of the Fermi sea. This is the case of the trimerons of reference  
\cite{Parish}, and of the dimerons at $a<0$ on a narrow Feshbach resonance \cite{TrefzgerCastin,Massignan,Zhai}; in the latter case see the
conclusion of \cite{TrefzgerCastin} for a physical interpretation.

Let us restrict ourselves here and in what follows to the monomeronic branch, and let us consider the case of a negative scattering length $a<0$
on a broad Feshbach resonance (therefore of negligible true and effective ranges); the dimeronic branch is then an excited branch \cite{Svistunov}
that is unstable \cite{BruunMassignan}\footnote{We recall that the energy of the dimer tends to $-\infty$ as $-\hbar^2/(2\mu a^2)$ when $a\to 0^+$,
$\mu$ being the reduced mass of the impurity and a fermion, and that there is no dimer for $a<0$.}.
Up to now the single impurity problem with resonant interaction has been treated analytically, essentially at zero wave vector $\KK=\mathbf{0}$,
using a non-perturbative variational approach that truncates the Hilbert space by keeping at most $n$ 
particle-hole pairs, but without
any constraint on their possible states.
This approach was initiated by reference \cite{Chevy} (see also \cite{Combescot_varia}), with  $n=1$; at $\KK=\mathbf{0}$ the energy 
$\Delta E(\mathbf{0})$ is real and the predicted approximate value $\Delta E^{[1]}(\mathbf{0})$ gives an upper bound to $\Delta E(\mathbf{0})$,
which was sufficient to establish the existence of a Fermi ``liquid" phase in a strongly polarized gas at the unitary limit $1/a=0$ \cite{Chevy}.
In this rather spectacular strongly interacting case, there is {\sl a priori} no small parameter allowing us to control the precision of the variational 
ansatz \cite{Chevy}; the fact that the result is identical to the one obtained with the non-perturbative (therefore non-systematic) use of the $T$-matrix 
formalism (in the ladder approximation) \cite{Combescot_varia,these_Giraud} does not prove that the result is reliable.
However, it was finally understood that a semi-analytical systematic study for increasing $n$ (in practice limited to $n=2$) is a winning strategy,
allowing one to explicitly verify the rapid convergence of the series $\Delta E^{[n]}(\mathbf{0})$ \cite{Combescot_deux,these_Giraud}
towards the numerical diagrammatic quantum Monte Carlo results of reference \cite{Svistunov}.

We shall devote this work to the more original case of a non zero total momentum, $\KK\neq \mathbf{0}$, still not {\yvan fully understood}
(see however reference \cite{Nishida} at large $K$).
The problem then changes nature and the monomeron becomes a resonance of complex energy $\Delta E(\KK)$ \cite{BishopNucl};
the moving quasi-particle is indeed unstable with respect to particle-hole pairs emission \cite{StringariFermi} since the kinetic energy and the 
momentum carried away by the pairs can be, let us recall it, as close to zero as we want.   
The variational approach also changes status: Not only it no longer provides an upper bound to the real part of the energy $\Re \Delta E(\KK)$,
but it also predicts a non-physical interval of $K$-values, starting at $0$, within which the imaginary part of the energy $\Im \Delta E(\KK)$ is exactly
zero \cite{lettre}\footnote{The same phenomenon occurs for the dimeron in two dimensions \cite{Levinsen}.}. This explicitly contradicts the perturbative result of reference \cite{BishopNucl} obtained for $M=m$ up to second order
included  in $\kf a$, which gives an $\Im \Delta E(\KK)$ continuously vanishing as $K^4$ when $\KK\to \mathbf{0}$,
and  more generally it disagrees with the Fermi ``liquid" theory of \cite{StringariFermi}. 

This failure of the variational approach at $\KK\neq \mathbf{0}$ is readily understood if one considers the one-impurity problem 
in the general context of a discrete state coupled to a continuum \cite{CCT}, as it was done in reference \cite{TrefzgerCastin}.
The discrete state corresponds to the impurity with wave vector $\KK$ in the presence of the unperturbed Fermi sea;
its energy (counted from the Fermi sea energy) is thus the impurity kinetic energy $\hbar^2 K^2/(2M)$.
The continuum is made of particle-hole pairs of any momenta in the Fourier space, in the presence of the impurity in the suitable 
wave vector state (such as to conserve the total momentum).
(i) Within an exact treatment of the problem, it is clear that the continuum contains in particular a monomeron of arbitrarily small total momentum,
thus of energy $\simeq \Delta E(\mathbf{0})$, that may be thought of as a relatively localized perturbation of the fermions gas in the neighborhood of a point
in real space \cite{polzg2}, in the presence of particle-hole pairs that are radiated at infinity and that carry away the missing momentum,
without necessarily costing a significant amount of energy.
A particle and a hole respectively of wave vector $\kk$ and $\qq$ can indeed carry away momentum of modulus up to $2\kf$ with a positive 
kinetic energy $\hbar^2 (k^2-q^2)/(2m)$ that is negligible when $k\to \kf^+$ and $q\to \kf^-$.
The lower bound of the continuum then corresponds to the exact (here negative) energy $\Delta E(\mathbf{0})$.
The coupling between the discrete state and the continuum leads in general the former to dilute into the latter to give birth to 
a resonance, and $\Im \Delta E(\KK)<0$ at $K>0$.
(ii) Within the variational treatment of the problem, limited to $n$ particle-hole pairs, one expects the continuum to 
start at the energy $\Delta E^{[n-1]}(\mathbf{0})$, since at least one pair must be radiated at infinity to bring the monomeron at rest.
Now this is strictly larger than $\Delta E^{[n]}(\mathbf{0})$, according to the usual variational reasoning; at $\KK=\mathbf{0}$, the coupling
of the discrete state to the continuum then seems to give rise to a discrete state of energy $\Delta E^{[n]}(\mathbf{0})$ separated from the 
continuum by an artificial energy gap of width $\Delta E^{[n-1]}(\mathbf{0})-\Delta E^{[n]}(\mathbf{0})$. 
If this is the case $\Delta E^{[n]}(\KK)$ should remain exactly real in a region around $\KK=\mathbf{0}$, 
that shrinks at larger $n$ but that has no physical meaning.
This plausible scenario is confirmed by the explicit calculation made for $n=1$ in reference \cite{TrefzgerCastin}, 
provided the interaction is sufficiently weak: For fixed and finite $M/m$, we find indeed that the continuum starts at the energy
$\Delta E^{[0]}(\mathbf{0})=0$, if $\kf a$ (negative) is sufficiently close to zero not to satisfy equation (24) of that reference. 
This has been used to estimate the non-physical value of the modulus of $\KK$ below which $\Im \Delta E^{[1]}(\KK)=0$ \cite{lettre}.

With the variational approach being disqualified at $\KK\neq \mathbf{0}$, the theory toolbox apart from numerics is rather empty.
Thus, we shall use the only reliable and systematic method, the perturbative approach, here up to the second order included 
in $\kf a$, in the spirit of the pioneering articles \cite{Walecka,BishopNucl,Bishop}. However, instead of a hard-sphere interaction we will use a 
Hubbard-type cubic lattice model of lattice spacing $b$, with a (here attractive) on-site interaction of bare coupling constant $g_0$ adjusted, as a 
function of $b$, to reproduce exactly the desired scattering length. This model was initially introduced for the case of a weakly interacting bosonic gas, 
in a tentative form in \cite{Cartago} then in its final form in \cite{Mora}, and since then it has witnessed some success for the case of spin $1/2$
fermions, even in the strongly interacting regime \cite{modele_sur_reseau1,modele_sur_reseau2,modele_sur_reseau3}.
For a fixed $b$ we expand the self-energy $\Sigma(\KK,\omega)$ up to order two in $\kf a$, then we take the continuous space limit
$b\to 0$ in the coefficients of the expansion. The key point is that all the corresponding integrals in the Fourier space can be calculated
analytically so that explicit expressions can be obtained for $\Sigma^{(2)}(\KK,\omega)$. We shall give these expressions for any $\omega$
and for any impurity-to-fermion mass ratio $M/m$.
The opposite order of taking limits ($b\to 0$ for a fixed scattering length, then $\kf a\to 0^-$) would lead to the same result 
by virtue of the one-parameter universality mentioned above and of the absence at $a<0$ of essentially non-perturbative effects
(as e.g.\ the emergence of a dimer at the zero-range limit for $a>0$), but one would have to resort to the non-perturbative resummation of
ladder diagrams as in \cite{Bishop}.

To {\sl second} order in $\kf a$, we find for $\Delta E(\mathbf{0})$ in the zero range limit 
exactly the same results as the ones of Bishop \cite{Bishop} for the hard-sphere interaction, after their
direct transposition from the case $a>0$ to the case $a<0$. This is without too much of surprise, since
it is well known that the non-zero range (of order $a$) of the hard-sphere interaction appears only at the  
next order: The two-body scattering amplitude $f_k=-1/[a^{-1}+ik-k^2 r_e/2+O(k^3 a^2)]$ for the hard-sphere
potential, of effective range $r_e=2 a/3$, differs from the one $f_k=-1/(a^{-1}+ik)$ of the zero range interaction
by terms at least of order {\sl three}, in $O(a^3 k^2)$, when $a\to 0$ for a fixed relative wave number $k$
between a fermion and the impurity.

On the other hand, at $\KK\neq\mathbf{0}$ our expressions of $\Delta E^{(2)}(\KK)$ are new.
They were already briefly presented in reference \cite{lettre}, which shows their experimental
observability by radio-frequency spectroscopy of cold atoms \cite{Zaccanti} but gives no detail of
derivation, contrarily to this work in which this is one of the motivations.  
The knowledge of $\Sigma^{(2)}(\KK,\omega)$ as a function of $\omega$ even allows us also
to go further and, thanks to a heuristic self-consistent equation for the complex energy $\Delta E(\KK)$,
to regularise the logarithmic divergence of the second order derivative with respect to $K$ of $\Delta E(\KK)$,
predicted by the perturbative theory for equal masses ($M=m$) at the Fermi surface ($K=\kf$) \cite{lettre}.
{\yvan Note that to second order in $\kf a$ our expressions of $\Delta E^{(2)}(\KK)$ obtained for $a<0$ can be 
directly extended to the case of the repulsive monomeron~\cite{ZhaiRepulsive} where $a>0$~\footnote{{\yvan 
In the case $a > 0$ the repulsive monomeron is an
excited branch which can decay to the dimeronic and to the attractive
monomeronic branches as discussed in reference~\cite{BruunDecay}.}}.}

This long article is organized as follows. After a formal and maybe unusual writing of the self-energy
$\Sigma(\KK,\omega)$ in terms of the resolvent of the Hamiltonian in section \ref{sec:ddselalrdh},
we shall expand it up to second order included in the coupling constant $g$ and we will readily express
the result as a single integral, see section \ref{sec:calcul_sigma2}, that we are able to calculate explicitly 
at the end of a rather technical section \ref{sec:cedsigdlcg}.   
We shall then reap rather formal benefits in section \ref{sec:les_singus}, by identifying in the plane $(K,\omega)$
the singularities of the third order derivatives of the self-energy (truncated to second order in $g$), which are the
counterparts at $\omega\neq 0$ to the singularities of the derivatives with respect to $K$ of the complex 
energy (truncated to second order in $g$) of reference \cite{lettre}.
We have kept the physical applications for section \ref{sec:qap}: After having recovered
the perturbative results of reference \cite{lettre}, we explicitly implement the aforementioned self-consistent 
approach and we predict a non-perturbative scaling law for the behavior of $\frac{d^2}{dK^2} \Delta E(\KK)$ in the neighborhood of the Fermi surface
for equal masses and when $g\to 0^-$. {\yvan En passant, we check in subsection \ref{subsec:finiteT} that
the impurity complex energy $\Delta E^{(2)}(\KK)$ is a smooth function of $K$ at non-zero temperature.}
In another perspective, we will show how our techniques of integral calculus 
allow us to access all moments of the momentum of the particle-hole pair emitted by the impurity into the Fermi sea 
within Fermi's golden rule approximation; this gives not only the damping rate of the impurity 
momentum in the spirit of \cite{StringariFermi}, but also its diffusion coefficient in the spirit of \cite{DavidHuse}.
Our results are of course limited to second order in $g$ but, contrarily to references \cite{StringariFermi,DavidHuse}, they
apply for any
momentum. We shall conclude in section \ref{sec:conclusion}.

\section{Definition of $\Sigma(\KK,\omega)$ and relation with the resolvent of the Hamiltonian}
\label{sec:ddselalrdh}

We start here with a few general reminders on the well known $N$-body Green's function approach \cite{Fetter},
for a spatially homogeneous system with periodic boundary conditions, independently on the 
interaction model, then we establish the relation, maybe less well known, of this formalism with the resolvent 
and the notions of effective Hamiltonian and displacement operator, more usual in atomic physics \cite{CCT}.
Variables and operators of the impurity will be distinguished from those of fermions by the use of capital letters
for the former, and small letters for the latter.  

The case considered here is the one at zero temperature. The single impurity Feynman Green's function is then defined
by \cite{Fetter} 
\be
\mathcal{G}(\RR,t;\RR',t') \equiv (i\hbar)^{-1}  \langle \phi_0 | \hat{T} [\hat{\Psi}(\RR,t) \hat{\Psi}^\dagger(\RR',t')] |\phi_0\rangle
\ee
where the state vector $|\phi_0\rangle$ is the ground state of $N$ fermions in the absence of impurity
(a simple Fermi sea), $\hat{\Psi}(\RR,t)$ is the impurity field operator at position $\RR$ and time $t$ in the
Heisenberg picture, and the operator $\hat{T}$, called T-product, orders the factors in the chronological
order from right to left, with the multiplication by the sign of the corresponding permutation if the field $\hat{\Psi}$
is fermionic. Since there is only one impurity, it is clear that its quantum statistic is irrelevant and that the
Green's function is zero for $t<t'$, so that it is both a Feynman and a retarded Green's function:
\be
\mathcal{G}(\RR,t;\RR',t') = (i\hbar)^{-1} Y(t-t') \langle \phi_0 |  \hat{\Psi}(\RR,t) \hat{\Psi}^\dagger(\RR',t') |\phi_0\rangle
\ee
with $Y$ the usual Heaviside step function. As the second member depends only on $\RR-\RR'$ and $t-t'$,
by virtue of spatial homogeneity and stationarity of the Fermi sea under free evolution, we take its spatiotemporal
Fourier transform\footnote{Our convention is that the spatiotemporal Fourier
transform of $f(x,t)$ defined on $\mathbb{R}^2$ is $\tilde{f}(k,\omega)=\int dt \int dx f(x,t) \exp[-i(kx -\omega t)]$.}
with respect to $\RR-\RR'$ and $t-t'$,
with 
the usual regularisation $\exp[-\epsilon (t-t')/\hbar]$, $\epsilon\to 0^+$, to obtain the propagator $\mathcal{G}(\KK,\omega)$.
By definition of the self-energy $\Sigma$, called proper self-energy in \cite{Fetter}, on one hand we have the Dyson equation
\be
\mathcal{G}(\KK,\omega) = [\hbar\omega +i\epsilon - E_\KK - \Sigma (\KK,\omega)  ]^{-1}
\label{eq:Gkov1}
\ee
with the impurity kinetic energy function,
\be
E_\KK = \frac{\hbar^2 K^2}{2M}
\label{eq:EK}
\ee
On the other hand, the evolution operator during $t$ of the full system of Hamiltonian $\hat{H}$ is
$\exp(-i \hat{H} t/\hbar)$, so that we obtain from an explicit calculation
\be
\mathcal{G}(\KK,\omega) = \langle \psi_\KK^0| \hat{G}(\hbar\omega +i \epsilon + e_0(N)) |\psi_\KK^0\rangle 
\ \ \ \mbox{where} \ \ \ |\psi_\KK^0\rangle = \hat{C}^\dagger_\KK |\phi_0 \rangle
\label{eq:Gkov2}
\ee
with $e_0(N)$ the ground state energy of the unperturbed $N$ fermions, $\hat{C}^\dagger_\KK$ the creation
operator of one impurity with wave vector $\KK$ and $\hat{G}(z)\equiv (z\hat{1}-\hat{H})^{-1}$ the resolvent
operator of the full Hamiltonian $\hat{H}$.

The link established by equations (\ref{eq:Gkov1},\ref{eq:Gkov2}) between the self-energy and the resolvent
is a link between two worlds, the $N$-body problem in condensed matter physics and the one of atomic physics,
where we rather speak about effective Hamiltonians and complex energy shifts. This link is made explicit by the
method of projectors \cite{CCT}. Let $\hat{P}$ be the orthogonal projector on $|\psi_\KK^0\rangle$, that is on the 
unperturbed state of the impurity with wave vector $\KK$ and the Fermi sea.
In the corresponding subspace of dimension one we define the non-hermitian effective Hamiltonian, that parametrically
depends, as the resolvent,
on a complex energy $z$, by the general exact relations   
\be
\hat{P} \hat{G}(z) \hat{P} = \frac{\hat{P}}{z \hat{P} - \hat{H}_{\rm eff}(z)} \ \mbox{and}\ \hat{H}_{\rm eff}(z)
\equiv\hat{P}\hat{H}\hat{P} +\hat{P}\hat{H} \hat{Q} \frac{\hat{Q}}{z\hat{Q}-\hat{Q}\hat{H}\hat{Q}}\hat{Q}\hat{H}\hat{P}
\label{eq:PGHetHeff}
\ee
where $\hat{Q}=\hat{1}-\hat{P}$ is the complementary projector to $\hat{P}$.
Here, $\hat{H}=\hat{H}_0+\hat{V}$, where $\hat{H}_0$, the kinetic Hamiltonian of the particles, commutes
with $\hat{P}$, and $\hat{V}$ is the impurity-fermion interaction Hamiltonian. We finally obtain an explicit 
operator-expression of the self-energy, in terms of the displacement operator $\hat{R}$ of reference \cite{CCT} readily
replaced here by its definition:
\be
\Sigma(\KK,\omega) = \langle \psi_\KK^0| \hat{V}|\psi_\KK^0\rangle +
\langle \psi_\KK^0 | \hat{V}\hat{Q}  \frac{\hat{Q}}{\hbar\omega+i\epsilon + e_0(N) -\hat{Q}\hat{H}\hat{Q}}\hat{Q}\hat{V} |\psi_\KK^0\rangle
\label{eq:sigma_expli}
\ee
This will allow us to expand $\Sigma(\KK,\omega)$ in powers of $\hat{V}$ without using
a diagrammatic representation.

\section{Expression of $\Sigma(\KK,\omega)$ up to second order in $g$ as a single integral}
\label{sec:calcul_sigma2}

\subsection{The lattice model and the result as a multiple integral}

To describe a zero-range interaction of fixed scattering length $a$ between the impurity and a fermion, it is not possible in three
dimensions to directly take the usual Dirac delta model, $V_\delta=g \delta(\RR-\rr)$, of effective coupling constant
\be
g=\frac{2\pi \hbar^2 a}{\mu},
\label{eq:lienga}
\ee
$\mu=m M/(m+M)$ being the reduced mass, except for a treatment limited to the Born approximation. Typically one makes this model
meaningful by introducing a cutoff on the relative wave vectors of the two colliding particles (not on the wave vectors of each
particle \cite{WernerGenLong}), and then one takes the infinite cutoff limit \cite{these_Giraud}. However, we shall adopt here a more
physical approach: We replace the Dirac delta by the Kronecker symbol, the latter also noted $\delta$, that is we use
the cubic lattice model described in detail in reference \cite{modrescours}, the space being discretised along 
each Cartesian direction with a lattice spacing $b$, submultiple of the period $L$ setting the periodic boundary conditions.     
The wave vectors of the particles then have a meaning modulo $2\pi/b$ in each direction, which allows us to restrict them to the first Brillouin zone
of the lattice, $\mathrm{FBZ}=[-\pi/b,\pi/b[^3$, and which provides a natural cutoff; thus the wave vectors span the set $\mathcal{D}=\mathrm{FBZ}\cap (2\pi/L)\mathbb{Z}^3$.
The full Hamiltonian $\hat{H}$ is the sum of the kinetic energy of the particles $\hat{H}_0$ and of the on-site interaction $\hat{V}$.
On one hand,   
\be
\hat{H}_0= \sum_{\kk\in \mathcal{D}} \epsilon_\kk \hat{c}_\kk^\dagger \hat{c}_\kk +\sum_{\KK\in\mathcal{D}} E_\KK \hat{C}_\KK^\dagger \hat{C}_\KK
\ee
where the kinetic energy of a fermion of wave vector $\kk$,
\be
\epsilon_\kk=\frac{\hbar^2 k^2}{2m}
\ee
and the annihilation operator $\hat{c}_\kk$ of a fermion, subject to the canonical anticommutation relations 
$\{ \hat{c}_\kk,\hat{c}_{\kk'}^\dagger\}= \delta_{\kk,\kk'}$, are the fermionic counterpart of the energy $E_\KK$ and of the
operator $\hat{C}_\KK$ introduced for the impurity in the previous section.
On the other hand,
\be
\hat{V}= \!\!\!\!\sum_{\rr\in [0,L[^3\cap b\mathbb{Z}^3}\!\!\!\! b^3 g_0\hat{\Psi}^\dagger(\rr) \hat{\psi}^\dagger(\rr) \hat{\psi}(\rr) \hat{\Psi}(\rr)
=\!\!\!\!\!\!\sum_{\kk,\kk',\KK,\KK'\in\mathcal{D}}\!\! \frac{b^3 g_0}{L^3} \delta^{\rm mod}_{\kk+\KK,\kk'+\KK'} 
\hat{C}_{\KK'}^\dagger \hat{c}_{\kk'}^\dagger
\hat{c}_\kk \hat{C}_\KK
\label{eq:defV}
\ee
where the field operator $\hat{\psi}(\rr)$, such that $\{\hat{\psi}(\rr),\hat{\psi}^\dagger(\rr')\}=\delta_{\rr,\rr'}/b^3$, is the fermionic counterpart
of the impurity field $\hat{\Psi}(\RR)$, $\delta^{\rm mod}$ is a Kronecker $\delta$ modulo a vector of the reciprocal lattice $(2\pi/b)\mathbb{Z}^3$,
and the bare coupling constant $g_0$ is adjusted so that the scattering length of the impurity on a fermion, defined of course for the infinite
lattice ($L=\infty$), has the arbitrary desired value in $\mathbb{R}$ \cite{modrescours}: 
\be
g_0^{-1} = g^{-1} - \int_{\mathrm{FBZ}} \frac{d^3k}{(2\pi)^3} \frac{2\mu}{\hbar^2 k^2}
\ee

Let us determine the self-energy $\Sigma(\KK,\omega)$ perturbatively up to second order included in $g$, for $g<0$, as explained
in the introduction. For a {\sl fixed} lattice spacing $b$, we shall make $g$ tend to zero from negative values. Thus $g_0$ tends to zero,
\be
g_0 \stackrel{a/b\to 0}{=} g + g^2 \int_{\mathrm{FBZ}} \frac{d^3k}{(2\pi)^3} \frac{2\mu}{\hbar^2 k^2} + O(g^3)
\label{eq:g0_dev}
\ee
In the relation (\ref{eq:sigma_expli}), at this order we can neglect $\hat{V}$ in the denominator. The action of $\hat{V}$ on the 
non-perturbed state $|\psi_\KK^0\rangle$ creates a hole of wave vector $\qq$ in the Fermi sea by promoting a fermion to the 
wave vector $\kk$; the impurity takes the momentum change and acquires the wave vector $\KK-\kk +\qq$ (modulo a vector of the reciprocal lattice).
In the obtained expression of $\Sigma(\KK,\omega)$, we replace $g_0$ with its expansion (\ref{eq:g0_dev}) {\sl then} we take the
continuous space limit $b\to 0$ for a fixed $g$.    
It remains to take the thermodynamic limit to obtain the exact perturbative expansion: 
\be
\label{eq:devsig}
\Sigma(\KK,\omega) = \Sigma^{(1)}(\KK,\omega) + \Sigma^{(2)}(\KK,\omega) + O(g^3)
\ee
with, up to second order:
\bea
\Sigma^{(1)}(\KK,\omega) &=& \rho g \\
\Sigma^{(2)}(\KK,\omega) &=& g^2\!\! \int_{q<\kf}\!\! \frac{d^3q}{(2\pi)^3}\! \int_{\mathbb{R}^3}\!\! \frac{d^3k}{(2\pi)^3} \!\!
\left[\frac{2\mu}{\hbar^2 k^2}-\frac{Y(k-\kf)}{F_{\kk,\qq}(\KK,\omega)}\right]
\label{eq:sig2}
\eea
This writing was made more compact as in \cite{lettre} thanks to the notation
\be
F_{\kk,\qq}(\KK,\omega) \equiv E_{\KK-\kk+\qq} +\epsilon_\kk -\epsilon_\qq -\hbar\omega -i\epsilon
\label{eq:defF}
\ee
Unsurprisingly, the contribution of order one reduces to the mean-field shift, which involves the average 
density of fermions $\rho$ or their Fermi wave number $\kf$: 
\be
\rho=\int_{q<\kf} \frac{d^3q}{(2\pi)^3} = \frac{\kf^3}{6\pi^2}
\ee
Remarkably, hereafter we will show that the sextuple integral in the second order contribution can be evaluated analytically in an explicit way. 

As a side remark, one may further expand $g_0$ and $\Sigma(\KK,\omega)$
in powers of $g$, at the cost of obtaining integrals that may be difficult to calculate analytically.
We give here as an example the result at order three:
\bea
\nonumber
\Sigma^{(3)}(\KK,\omega) &=& g^3 \!\! \int_{q<\kf}\!\! \frac{d^3q}{(2\pi)^3}\! \left\{
\int_{\mathbb{R}^3}\!\! \frac{d^3k}{(2\pi)^3} \!\! \left[\frac{2\mu}{\hbar^2 k^2}-\frac{Y(k-\kf)}{F_{\kk,\qq}(\KK,\omega)}\right]\right\}^2\\
&-&g^3 \int_{\mathbb{R}^3}\!\! \frac{d^3k}{(2\pi)^3} \!\! \left[\int_{q<\kf}\!\! \frac{d^3q}{(2\pi)^3} \frac{Y(k-\kf)}{F_{\kk,\qq}(\KK,\omega)}
\right]^2 -\frac{\rho g}{\hbar} \partial_\omega \Sigma^{(2)}(\KK,\omega)
\label{eq:sig3}
\eea
resulting from a triple action of $\hat{V}$ on the non-perturbed state $|\psi_\KK^0\rangle$, with forced return to $|\psi_\KK^0\rangle$.
The first action of $\hat{V}$ creates a particle-hole pair of wave vectors $\kk$ and $\qq$. The second action of $\hat{V}$ can bring neither back
to the initial state [due to the projector $\hat{Q}$ into equation (\ref{eq:sigma_expli})], nor forth to the state with two particle-hole pairs
(since the third action of $\hat{V}$ cannot destroy two pairs). It then (i) scatters the excited fermion from $\kk$ to $\kk'$ with an amplitude
$g_0$, or (ii) it scatters the hole from $\qq$ to $\qq'$ with an amplitude $-g_0$, by collision with the impurity, or (iii) it does not change anything 
at all [term $\kk'=\kk$ and $\KK'=\KK$ in equation (\ref{eq:defV})]. This gives rise respectively to the first, the second and the third
term of equation (\ref{eq:sig3}); the integrals over $\kk$ and $\kk'$, or over $\qq$ and $\qq'$, which have a symmetric integrand with
respect to the exchange of wave vectors, lead to a square of an integral over $\kk$, or over $\qq$.

\subsection{From a sextuple integral to a single integral for $\Sigma^{(2)}(\KK,\omega)$}
\label{subsec:pisis}

We detail here, step by step, the reduction of the multidimensional integral giving $\Sigma^{(2)}(\KK,\omega)$ in 
equation (\ref{eq:sig2}). 

First of all it is convenient to use dimensionless quantities, by expressing wave vectors in units of the Fermi wave number,
the energy difference between $\hbar \omega$ and the impurity kinetic energy $E_\KK$
in units of the Fermi energy $\Ef\equiv\hbar^2\kf^2/(2m)$ of the fermions,   
\be
\label{eq:adim1}
\bar{\KK} \equiv \frac{\KK}{\kf}, \ \ \bar{\qq}\equiv\frac{\qq}{\kf}, \ \ \bar{\kk}\equiv\frac{\kk}{\kf},
\ \ \varepsilon\equiv \frac{E_\KK-\hbar\omega}{\Ef}
\ee
and the $g^2$-component of the self-energy in units of $(\rho g)^2/\Ef$:
\be
\label{eq:sig2adim}
\Sigma^{(2)}(\KK,\omega) \equiv \frac{(\rho g)^2}{\Ef} \bar{\Sigma}^{(2)}(\bar{K},\varepsilon)
\ee
Regarding the impurity mass, we express it in units of the mass of a fermion by the dimensionless number
\be
r\equiv \frac{M}{m}
\ee
Then, the rotational invariance of $\KK\mapsto \Sigma^{(2)}(\KK,\omega)$, already taken into account in the writing (\ref{eq:sig2adim}),
allows us to average equation (\ref{eq:sig2}) on the direction $\hat{\KK}$ of the impurity wave vector. At fixed $\kk$ and $\qq$, the
expansion of $F_{\kk,\qq}(\KK,\omega)$ in powers of $\KK$ leads us to introduce spherical coordinates of polar axis given by
the direction of $\kk-\qq$; then the integrand depends only on the cosine $w$ of the polar angle of $\hat{\KK}$, so that:
\bea
\label{eq:moychapK}
\langle \frac{\hbar^2\kf^2/(2M)}{F_{\kk,\qq}(\KK,\omega)}\rangle_{\hat{\KK}} &=& \!\!\!\!\int_{-1}^1 \frac{dw}{2} \frac{1}{x-yw-i\epsilon} \\
&\stackrel{\epsilon\to 0^+}{\to}& \!\frac{1}{2y} \ln\frac{|x+y|}{|x-y|}\! +\!\frac{i\pi}{2y}[Y(x+y)\!-\!Y(x-y)] \equiv f(x,y)
\label{eq:deff}
\eea
where 
\be
\lambda\equiv |\bar{\kk}-\bar{\qq}|, \ \ x\equiv \lambda^2+r(\bar{k}^2-\bar{q}^2+\varepsilon)\in\mathbb{R}, \ \ y\equiv 2\bar{K} \lambda\in \mathbb{R}^+ 
\label{eq:deflamxy}
\ee
and we used the primitive $\frac{1}{2} \ln (v^2 +\epsilon^2) + i\arctan (v/\epsilon)$ of the function 
$v\mapsto (v-i\epsilon)^{-1}$ on $\mathbb{R}$.

Similarly, in the integration over $\kk$ at fixed $\qq$ we choose the polar axis of direction $\qq$, so that 
the integrand depends only on the polar angle $\theta$ between $\kk$ and $\qq$, not on the azimuthal angle.
In the polar integral, we use the variable $\lambda$ of equation (\ref{eq:deflamxy}) rather than $\theta$
itself, with
\be
\sin \theta \, d\theta = \frac{\lambda d\lambda}{\bar{k}\bar{q}}
\ee
Finally, in the integral over $\qq$, that is the most external one, the integrand does not depend anymore on the direction
of $\qq$, which brings out the usual $4\pi$ solid angle factor. At this point we are easily reduced to a triple integral:
\be
\bar{\Sigma}^{(2)}(\bar{K},\omega)=\frac{9r}{2}\!\int_0^1\!\! \bar{q} d\bar{q} \!\int_0^{+\infty}\!\!\!\!\!\bar{k} d\bar{k} \!\int_{|\bar{k}-\bar{q}|}^{\bar{k}+\bar{q}} \!\!\lambda d\lambda
\left[\frac{1}{(1+r)\bar{k}^2}-Y(\bar{k}-1) f(x,y)\right]
\ee
where the function $f$ is the one of equation (\ref{eq:deff}).
The integration over $\lambda$, although feasible, is tricky since $\lambda$ appears in $f(x,y)$ under a trinomial form, and the complexity
of the result compromises further integration; instead, $\bar{k}$ and $\bar{q}$ appear only by their square. Hence the idea to reverse the order
of integration as in reference \cite{polzg2}: We perform separately the complete integration over the domain $\bar{k}<1$, and 
otherwise we use 
\be
\int_0^1 d \bar{q} \int_1^{+\infty} d\bar{k} \int_{|\bar{k}-\bar{q}|}^{\bar{k}+\bar{q}} d\lambda = 
\int_0^{+\infty} d\lambda \int_{\max(1-\lambda,0)}^1 d\bar{q} \int_{\max(\lambda-\bar{q},1)}^{\lambda+\bar{q}} d\bar{k}
\ee
All this leads to
\be
\bar{\Sigma}^{(2)}(\bar{K},\varepsilon) = \frac{3r}{1+r} -\frac{9r}{2} \int_0^{+\infty} d\lambda
\left[\frac{\psi^+(\lambda)-\psi^-(\lambda)}{4\bar{K}}-\chi(\lambda)\right]
\label{eq:sig2opera}
\ee
in terms of the auxiliary functions
\bea
\label{eq:defpsipm}
\psi^{\pm}(\lambda) &\equiv& \int_{\max(1-\lambda,0)}^1 \bar{q}d\bar{q} \int_{\max(\lambda-\bar{q},1)}^{\lambda+\bar{q}}
\bar{k}d\bar{k}\, u[\lambda^2\pm 2\bar{K}\lambda +r(\bar{k}^2-\bar{q}^2+\varepsilon)]\\
\label{eq:defchi}
\chi(\lambda) &\equiv& \int_{\max(1-\lambda,0)}^1 \bar{q}d\bar{q} \int_{\max(\lambda-\bar{q},1)}^{\lambda+\bar{q}}
\bar{k}d\bar{k} \frac{\lambda}{(1+r) \bar{k}^2}
\eea
where $\psi^-(\lambda)$ is obtained from $\psi^+(\lambda)$ simply by changing $\bar{K}$ to $-\bar{K}$, and the useful function
\be
\label{eq:defu}
u(X) \equiv \ln |X| + i\pi Y(X),
\ee
naturally introduced by the property $2y f(x,y)=u(x+y)-u(x-y)$, will also intervene 
{\sl via} its primitives $u^{[n]}(X)$ of order $n$, which vanish at zero as well as all their derivatives up to order $n-1$:
\be
\label{eq:un}
u^{[n]}(X) =\frac{X^n}{n!} \left[u(X)-\sum_{s=1}^{n} \frac{1}{s}\right]
\ee
$u(X)$ is actually the limit of the usual branch of the complex logarithm $\ln z$
when $z$ tends to $-X \in \mathbb{R}$ from the upper half of the complex plane. 

Let us outline how to compute $\psi^\pm(\lambda)$. The integration over $\bar{k}$ is trivial provided we take $\bar{k}^2$ as the integration variable.
It directly leads to the primitive $u^{[1]}$, evaluated at points of the form $A \bar{q}^2+B$ or $A \bar{q}+B$, where the coefficients 
$A$ and $B$ do not depend on $\bar{q}$. The integration over $\bar{q}$ is then either of the form $\int d\bar{q}\, \bar{q}\, u^{[1]}(A \bar{q}^2 +B)$,
in which case we take $\bar{q}^2$ as the integration variable and $u^{[2]}$ appears, or of the form $\int d\bar{q} \,\bar{q}\, u^{[1]}(A \bar{q} +B)$,
in which case we use the integration by parts (taking the derivative of the factor $\bar{q}$) that leads to $u^{[2]}$ and to $u^{[3]}$.
In practice, we are led to distinguish between the case (i) $0<\lambda <1$, of lower boundaries $1-\lambda$ and $1$ in the integrals
over $\bar{q}$ and over $\bar{k}$ respectively, 
(ii) $1< \lambda <2$, of lower boundaries $0$ for $\bar{q}$ and $\lambda-\bar{q}$ (or $1$) for $\bar{k}$
depending on $\bar{q}$ being lower (or higher) then $\lambda-1$, and (iii) $\lambda >2$, of lower boundaries $0$ and $\lambda-\bar{q}$.
However, we note that the first two cases lead to the same expressions\footnote{When $\lambda\in [1,2]$, after integration over $\bar{k}$
we obtain for $\psi^\pm(\lambda)$ an expression of the form $\int_0^{\lambda-1} d\bar{q} [f(\bar{q})+f(-\bar{q})] + 
\int_{\lambda-1}^1 d\bar{q} [f(\bar{q})-g(\bar{q})]$, which we transform by the change of variable $\bar{q}\to -\bar{q}$ in the 
parts containing $f(-\bar{q})$ and $g(\bar{q})$. After collecting the different pieces and using the odd parity of $g(\bar{q})$,
which implies that $\int_{-1}^{+1} d\bar{q}\, g(\bar{q})=0$, we end up with the expression $\int_{1-\lambda}^1 d\bar{q} [f(\bar{q})-g(\bar{q})]$
which is exactly the same one as $\psi^\pm(\lambda)$ over $[0,1]$.}, so that it is sufficient to distinguish the interval $[0,2]$, over which
\be
\psi^\pm(\lambda) = \frac{u^{[2]}[P^\pm_\gamma(\lambda)]}{4r^2} + \frac{u^{[2]}[P^\pm_\alpha(\lambda)]\!-\!u^{[2]}[P^\pm_\beta(\lambda)]}{4r^2 \lambda} 
- \frac{u^{[3]}[P^\pm_\alpha(\lambda)]\!-\!u^{[3]}[P^\pm_\beta(\lambda)]}{8r^3\lambda^2}
\label{eq:forme_inf}
\ee
and the interval $[2, +\infty[$, over which
\be
\psi^\pm(\lambda) = \frac{u^{[2]}[P^\pm_\alpha(\lambda)]+u^{[2]}[P^\pm_\delta(\lambda)]}{4r^2\lambda}
-\frac{u^{[3]}[P^\pm_\alpha(\lambda)]-u^{[3]}[P^\pm_\delta(\lambda)]}{8r^3 \lambda^2}
\label{eq:forme_sup}
\ee
We have introduced here the trinomials appearing in the expression of $\psi^\pm(\lambda)$:
\bea
\label{eq:pa}
P^\pm_{\alpha}(\lambda) &=& (1+r) \lambda^2 +2 (r\pm\bar{K}) \lambda + r\varepsilon \\
\label{eq:pb}
P^\pm_{\beta}(\lambda) &=& (1-r) \lambda^2 +2 (r\pm\bar{K}) \lambda + r\varepsilon \\
\label{eq:pc}
P^\pm_{\gamma}(\lambda) &=& \lambda^2 \pm 2 \bar{K}\lambda + r\varepsilon \\
\label{eq:pd}
P^\pm_{\delta}(\lambda) &=& (1+r) \lambda^2 +2 (\pm\bar{K}-r) \lambda + r\varepsilon
\eea
The four trinomials corresponding to the function $\psi^-(\lambda)$ 
are deduced of course from those associated to $\psi^+(\lambda)$ by changing $\bar{K}$ to $-\bar{K}$. 
They obey the duality relations that we use later in this paper:
\be
\label{eq:dualite}
P^-_\alpha(\lambda)= P^+_\delta(-\lambda) \ \ \mbox{and}\ \ \ P^-_\delta(\lambda)=P^+_\alpha(-\lambda)
\ee
Notice that $P^\pm_{\beta}(\lambda)$ is actually of degree one in the special case ($r=1$) where the impurity and the fermions
have the same mass, $M=m$. 

As for the integral (\ref{eq:defchi}), an elementary calculation leads to\footnote{This allows us to verify that the integrand of 
(\ref{eq:sig2opera}) is $O(1/\lambda^2)$ and that the integral converges at large $\lambda$.}
\bea
\label{eq:chi_inf}
\chi(\lambda) = \frac{\lambda}{2(1+r)} \left[(1-\lambda^2)\ln(1+\lambda)+\lambda\left(\frac{3}{2}\lambda-1\right)\right] && \ \ \forall \lambda\in [0,2] \\
\label{eq:chi_sup}
\chi(\lambda) = \frac{\lambda}{2(1+r)} \left[(1-\lambda^2) \ln \frac{\lambda+1}{\lambda-1} + 2\lambda\right]  && \ \ \forall \lambda\in [2,+\infty[
\eea

\section{Explicit calculation of $\Sigma^{(2)}(\KK,\omega)$ in the general case}
\label{sec:cedsigdlcg}

In the previous section we expressed the contribution of order $g^2$ to the impurity self-energy 
as a single integral, see the integral of $\psi^+(\lambda)-\psi^-(\lambda)$ over $\lambda$ in equation (\ref{eq:defpsipm}),
in which we now temporarily introduce a finite upper bound $\Lambda>2$. The evaluation of this integral can be done
explicitly. Let us give here the main steps. 

\subsection{Expression in terms of two functionals $I[P]$ and $J[P]$}
\label{subsec:eetddf}

The first step consists in reducing the number of types of terms in the integrand. In equations (\ref{eq:forme_inf}) and (\ref{eq:forme_sup})
there are {\sl a priori} three distinct types, according to the power $0$, $1$ or $2$ of $\lambda$ in the denominator. However, it suffices
to integrate the terms of the third type by parts (integrating the factor $1/\lambda^2$), to transform them into terms of the first two types; 
this holds over each interval of integration $[0,2]$ and $[2,\Lambda]$. Also, the all-integrated terms in $\lambda=2$ cancel exactly since  
$P^\pm_\beta(2)=P^\pm_\delta(2)$, and the all-integrated term in $\lambda=0$ is zero since in addition 
$\frac{d}{d\lambda} P_\alpha^\pm (0) = \frac{d}{d\lambda} P_\beta^\pm (0)$. Writing the integrals over $[2,\Lambda]$ as the difference 
of the integrals over $[0,\Lambda]$ and over $[0,2]$ with the same integrand, we are reduced to zero lower integration bounds, and finally
to the only two functionals 
\bea
\label{eq:defI}
I[P](\lambda) &\equiv& \int_0^\lambda dt \, u^{[2]}[P(t)] \\
\label{eq:defJ}
J[P](\lambda) &\equiv& \int_0^\lambda dt \, \frac{u^{[2]}[P(t)]-u^{[2]}[P(0)]}{t}
\eea
where the polynomial $P(\lambda)$ is in practice one of the eight trinomials $P_\sigma^\pm(\lambda)$, $\sigma\in \{\alpha,\beta,
\gamma,\delta\}$.
The term subtracted in the numerator of the integrand in $J[P]$ assures the convergence of the integral 
in its lower bound without introducing any remainder in the final result since all the trinomials have the
same value in zero. We finally get
\bea
\int_0^\Lambda d\lambda\, \psi^{\pm}(\lambda) = C^\pm(\Lambda) 
-\frac{(1+r)\{I^\pm_\alpha(\Lambda)-I^\pm_\delta(\Lambda)\} 
\pm\bar{K}\{J^\pm_\alpha(\Lambda)-J^\pm_\delta(\Lambda)\}}{4 r^3} \nonumber\\
+ \frac{I^\pm_\gamma(2)-I^\pm_\beta(2)-I^\pm_\delta(2)}{4r^2}
+\frac{I^\pm_\beta(2)-I^\pm_\delta(2)\pm\bar{K}\{J^\pm_\beta(2)-J^\pm_\delta(2)\}}{4 r^3}
\label{eq:avec_coupure}
\eea
with the shorthand notations
\be
\label{eq:raccourcis}
I^\pm_\sigma(\lambda)\equiv I[P^\pm_\sigma](\lambda) \ \ \ \mbox{and}\ \ \ 
J^\pm_\sigma(\lambda)\equiv J[P^\pm_\sigma](\lambda)
\ee
for the functionals $I$ and $J$ evaluated at the eight trinomials. The contribution of the all-integrated term
in $\Lambda$ resulting from the integration by parts is
\be
C^{\pm}(\Lambda) = \frac{u^{[3]}[P^\pm_\alpha(\Lambda)]-u^{[3]}[P^\pm_\delta(\Lambda)]}{8 r^3 \Lambda}
\ee

\subsection{Explicit value of the functionals}

The second step consists in calculating the functionals $I[P]$ and $J[P]$, where $P$ is a polynomial.
We usefully write it in its factorised form,
\be
\label{eq:decomp}
P(\lambda) = A_P \prod_{\lambda_0\in \mathrm{Root}\, P} (\lambda-\lambda_0)
\ee
where $A_P$ is the leading coefficient and $\mathrm{Root}\, P$ is the set of roots of $P(\lambda)$ counted with
their multiplicity. According to equations (\ref{eq:pa},\ref{eq:pb},\ref{eq:pc},\ref{eq:pd}), it is sufficient here to restrict to the 
polynomials $P(\lambda)$ of degree at most two with real coefficients, which includes the special case $P_\beta^{\pm}(\lambda)$
(of degree one) for a mass ratio $r=1$. Then we need to consider two distinct cases, the one where all roots of $P(\lambda)$ are
real, and the one where the two roots are complex conjugate. 

Let us first tackle with the {\bf evaluation of the functional} $I[P]$. In order to make notations more compact and the result reusable 
in the calculation of $J[P]$, we introduce the auxiliary polynomial $Q_I$ of $P$ for the functional $I[P]$:
\be
\label{eq:defQI}
Q_I(\lambda) = \frac{1}{2} [P(\lambda)]^2
\ee
Thus, taking into account (\ref{eq:un}) and (\ref{eq:defI}), we get the useful rewriting
\be
\label{eq:Iut}
I[P](\lambda)=\int_0^\lambda dt\, Q_I(t) \left[\ln |P(t)| + i\pi Y[P(t)] -\frac{3}{2}\right]
\ee
In this integral, the contribution of the logarithm is obtained by simply proposing a primitive $\mathcal{F}(t)$ of the function
$t\mapsto Q_I(t) \ln |P(t)|$, which the reader may check by calculating the derivative $\mathcal{F}'(t)$.
If $P$ has real roots we choose
\bea
\label{eq:Fr}
\mathcal{F}(t) &=& Q_I^{[1]}(t)\left[\ln |A_P|-\frac{3}{2}\right] + 
\sum_{\lambda_0\in \mathrm{Root}\, P} \Big\{\left[Q_I^{[1]}(t)-Q_I^{[1]}(\lambda_0)\right] \ln |t-\lambda_0| - \int_0^t d\tau \frac{Q_I^{[1]}(\tau)-Q_I^{[1]}(\lambda_0)}{\tau-\lambda_0}\Big\}
\eea
where the polynomial $Q_I^{[1]}(t)$ is the primitive of the polynomial $Q_I(t)$ vanishing in $t=0$, in agreement with the
notation introduced previously for the function $u$. The only potential singularities of the function $\mathcal{F}(t)$, located
in the values of the roots $\lambda_0$, come from the second contribution, which remains, however, continuous since
the polynomial prefactor of $\ln |t-\lambda_0|$ vanishes in $t=\lambda_0$; the other contributions are polynomials in $t$.
If $P$ has complex roots, we take instead the primitive 
\bea
\label{eq:Fi}
\mathcal{F}(t) &=& Q_I^{[1]}(t) \left[\ln |P(t)|-\frac{3}{2}\right]+
\sum_{\lambda_0\in \mathrm{Root}\, P} \Big\{ Q_I^{[1]}(\lambda_0)[\ln (-\lambda_0) - \ln (t-\lambda_0)] - \int_0^t d\tau \frac{Q_I^{[1]}(\tau)-Q_I^{[1]}(\lambda_0)}{\tau-\lambda_0}\Big\}
\eea
where the logarithm $\ln z$ in the complex plane is defined with its principal branch $\Im \ln z\in ]-\pi,\pi[$
corresponding to a branch cut on the real negative half-axis. The function $\mathcal{F}(t)$ is smooth (infinitely differentiable)
on the real axis. As it vanishes in $t=0$ and as its derivative is real, since $Q_I$ has real coefficients, it is real-valued.
Moreover, we notice that it can be deduced from (\ref{eq:Fr}) up to an additive constant by changing 
$|t-\lambda_0|$ to $(t-\lambda_0)$ in the logarithm.

Let us now consider the contribution of the Heaviside function to the integral (\ref{eq:Iut}).
If $P$ has complex roots, we can replace $Y[P(t)]$ with $Y[P(\lambda)]$ in the integrand, since $P(t)$
has a constant sign on the real axis; the contribution $i\pi Y[P(\lambda)] Q_I^{[1]}(\lambda)$ appears
after integration.
If $P$ has real roots, we integrate by parts according to the theory of distributions, by integrating the polynomial factor
$Q_I(t)$ and by taking the derivative of the factor containing the Heaviside function:    
\be
\label{eq:derivY}
\frac{d}{dt} Y[P(t)] = P'(t) \delta[P(t)] = \sum_{\lambda_0\in \mathrm{Root}\, P} 
\frac{P'(\lambda_0)}{|P'(\lambda_0)|} \delta(t-\lambda_0)
\ee
according to the well-known properties of $Y$ and of the Dirac $\delta$ distribution.
The all-integrated term contains the already mentioned contribution $i\pi Y[P(\lambda)] Q_I^{[1]}(\lambda)$,
and the remaining integral is elementary, given that $\int_0^\lambda dt\, \delta(t-\lambda_0)=
Y(\lambda-\lambda_0) -Y(-\lambda_0)$. Note that the prefactor of $\delta$ in (\ref{eq:derivY}) is a pure sign,
which is the one of the leading coefficient $A_P$ for the largest of the roots and its opposite for the smallest 
of the roots\footnote{In the case where $P$ has actually a double root, which can be seen as the convergence of two single roots
towards a common value $\lambda_0$, the contributions of the two roots to the remaining integral cancel out and only the
all-integrated term survives.}. 

It remains to give the {\bf final expression} of $I[P](\lambda)$ on the real axis, valid, let us stress it, for $P$
of degree at most two with real coefficients, but for any value of the polynomial $Q_I$, not at all limited to (\ref{eq:defQI}),
as it is apparent in the description of our calculations. If $P$ has real roots,   
\bea
\label{eq:IPrr}
I[P](\lambda) = Q_I^{[1]}(\lambda) \left\{u[P(\lambda)]-\frac{3}{2}\right\} 
+\sum_{\lambda_0\in \mathrm{Root}\, P} \left\{ -\int_0^\lambda dt \frac{Q_I^{[1]}(t)-Q_I^{[1]}(\lambda_0)}{t-\lambda_0}
\right.\nonumber \\
\left. 
-Q_I^{[1]}(\lambda_0) \left[
\ln|\lambda-\lambda_0| -\ln |\lambda_0| + i\pi \frac{P'(\lambda_0)}{|P'(\lambda_0)|}
[Y(\lambda-\lambda_0)-Y(-\lambda_0)]
\right]
\right\} 
\eea
If $P$ has complex roots,
\bea
\label{eq:IPri}
I[P](\lambda) = Q_I^{[1]}(\lambda) \left\{u[P(\lambda)]-\frac{3}{2}\right\}  
+\sum_{\lambda_0\in \mathrm{Root}\, P} \left\{ -\int_0^\lambda dt \frac{Q_I^{[1]}(t)-Q_I^{[1]}(\lambda_0)}{t-\lambda_0}
\right.\nonumber \\
\left. 
-Q_I^{[1]}(\lambda_0) \left[
\ln(\lambda-\lambda_0) -\ln (-\lambda_0)
\right]
\right\}
\eea
bearing in mind that if $Q_I(\lambda)$ has real coefficients the imaginary part of $I[P](\lambda)$ originates only
from the one of $u[P(\lambda)]$, therefore from the first term.

Let us now tackle with the {\bf evaluation of the functional} $J[P]$, under the same hypothesis of a polynomial $P(\lambda)$
of degree at most two with real coefficients. The auxiliary polynomial has to be defined as follows: 
\be
\label{eq:defQJ}
Q_J(\lambda) = \frac{Q_I(\lambda)-Q_I(0)}{\lambda} = \frac{[P(\lambda)]^2-[P(0)]^2}{2\lambda}
\ee
to lead to the useful splitting in two sub-functionals,
\be
\label{eq:j1pj2}
J[P](\lambda) = J_1[P](\lambda) + \frac{[P(0)]^2}{2} J_2[P](\lambda),
\ee
with
\bea
\label{eq:defJ1}
J_1[P](\lambda) &=& \int_0^\lambda dt\, Q_J(t) \left\{u[P(t)]-\frac{3}{2}\right\}  \\
J_2[P](\lambda) &=& \int_0^\lambda dt\, \frac{u[P(t)]-u[P(0)]}{t}
\label{eq:defJ2}
\eea
As the expressions (\ref{eq:IPrr}) and (\ref{eq:IPri}) are valid for any polynomial $Q_I$, the functional $J_1[P](\lambda)$
is obtained by replacing $Q_I$ with $Q_J$, and thus $Q_I^{[1]}$ with $Q_J^{[1]}$. In the functional $J_2[P](\lambda)$,
the imaginary part of $u[P(t)]-u[P(0)]$ in the numerator of the integrand is zero if $P$ has complex roots since
$P$ has then a constant sign. Otherwise its contribution is evaluated by integrating by parts as for $I[P]$.
In the real part of $u[P(t)]-u[P(0)]$, one uses the factorization (\ref{eq:decomp}) for $P$; dividing by $t$, one obtains
the function $t\mapsto \ln[|t-\lambda_0|/|\lambda_0|]/t$ for each {\yvan real root $\lambda_0$ of $P(\lambda)$,
or $t\mapsto \ln[(t-\lambda_0)/\lambda_0]/t$ for each complex root}, whose integral is expressible
in terms of the dilogarithm $\mathrm{Li}_2$. Note that $\mathrm{Li}_2$ is also called polylogarithm of order two or Jonqui\`ere's function
of parameter equal to two, and it satisfies $\mathrm{Li}_2'(z)=-\ln(1-z)/z$ and $\mathrm{Li}_2(0)=0$.
If $P$ has real roots we finally find
\bea
J_2[P](\lambda) = i\pi \left\{ Y[P(\lambda)] -Y[P(0)] \right\} \ln |\lambda| + \sum_{\lambda_0\in \mathrm{Root}\, P}
\Big\{-\bar{\mathrm{Li}}_2(\lambda/\lambda_0)  \nonumber \\
-i\pi \ln |\lambda_0| \frac{P'(\lambda_0)}{|P'(\lambda_0)|}[Y(\lambda-\lambda_0)-Y(-\lambda_0)]\Big\}
\label{eq:J2Prr}
\eea
where the function $\bar{\mathrm{Li}}_2$ is real-valued on the real axis,
\be
\label{eq:defLi2bar}
\bar{\mathrm{Li}}_2(\lambda) = \lim_{\epsilon\to 0^+} \frac{\mathrm{Li}_2(\lambda+i\epsilon)+\mathrm{Li}_2(\lambda-i\epsilon)}{2}
=-\int_0^\lambda \frac{dt}{t} \ln|t-1|
\ee
it coincides with $\mathrm{Li}_2$ for $\lambda<1$ but gives for $\lambda>1$ the average of the values 
of $\mathrm{Li}_2$ just above and just below its branch cut $[1,+\infty[$.
If $P$ has complex roots, we obtain the real-valued result
\be
J_2[P](\lambda) = -\sum_{\lambda_0\in \mathrm{Root}\, P} \mathrm{Li}_2(\lambda/\lambda_0)
\label{eq:J2Pri}
\ee

\subsection{For an infinite cutoff}

The third step consists in taking the infinite cutoff limit, $\Lambda\to +\infty$.
The various terms depending on $\Lambda$ in (\ref{eq:avec_coupure}) and the integral 
of $\chi(\lambda)$ over $[0,\Lambda]$ in (\ref{eq:sig2opera}) diverge if they are considered 
individually. However, their divergent contributions have to cancel exactly in the final
result for $\bar{\Sigma}^{(2)}(\bar{K},\varepsilon)$, since the integral in (\ref{eq:sig2opera})
is convergent, and it is useless to evaluate them one by one.
To eliminate them in a simple but systematic way, let us write each term of (\ref{eq:avec_coupure})
asymptotically in the canonical form: 
\be
\mathcal{I}(\Lambda) \stackrel{\Lambda\to+\infty}{=}\sum_{(i,j)\in\mathbf{N}^2} a_{i,j} \Lambda^i
(\ln \Lambda)^j + o(1)
\label{eq:canon}
\ee
with a finite number of nonzero coefficients $a_{i,j}$. The unicity of writing of this form 
allows us to uniquely define the {\sl partie finie} of $\mathcal{I}(\Lambda)$ in $+\infty$:
\be
\mathrm{Pf}\, \mathcal{I}(+\infty) \equiv a_{0,0} =\lim_{\Lambda\to+\infty} \Big[
\mathcal{I}(\Lambda) - \sum_{(i,j)\in\mathbf{N}^{2*}} a_{i,j} \Lambda^i (\ln \Lambda)^j
\Big]
\ee
An explicit calculation of the integral of $\chi(\lambda)$ over $[0,2]$ and over $[2,\Lambda]$,
starting from (\ref{eq:chi_inf},\ref{eq:chi_sup}) and using the integration by parts to eliminate 
the logarithm, leads to only one divergent term, linear in $\Lambda$ thus of the form (\ref{eq:canon}),
and finally to
\be
\mathrm{Pf}\, \int_0^{+\infty} d\lambda\, \chi(\lambda) = -\frac{2}{3(1+r)}
\ee
This, by the way, cancels exactly the constant term $\frac{3r}{1+r}$ in equation (\ref{eq:sig2opera}).
Also, it is clear from equations (\ref{eq:IPrr}) and (\ref{eq:IPri}) that the functional $I[P](\Lambda)$ obeys
the form (\ref{eq:canon}): $u[P(\Lambda)]$ and $\ln|\Lambda-\lambda_0|$ or $\ln(\Lambda-\lambda_0)$
give contributions in $\ln \Lambda + O(1)$, and the other factors or terms give purely polynomial
divergent contributions. The calculation of its {\sl partie finie} is then trivial if we realize that, for all polynomials
$R(\lambda)=\sum_{n\geq 1} b_n \lambda^n$ vanishing in zero thus of zero {\sl partie finie} in $+\infty$,
such as the polynomial $Q_I(\lambda)$ or the contribution of $\int_0^\lambda$ in (\ref{eq:IPrr}) and (\ref{eq:IPri}),
we have
\be
\label{eq:interm}
\mathrm{Pf}\, [R(\Lambda) \ln (\Lambda-\lambda_0)]_{\Lambda\to +\infty}=
-\sum_{n\geq 1} b_n\frac{\lambda_0^n}{n} 
=-\int_0^{\lambda_0} d\lambda \frac{R(\lambda)}{\lambda}
\ee
given the asymptotic expansion of $\ln(\lambda-\lambda_0)=\ln \lambda-\sum_{n\geq 1} (\lambda_0/\lambda)^n/n$
in $+\infty$. This gives for $P(\lambda)$ with real roots:
\be
\mathrm{Pf}\, I[P](+\infty)\!=\!\!\!\!\!\!\sum_{\lambda_0\in\mathrm{Root}\, P}\!\!\!\!\!\Big\{
Q_I^{[1]}(\lambda_0) \Big[\ln |\lambda_0| -i\pi Y(\lambda_0) \frac{P'(\lambda_0)}{|P'(\lambda_0)|}\Big]
-\!\!\int_0^{\lambda_0} \!\!\!\!\! d\lambda \frac{Q_I^{[1]}(\lambda)}{\lambda}
\Big\}
\ee
and for $P(\lambda)$ with complex roots:
\be
\mathrm{Pf}\, I[P](+\infty)=\sum_{\lambda_0\in\mathrm{Root}\, P} \Big\{
Q_I^{[1]}(\lambda_0) \ln(-\lambda_0)
-\int_0^{\lambda_0} d\lambda \frac{Q_I^{[1]}(\lambda)}{\lambda} 
\Big\}
\ee
These considerations and expressions extend directly to the functional $J_1[P](\Lambda)$, since it is sufficient to replace
the polynomial $Q_I^{[1]}(\lambda)$ with the polynomial $Q_J^{[1]}(\lambda)$.
As for the functional $J_2[P](\lambda)$, the properties of the dilogarithm function, or simply a direct reasoning
on the integrals that led to it\footnote{\label{note:Li2_asymptote}We obtain on the real axis $\bar{\mathrm{Li}}_2(x)=\frac{\pi^2}{12}-\frac{1}{2}(\ln |x|)^2+\frac{\pi^2}{4}\frac{x}{|x|}
-\frac{1}{x}+O(\frac{1}{x^2})$ for $x\to \pm\infty$, and out of the real axis,
$\mathrm{Li}_2(z)=-\frac{1}{2}[\ln(-z)]^2-\frac{\pi^2}{6}+
O(\frac{1}{z})$, with the principal branch of the complex logarithm.}, give for $P(\lambda)$ with real roots a result that may be complex
\bea
\mathrm{Pf}\, J_2[P](+\infty) = \sum_{\lambda_0\in\mathrm{Root}\, P} \Big[\frac{(\ln |\lambda_0|)^2}{2} -\frac{\pi^2}{3} Y(\lambda_0)
+\frac{\pi^2}{6} Y(-\lambda_0)-i\pi \ln |\lambda_0| Y(\lambda_0) \frac{P'(\lambda_0)}{|P'(\lambda_0)|}\Big]
\eea
and for $P(\lambda)$ with complex roots a real result
\be
\mathrm{Pf}\, J_2[P](+\infty) = \frac{\pi^2}{3} + \sum_{\lambda_0\in\mathrm{Root}\, P} \frac{1}{2} [\ln(-\lambda_0)]^2
\ee
It remains to deal with the quantity $C^\pm(\Lambda)$ in equation (\ref{eq:avec_coupure}). For our usual 
generic polynomial $P(\lambda)$ of degree at most two with real coefficients, we find, independently on the
fact that its roots are real or complex, the expression
\bea
\label{eq:boutC}
\mathrm{Pf}\, \left[\frac{u^{[3]}[P(\Lambda)]}{\Lambda}\right]_{\Lambda\to+\infty} &=& Q_C(0) [\ln |A_P|-\frac{11}{6} +i\pi Y(A_P)]-\sum_{\lambda_0\in\mathrm{Root}\,P} \int_0^{\lambda_0} d\lambda \frac{Q_C(\lambda)-Q_C(0)}{\lambda}
\eea
in terms of the leading coefficient $A_P$ of $P$ and of the auxiliary polynomial $Q_C(\lambda)\equiv 
\{[P(\lambda)]^3-[P(0)]^3\}/(3! \lambda)$ which is {\sl a priori} nonzero in zero. 
We have used (\ref{eq:interm}) with $R(\lambda)=[Q_C(\lambda)-Q_C(0)]/\lambda$.

Fortunately, the obtained results for the {\sl partie finie} when the cutoff $\Lambda\to +\infty$ can be largely simplified
thanks to the duality relations (\ref{eq:dualite}). Indeed, only the polynomials $P^\pm_\alpha(\lambda)$ and $P^\pm_\delta(\lambda)$
appear in the $\Lambda$-dependent terms of equation (\ref{eq:avec_coupure}), and only the difference $\psi^+(\lambda)-\psi^-(\lambda)$ matters in the
final result (\ref{eq:sig2opera}). To the generic polynomial $P(\lambda)$ of degree two we then associate its dual
\be
\check{P}(\lambda)\equiv P(-\lambda)
\ee
Of course, the roots of $\check{P}(\lambda)$ are the opposite of the roots of $P(\lambda)$, while the two polynomials 
have the same value in zero and the same leading coefficient.
One simply substitutes the polynomial $P(\lambda)$ and its roots $\lambda_0$, its auxiliary polynomials $Q_I(\lambda)$, $Q_J(\lambda)$ 
defined by (\ref{eq:defQI}) and (\ref{eq:defQJ}),
and their primitives $Q_I^{[1]}(\lambda)$, $Q_J^{[1]}(\lambda)$ (that appear in the expression
of the {\sl partie finie} of the functionals $I[P](\Lambda)$ and $J[P](\Lambda)$ in $\Lambda=+\infty$), with 
the dual polynomial $\check{P}(\lambda)\equiv P(-\lambda)$ and its roots $\check{\lambda}_0=-\lambda_0$,
its auxiliary polynomials $\check{Q}_I(\lambda)=Q_I(-\lambda)$, $\check{Q}_J(\lambda)=-Q_J(-\lambda)$ and their
primitives $\check{Q}_I^{[1]}(\lambda)=-Q_I^{[1]}(-\lambda)$, $\check{Q}_J^{[1]}(\lambda)=Q_J^{[1]}(-\lambda)$.
One then obtains the expression of the {\sl partie finie} of $I[\check{P}](\Lambda)$ and $J[\check{P}](\Lambda)$ in $\Lambda=+\infty$.
Moreover, in equation (\ref{eq:boutC}), replacing $P(\lambda)$ with $\check{P}(\lambda)$ amounts to replacing 
$Q_C(\lambda)$ with $\check{Q}_C(\lambda)=-Q_C(-\lambda)$.

Thus we obtain a series of simplified relations. First, $\mathrm{Pf}\,\{[u^{[3]}[P(\Lambda)]+u^{[3]}[\check{P}(\Lambda)]]/\Lambda\}=0$ 
in $\Lambda=+\infty$ so that
\be
\mathrm{Pf}\, [C^+-C^-](+\infty)=0
\ee
and the terms $C^{\pm}(\Lambda)$ of equation (\ref{eq:avec_coupure})
give no contribution to $\bar{\Sigma}^{(2)}(\bar{K},\varepsilon)$.
Then, for the functional $I$, whose values at $P$ and $\check{P}$ have to be summed up, in the case realized
in practice of a coefficient $A_P>0$ in $P(\lambda)$, we have
\be
\label{eq:symI}
\mathrm{Pf}\, \{I[P]+I[\check{P}]\}(+\infty) \stackrel{A_P>0}{=} 
-i\pi [Q_I^{[1]}(\lambda_2) - Q_I^{[1]}(\lambda_1)]
=-\frac{i\pi A_P^2}{60} (\lambda_2-\lambda_1)^5
\ee
where the roots $\lambda_1$, $\lambda_2$ of $P(\lambda)$ are sorted by increasing order if they are real,
and by increasing order of their imaginary part if they are complex, so that
\be
\lambda_2-\lambda_1=\frac{\Delta_P^{1/2}}{|A_P|} \ \mbox{(real roots)} \ \ \mbox{or}\ \ \frac{i(-\Delta_P)^{1/2}}{|A_P|}
 \ \mbox{(complex roots)},
\ee
$\Delta_P$ being the discriminant of the polynomial $P(\lambda)$\footnote{In the case of complex roots, we use the property 
$\ln(\lambda_2)-\ln(-\lambda_2)=\ln(-\lambda_1)-\ln(\lambda_1)=i\pi$.}. 
We proceed in a similar way for the functional $J$, whose values at $P$ and $\check{P}$ have to be subtracted
since $J$ is multiplied by $\bar{K}$ in equation (\ref{eq:avec_coupure}).
Taking into account the splitting (\ref{eq:j1pj2}), let us first writ, here also for $A_P>0$,
\bea
\mathrm{Pf}\, \{J_1[P]-J_1[\check{P}]\}(+\infty) \stackrel{A_P>0}{=}  
-i\pi [Q_J^{[1]}(\lambda_2)-Q_J^{[1]}(\lambda_1)]
\nonumber\\
=-\frac{i\pi A_P^2}{24}(\lambda_2-\lambda_1) (\lambda_1+\lambda_2) [(\lambda_1+\lambda_2)^2-10 \lambda_1 \lambda_2]
\label{eq:symJ1}
\eea
where we recall that the sum $\lambda_1+\lambda_2$ and the product $\lambda_1 \lambda_2$ of the roots are respectively
the coefficients of the terms of order one in $\lambda$ and of order zero in $\lambda$ of the normalized dual polynomial 
$\check{P}(\lambda)/A_P$. Then, to give the remaining term, let us consider first the case of $P(\lambda)$ with real roots:
\be
\label{eq:symJ2R}
\mathrm{Pf}\, \{J_2[P]-J_2[\check{P}]\}(+\infty) \stackrel{A_P>0}{=} -\pi^2 Y(\lambda_1 \lambda_2)\frac{\lambda_1+\lambda_2}{|\lambda_1+\lambda_2|}
-i\pi \ln \frac{|\lambda_2|}{|\lambda_1|},
\ee
where we have used $A_P>0$, $Y(x)-Y(-x)=\mathrm{sgn}\, x$, 
$\mathrm{sgn}\,x+\mathrm{sgn}\, y = 2 Y(xy)\, \mathrm{sgn}\, (x+y)$ for any real $x$ and $y$.
In the case of $P(\lambda)$ with complex roots, without any hypothesis on the sign of $A_P$, it reads
\bea
\label{eq:symJ2I}
\mathrm{Pf}\, \{J_2[P]-J_2[\check{P}]\}(+\infty) = -i\pi \ln \left(\frac{-\lambda_2}{\lambda_1}\right)
= -2\pi \arcsin\frac{\lambda_1+\lambda_2}{2|\lambda_1\lambda_2|^{1/2}}
\eea
This concludes our calculation of $\bar{\Sigma}^{(2)}(\bar{K},\varepsilon)$.

\subsection{A compact form of the final result}

To conclude this section, let us give (under a compact form as in \cite{lettre}) the contribution of order
$g^2$ to the self-energy of the impurity, in dimensionless units as in (\ref{eq:sig2adim}): 
\be
\bar{\Sigma}^{(2)}(\bar{K},\varepsilon) = \frac{9}{32 r^2\bar{K}} 
[\mathcal{S}^+-\mathcal{S}^-]
\label{eq:sig2scalli}
\ee
where the quantities $\mathcal{S}^\pm$ defined as
\be
\mathcal{S}^\pm\! \equiv\! (1+r) [i_\alpha^\pm +I_\delta^\pm(2)]  -(1-r) I_\beta^\pm(2) -r I_\gamma^\pm(2) 
\pm\bar{K} [j_\alpha^\pm -J_\beta^{\pm}(2)+J_\delta^\pm(2)]
\label{eq:scalli}
\ee
are mutually interchanged by changing $\bar{K}$ to $-\bar{K}$,  
which does not cause any problem since the property $\bar{K}>0$ was never used in the two previous subsections. In equation (\ref{eq:scalli}),
the quantities $I_\sigma^\pm(\lambda)$ and $J_\sigma^\pm(\lambda)$, related to the trinomials
(\ref{eq:pa},\ref{eq:pb},\ref{eq:pc},\ref{eq:pd}) by equation (\ref{eq:raccourcis}), can be evaluated
explicitly for $\lambda=2$ thanks to the expressions (\ref{eq:IPrr},\ref{eq:IPri})  
and (\ref{eq:J2Prr},\ref{eq:J2Pri}), given the splitting (\ref{eq:j1pj2}) and the link
between the functionals $J_1[P]$ and $I[P]$ indicated just below equation (\ref{eq:defJ2});
the new quantities introduced in (\ref{eq:scalli}),
\be
i_\alpha^\pm \equiv \mathrm{Pf}\,(I^\pm_\alpha +I^\mp_\delta)(+\infty) \ \ \ \mbox{and}\ \ \ 
j_\alpha^\pm \equiv \mathrm{Pf}\,(J^\pm_\alpha -J^\mp_\delta)(+\infty)
\ee
are deduced from the results (\ref{eq:symI}), (\ref{eq:symJ1}) and (\ref{eq:symJ2R},\ref{eq:symJ2I}) 
by setting $P(\lambda)=P^\pm_\alpha(\lambda)$ since we have the duality (\ref{eq:dualite}),
and by bearing in mind also the splitting (\ref{eq:j1pj2}).

\section{Singularities of the derivatives of $\bar{\Sigma}^{(2)}(\bar{K},\varepsilon)$}
\label{sec:les_singus}

The self-energy of the impurity to second order in the interaction,
$(\bar{K},\varepsilon)\mapsto\bar{\Sigma}^{(2)}(\bar{K},\varepsilon)$ in its dimensionless form,
calculated explicitly in section \ref{sec:cedsigdlcg}, is a smooth function on $\mathbb{R}^+\times \mathbb{R}$
except on certain singularity curves that we are going to study here.
To see it, it is sufficient to remark that each of the terms in the splitting 
(\ref{eq:sig2scalli},\ref{eq:scalli}) is a function of the leading coefficients $A_P$ and of the 
roots $\lambda_0$ of the polynomials $P_\sigma^{\eta}(\lambda)$, with $\sigma\in\{\alpha,\beta,\gamma,\delta\}$ and $\eta=\pm$
depending on whether we consider $\mathcal{S}^\pm$. While the $A_P$ are constant, the roots $\lambda_0$ are non-trivial
functions of $(\bar{K},\varepsilon)$. Our discussion considers here the general case $r\neq 1$, but it is readily adapted
to the particular case $r=1$.

\subsection{{\yvan Location} of the singularities in the plane $(\bar{K},\varepsilon)$}
\label{subsec:local}

\begin{figure}[htb]
\begin{center}
\includegraphics[width=10cm,clip=]{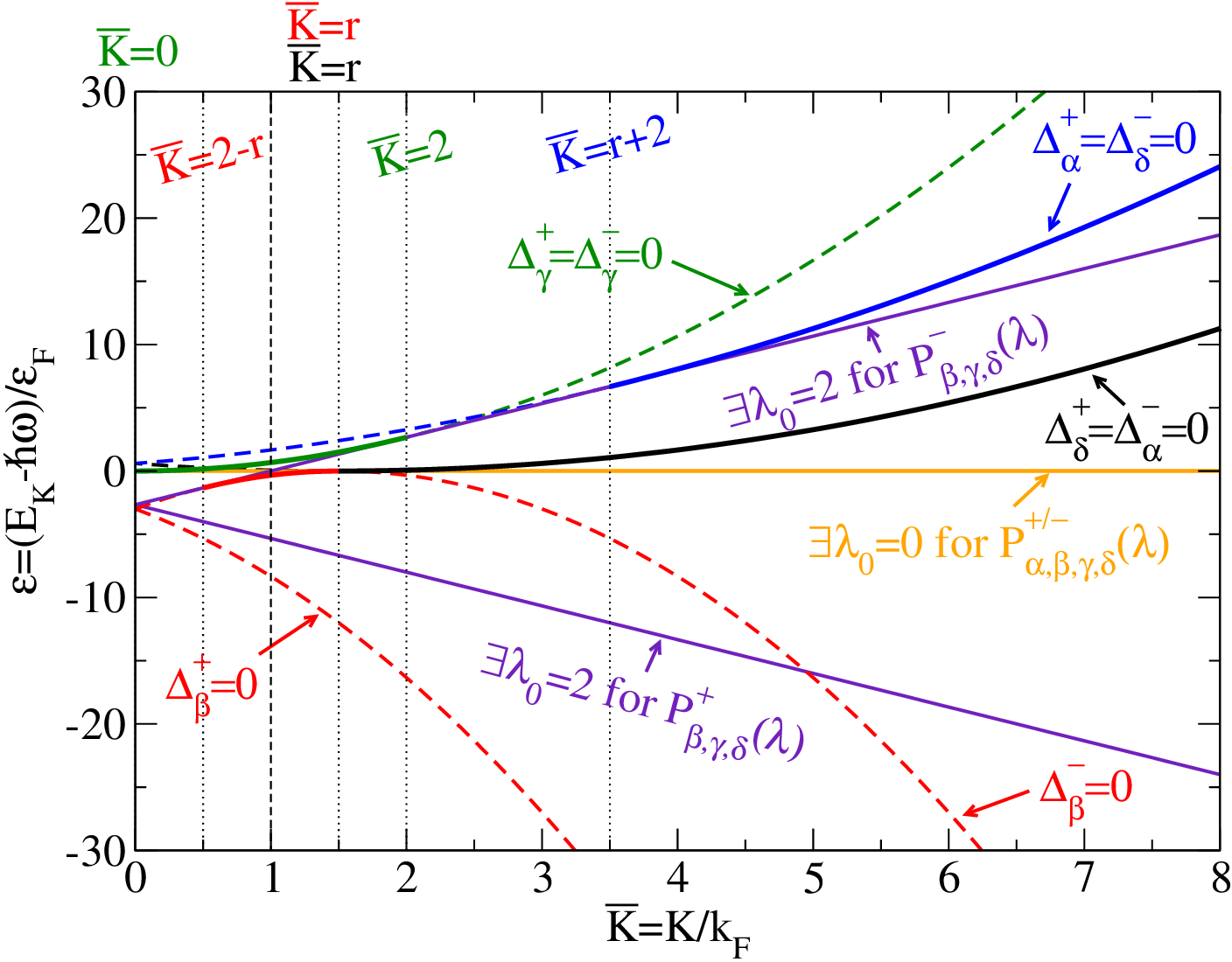}
\end{center}
\caption{The singularity curves of the derivatives with respect to 
$(\bar{K},\varepsilon)\in \mathbb{R}^+\times \mathbb{R}$ 
of the dimensionless self-energy of order two $\bar{\Sigma}^{(2)}(\bar{K},\varepsilon)$.
(i) On the parabolas $\Delta_{\sigma}^\eta=0$, where $\Delta_{\sigma}^\eta$
is the discriminant of the polynomial $P_\sigma^{\eta}(\lambda)$, $\sigma\in\{\alpha,\beta,\gamma,\delta\}$ and $\eta\in\{+,-\}$,
the roots of the polynomials are non-differentiable functions of $(\bar{K},\varepsilon)$; this actually leads 
to a singularity of the derivatives of $\bar{\Sigma}^{(2)}(\bar{K},\varepsilon)$ on the full line portions, not on the 
dashed line portions. (ii) $\bar{\Sigma}^{(2)}(\bar{K},\varepsilon)$
is not a smooth function of the roots on the horizontal half-line, where at least one (but also each)
polynomial $P_{\sigma}^{\pm}(\lambda)$, $\sigma\in\{\alpha,\beta,\gamma,\delta\}$, has
a root $\lambda_0=0$, and on the oblique half-line (downwards for $\eta=+$, upwards for $\eta=-$),
where at least one (but also each) polynomial $P_{\sigma}^{\eta}(\lambda)$, $\sigma\in\{\beta,\gamma,\delta\}$, has
one root $\lambda_0=2$.
Vertical dotted lines mark the abscissa of the tangent points of the parabolas with the singularity lines; 
the values of the abscissas are given explicitly as functions of the impurity-to-fermion mass ratio $r=M/m$,
in a color code and obliquity code identifying respectively 
the parabola and the line that are tangent. The vertical dashed line marks the abscissa $\bar{K}=1$ of the 
crossing point of the horizontal and the oblique upwards singularity lines. In the figure, $r=3/2$.
}
\label{fig:zones}
\end{figure}

A first source of singularity for the derivatives of $\bar{\Sigma}^{(2)}(\bar{K},\varepsilon)$ is 
the non-differentiability of the roots with respect to $(\bar{K},\varepsilon)$, which happens when the
discriminant $\Delta_{\sigma}^\eta$ of one of the polynomials $P_\sigma^{\eta}(\lambda)$ vanishes, so that the quantity 
$(\Delta_{\sigma}^\eta)^{1/2}$ in the expression of the roots is not anymore differentiable: 
\bea
\label{eq:doubleab}
\Delta_\alpha^{\eta}=0 \Leftrightarrow \varepsilon=\frac{(r+\eta \bar{K})^2}{r(1+r)} & \hspace{3cm} &
\Delta_\beta^{\eta}=0 \Leftrightarrow \varepsilon=\frac{(r+\eta \bar{K})^2}{r(1-r)}  \\
\label{eq:doublecd}
\Delta_\gamma^{\eta}=0 \Leftrightarrow \varepsilon=\frac{\bar{K}^2}{r} & \hspace{3cm} &
\Delta_\delta^{\eta}=0 \Leftrightarrow \varepsilon=\frac{(r-\eta\bar{K})^2}{r(1+r)}
\eea
Although there are eight distinct polynomials into play, the corresponding locus of points in the half-plane $(\bar{K}>0,\varepsilon)$
is composed of portions of five parabolas only, see figure \ref{fig:zones}, due to the duality relations
(\ref{eq:dualite}) and $P_\gamma^{-}(\lambda)=P_\gamma^{+}(-\lambda)$.
Note that {\yvan (for $r < 1$ if $\sigma=\beta$)} the polynomial $P_\sigma^{\eta}(\lambda)$ has real roots when $(\bar{K},\varepsilon)$ is below the parabola
$\Delta_\sigma^\eta=0$, it has a double root when $(\bar{K},\varepsilon)$ is on the 
parabola, and it has complex roots when $(\bar{K},\varepsilon)$ is above the parabola.

A second source of singularity for the derivatives of $\bar{\Sigma}^{(2)}(\bar{K},\varepsilon)$ is
the non-differentiability of $\bar{\Sigma}^{(2)}$ with respect to the roots $\lambda_0$ of the 
polynomials. As it appears on equations (\ref{eq:IPrr},\ref{eq:J2Prr}), taken for $\lambda=2$,
and on equation (\ref{eq:symJ2R}), this may be due to logarithmic singularities in the real part,
coming from the logarithm itself or from the behavior of the function $\bar{\mathrm{Li}}_2(x)$ in the neighborhood of $x=1$
and $x=\pm \infty$, and from the discontinuity of the Heaviside function in the imaginary part. This can happen
only if one of the roots $\lambda_0$ of the considered polynomial tends to zero or two\footnote{\label{note:bouts_lisses}
For any fixed $\lambda$, the quantities $Q_I^{[1]}(\lambda)$ and $Q_J^{[1]}(\lambda)$ are polynomials in the
coefficients of $P(\lambda)$, thus here $C^\infty$ functions in the roots $\lambda_{1,2}$ as well as
in $\bar{K}$ and $\varepsilon$. The quantity $Q_I^{[1]}(\lambda_0)$ is a polynomial in $\lambda_{1,2}$, 
thus is $C^\infty$ in $\lambda_{1,2}$, but not necessarily $C^\infty$ in $\bar{K}$ and $\varepsilon$ where the roots are not
$C^\infty$ functions of $\bar{K}$ and $\varepsilon$.
Also, the integral term in  (\ref{eq:IPrr}) and its counterpart in $J_1[P](\lambda)$ are polynomials in $\lambda_{1,2}$; but
after summing over $\lambda_0$, they become {\sl symmetric} polynomials in $\lambda_{1,2}$, that is, according
to a classical result, polynomials in $\lambda_1+\lambda_2$ and $\lambda_1 \lambda_2$, thus here 
$C^\infty$ functions of $\bar{K}$ and $\varepsilon$.}. 
We find that the condition of existence of a zero root is the same for the eight polynomials 
$P_\sigma^\eta$, where $\sigma\in\{\alpha,\beta,\gamma,\delta\}$ and $\eta=\pm$:
\be
\label{eq:exlam0}
\forall \sigma\in\{\alpha,\beta,\gamma,\delta\}, \forall \eta=\pm: P_\sigma^\eta(0)=0
\Leftrightarrow \varepsilon=0
\ee
which corresponds to a single line in the plane $(\bar{K},\varepsilon)$.
Asking if one of the roots is equal to two is useful only for the six polynomials
$P_\sigma^\eta$, with $\sigma\in\{\beta,\gamma,\delta\}$ and $\eta=\pm$, since
the index $\alpha$ in (\ref{eq:scalli}) appears only in the {\sl partie finie} $i_\alpha^\pm$ and $j_\alpha^\pm$; 
the expressions (\ref{eq:symI},\ref{eq:symJ1}) are indeed smooth functions of the roots, as well as (\ref{eq:symJ2R})
except in $\lambda_{1,2}=0$. Then we find that the condition of existence of a root $\lambda_0=2$
depends only on the sign of $\eta$, it is the same for the three polynomials
$P_\sigma^+(\lambda)$ from one hand, and for the three polynomials $P_\sigma^-(\lambda)$ from the other hand: 
\be
\label{eq:exlam2}
\forall \sigma\in\{\beta,\gamma,\delta\}, \forall \eta=\pm:  P_\sigma^\eta(2)=0 \Leftrightarrow 
r\varepsilon = -4 (1+\eta \bar{K})
\ee
which corresponds to two lines only in the plane $(\bar{K},\varepsilon)$,
that are interchanged by reflection with respect to the vertical axis, where they cross.

There exist points of the plane $(\bar{K},\varepsilon)$ that are {\sl doubly} singular, that combine the two sources
of singularity for a given polynomial $P_{\sigma}^{\eta}(\lambda)$: The polynomial there has a double root equal
to zero or two.
As shown in figure~\ref{fig:zones}, these are not only intersection points of the parabola $\Delta_\sigma^\eta=0$
and the line $\epsilon=0$ or $r\varepsilon = -4 (1+\eta \bar{K})$, but also tangent points, since $P_{\sigma}^{\eta}(\lambda)$,
with real coefficients, cannot have both a real root and a complex root. In this subsection, contrarily to the
subsection  \ref{subsec:regul} for $r=1$, we do not discuss the case of these doubly singular points, nor the case of
the points of intersection between singularity curves associated to different polynomials, and what follows will be valid
only for the other points, that is for the generic points, of the singularity curves.   

{\yvan We have not found any clear physical interpretation of the singularity curves, which
must be linked to the existence of a Fermi surface which
imposes sharp boundaries $k=q=\kf$ in the domain of variation of $\kk$ and $\qq$ in equation~(\ref{eq:sig2}).
At the Fermi surface, $\lambda=0$ and $\lambda=2$ correspond to a particle and a hole 
having equal and equal-and-opposite momenta, respectively.
Then, by using the expectation of reference \cite{lettre} that highly singular points should be obtained
when both $F_{\kk,\qq}(\KK,\omega)$ and its first order differential with respect to $\kk$ and $\qq$ vanish
at some point of the Fermi surface, we obtain the multiply-singular point $(\bar{K}=r,\varepsilon=0)$, for
$\lambda=0$.}

\subsection{On the parabolas: The effect of a double root}

Let us first study the singularities on the parabolas $\Delta_\sigma^\eta=0$, $\sigma\in\{\alpha,\beta,\gamma,\delta\}$ 
and $\eta=\pm$, in a point of the plane $(\bar{K},\varepsilon)$ where the polynomial $P_\sigma^\eta(\lambda)$
has a double root $\lambda_{0,0}^{(\eta\sigma)}$. To simplify, we will approach the parabola only from $\Delta_\sigma^\eta>0$,
where $P_\sigma^\eta(\lambda)$  has real roots, $\lambda_1^{(\eta\sigma)}<\lambda_2^{(\eta\sigma)}$, arbitrarily close
to $\lambda_{0,0}^{(\eta\sigma)}$.
As $\lambda_{0,0}^{(\eta\sigma)}$ is different from zero and from two, and given the footnote~\ref{note:bouts_lisses}, only the second line
in the equation (\ref{eq:IPrr}) and its counterpart in the functional $J$ can lead to singularities. Inspired by the footnote, we 
express the $Q_I^{[1]}(\lambda_{1,2}^{(\eta\sigma)})$ as linear combinations of their sum $S_I$ and of their difference $D_I$,
\be
\label{eq:SIDI}
S_I=Q_I^{[1]}(\lambda_{1}^{(\eta\sigma)})+Q_I^{[1]}(\lambda_{2}^{(\eta\sigma)})\ \ \mbox{and}\ \ 
D_I=Q_I^{[1]}(\lambda_{2}^{(\eta\sigma)})-Q_I^{[1]}(\lambda_{1}^{(\eta\sigma)})
\ee
The contribution of $S_I$ to the real part of $I_\sigma^\eta(2)$ contains as a factor the logarithm of expressions that,
like $S_I$, are symmetric functions of the roots, that is their product or the product of their deviation from two;
as the whole this is a smooth function of  $(\bar{K},\varepsilon)$.  
Remarkably, the contribution of $D_I$ to the real part of $I_\sigma^\eta(2)$ is also a smooth function of $(\bar{K},\varepsilon)$
at the double root point, by virtue of the following property that we will apply, given the explicit value (\ref{eq:symI}) of $D_I$,
to the function $f(\lambda_1,\lambda_2)=(\lambda_2-\lambda_1)^5
\ln\{\lambda_1(2-\lambda_2)/[\lambda_2(2-\lambda_1)]\}$: 

\noindent{\bf Property~1:} {\sl If $f(\lambda_1,\lambda_2)$ is a symmetric function of $(\lambda_1,\lambda_2)\in\mathbb{R}^2$
and a smooth function ($C^\infty$) in a neighborhood of $(\lambda_{0,0}^{\eta\sigma},\lambda_{0,0}^{\eta\sigma})$, then
$f(\lambda_1^{(\eta\sigma)},\lambda_2^{(\eta\sigma)})$ is a $C^\infty$ function of $(\bar{K},\varepsilon)$
at the point where $P_\sigma^\eta(\lambda)$ has a double root $\lambda_{0,0}^{\eta\sigma}$.}

To establish this property, let us remark that $\lambda_{1,2}^{(\eta\sigma)}-\lambda_{0,0}^{(\eta\sigma)}=
\mp (\Delta_{\sigma}^\eta)^{1/2}/(2|A_\sigma^\eta|)$ up to a single additive $C^\infty$ function of $(\bar{K},\varepsilon)$,
with $A_\sigma^\eta$ being the leading coefficient of $P_\sigma^\eta(\lambda)$.
As $f(\lambda_1,\lambda_2)$ is a symmetric function, the Taylor expansion of $f(\lambda_1^{(\eta\sigma)},\lambda_2^{(\eta\sigma)})$
in powers of $(\Delta_{\sigma}^\eta)^{1/2}$ produces only even powers, which are $C^\infty$ functions of $(\bar{K},\varepsilon)$, hence the result.
\qed

Likewise, we find that the real part of $J_\sigma^\eta(2)$ is a $C^\infty$ function of $(\bar{K},\varepsilon)$
at the considered point.
To study the piece $J_1$, in the splitting (\ref{eq:j1pj2}), we introduce $S_J$ and $D_J$ {\yvan [see equation~(\ref{eq:symJ1})]},
by replacing in (\ref{eq:SIDI}) the auxiliary polynomial $Q_I$ with $Q_J$.
In the case of the $J_2$ piece, which is multiplied by $[P(0)]^2/2$, a $C^\infty$ function of $(\bar{K},\varepsilon)$, 
we use the property~1 with $f(\lambda_1,\lambda_2)=\bar{\mathrm{Li}}_2(2/\lambda_1)+\bar{\mathrm{Li}}_2(2/\lambda_2)$,
which is legitimate since $2/\lambda_{1,2}$ is here in the neighborhood of neither $\pm\infty$ nor the edge of the branch cut $\Re z\geq 1$
of the function $\mathrm{Li}_2(z)$ in the complex plane.

Let us now consider the case of the imaginary part of $I_\sigma^\eta(2)$ and $J_\sigma^\eta(2)$. The Heaviside functions
of the roots in the neighborhood of the considered point take the same value for $\lambda_{1,2}$, it is their value in the double root,
which can be expressed in terms of a rectangular function, the indicator function of the interval $[0,2]$, which vanishes everywhere except 
on this interval where it takes the value one:  
\be
\Pi_{[0,2]}(x)\equiv Y(2-x)-Y(-x)
\ee
As $P'(\lambda_{1,2})/|P'(\lambda_{1,2})|$ have opposite signs, only the function $D_I$ and $D_J$ contribute.
Finally there appear antisymmetric functions of the roots, which do not obey property~1 and are not $C^\infty$ in $(\bar{K},\varepsilon)$
at the considered point. In particular, we collect the pieces $J_1$ and $J_2$ so as to avoid 
stronger intermediate singularities, which amounts to considering the function $(\lambda_2-\lambda_1) f(\lambda_1,\lambda_2)$
where $f(\lambda_1,\lambda_2)=\frac{\lambda_1+\lambda_2}{24}[(\lambda_1+\lambda_2)^2-10\lambda_1\lambda_2]
+\frac{(\lambda_1 \lambda_2)^2}{2} \ln(\lambda_2/\lambda_1)$ is $C^\infty$, symmetric, homogeneous of degree three
and such that $f(1+\delta,1-\delta)=\frac{4}{15}\delta^4 +O(\delta^5)$ when $\delta\to 0$. We obtain up to an additive
$C^\infty$ function of $(\bar{K},\varepsilon)$: 
\bea
\label{eq:singI_type1}
I_\sigma^\eta(2) \stackrel{\Delta^\eta_\sigma\to 0^+}{=} -\frac{i\pi \Pi_{[0,2]}(\lambda_{0,0}^{(\eta\sigma)})}{60 (A_\sigma^\eta)^3} (\Delta_\sigma^\eta)^{5/2} 
+ \mathrm{C}^\infty \\
\label{eq:singJ_type1}
J_\sigma^\eta(2) \stackrel{\Delta^\eta_\sigma\to 0^+}{=} -\frac{i\pi \Pi_{[0,2]}(\lambda_{0,0}^{(\eta\sigma)})}{60 \lambda_{0,0}^{(\eta\sigma)}
(A_\sigma^\eta)^3} [(\Delta_\sigma^\eta)^{5/2}
+O(\Delta_\sigma^\eta)^3] + \mathrm{C}^\infty
\eea
As a consequence, the {\bf third order} derivative of $\Im I_\sigma^\eta(2)$ and of $\Im J_\sigma^\eta(2)$
along the direction normal to the parabola $\Delta_\sigma^\eta=0$ diverges as the inverse of the square root of the distance
to the parabola, provided the double root of the polynomial $P_\sigma^\eta(\lambda)$ is in the interval $]0,2[$,
which happens on the portion of parabola between its tangent points with the lines $\exists \lambda_0^{(\eta\sigma)}=0$ and $\exists \lambda_0^{(\eta\sigma)}=2$
of equations (\ref{eq:exlam0}) and (\ref{eq:exlam2}).

In the case $\sigma=\alpha$, one must consider the symmetrized {\sl partie finie} $i_\alpha^\pm$ and $j_\alpha^\pm$.
In reality, their singularities interfere with those of $I_\delta^\mp(2)$ and $J_\delta^\mp(2)$, as can be seen on the expression
(\ref{eq:sig2scalli}) of $\bar{\Sigma}^{(2)}(\bar{K},\varepsilon)$, since the parabolas
$\Delta_\alpha^\pm=0$ and $\Delta_\delta^{\mp}=0$ coincide. One thus has to collect them to obtain
\bea
\label{eq:itat1}
i_\alpha^\eta-I_\delta^{-\eta}(2)\stackrel{\Delta^\eta_\alpha\to 0^+}{=}\!-\frac{i\pi(\Delta_\alpha^{\eta})^{5/2} }{60 (A_\alpha^\eta)^3} 
\left\{1-\Pi_{[0,2]}(-\lambda_{0,0}^{(\eta\alpha)})\right\}+ \mathrm{C}^\infty \\
\label{eq:jtat1}
j_\alpha^\eta+J_\delta^{-\eta}(2) \stackrel{\Delta^\eta_\alpha\to 0^+}{=} 
\!-\frac{i\pi [(\Delta_\alpha^{\eta})^{5/2}+O(\Delta_\alpha^\eta)^3]}
{60 \lambda_{0,0}^{(\eta\alpha)}(A_\alpha^\eta)^3} 
\left\{1-\Pi_{[0,2]}(-\lambda_{0,0}^{(\eta\alpha)})\right\}+\mathrm{C}^\infty
\eea
where the duality (\ref{eq:dualite})  implies $-\lambda_{0,0}^{(\eta\alpha)}=\lambda_{0,0}^{(-\eta\delta)}$.

It remains to add all contributions to $\bar{\Sigma}^{(2)}(\bar{K},\varepsilon)$ according to the compact
writing (\ref{eq:sig2scalli}), for each of the five possible distinct parabolas, at least to verify the absence 
of cancellations of the contributions of the functionals $I[P]$ and $J[P]$:  
\bea
\label{eq:ssp1}
\bar{\Sigma}^{(2)}(\bar{K},\varepsilon) \stackrel{\Delta_\beta^\eta\to 0^+}{=} 
\frac{3i\pi\Pi_{[0,2]}(\frac{\eta\bar{K}+r}{r-1})}{640 r(1-r)^2 \bar{K} (\bar{K}+\eta r)}
[(\Delta_\beta^\eta)^{5/2}+O(\Delta_\beta^\eta)^3] +\mathrm{C}^\infty\\
\label{eq:ssp2}
\bar{\Sigma}^{(2)}(\bar{K},\varepsilon)\stackrel{\Delta_\gamma^+=\Delta_\gamma^-\to 0^+}{=} \frac{3i\pi[\Pi_{[0,2]}(-\bar{K})-\Pi_{[0,2]}(\bar{K})]}{640 r \bar{K}}
(\Delta_\gamma^+)^{5/2}+\mathrm{C}^\infty \\
\label{eq:ssp3}
\bar{\Sigma}^{(2)}(\bar{K},\varepsilon)\stackrel{\Delta_\alpha^\eta=\Delta_\delta^{-\eta}\to 0^+}{=}
\!\!\frac{-3i\pi[1-\Pi_{[0,2]}(\frac{\eta\bar{K}+r}{1+r})]}{640 r (1+r)^2 \bar{K} (\bar{K}+\eta r)}
[(\Delta_\alpha^\eta)^{5/2}\!+\!O(\Delta_\alpha^\eta)^3]\!+\!\mathrm{C}^\infty
\eea
Note, in these expressions, the occurrence 
of a  prefactor that diverges at the tangent point of the considered parabola $\Delta_\sigma^\eta=0$ 
($\sigma\in\{\beta,\gamma,\alpha\}$) with the horizontal axis $\varepsilon=0$, as well as at the intersection point, of abscissa 
$\bar{K}=0$, of the parabolas of the same class $\sigma\in\{\alpha,\beta\}$ but of opposite $\eta$.
The portions of parabola where a divergence of the {\bf third order} derivative of $\bar{\Sigma}^{(2)}(\bar{K},\varepsilon)$ 
in the normal direction actually occurs  are represented in bold lines on figure \ref{fig:zones}; the other portions
are in {\yvan dashed} lines.

\subsection{On the horizontal line: The effect of a zero root}
\label{subsec:horizon}

Let us study now the singularities on the horizontal line $\varepsilon=0$, in a point of the plane $(\bar{K},\varepsilon)$
where one (and in practice each) of the polynomials $P_\sigma^\eta(\lambda)$, with $\sigma\in\{\alpha,\beta,\gamma,\delta\}$ and $\eta=\pm$,
has a zero root, the other root being different from zero and two.
In the neighborhood of this point, $P_\sigma^\eta(\lambda)$ has a real root $\lambda_0^{(\eta\sigma)}$
that linearly vanish in $\varepsilon$, therefore changing sign, as 
\be
\label{eq:devrac}
\lambda_0^{(\eta\sigma)} \stackrel{\varepsilon\to 0}{=} -\frac{r\varepsilon}{{P_\sigma^\eta}'(0)}
-\frac{A_\sigma^\eta}{[{P_\sigma^\eta}'(0)]^3}(r\varepsilon)^2 + O(\varepsilon^3),
\ee
the other root $\lambda_0'^{(\eta\sigma)}=r\varepsilon/(A_\sigma^\eta \lambda_0^{(\eta\sigma)})$ having a finite limit.

Let us first look at the {\sl partie finie} contributions $i_\alpha^\eta$ and $j_\alpha^\eta$ to the self-energy
of order two, see equation (\ref{eq:scalli}).
As the roots of the polynomials are $C^\infty$ functions of $(\bar{K},\varepsilon)$ in the neighborhood of the 
considered point, the quantity $i_\alpha^\eta$, as well as the contribution of the functional $J_1$ to $j_\alpha^\eta$,
are also $C^\infty$, by virtue of equations (\ref{eq:symI}) and (\ref{eq:symJ1}).
On the other hand, after multiplication of the contribution of the functional  $J_2$
by $[P_{\alpha}^{\eta}(0)]^2/2$, see (\ref{eq:symJ2R}),
we find that the {\sl second order} derivative of $j_\alpha^{\eta}$ with respect to $\varepsilon$ has a logarithmically divergent imaginary part
in $\varepsilon=0$, and a discontinuous real part. 

Can the contributions $I_\sigma^\eta(2)$ and $J_\sigma^\eta(2)$ lead to stronger singularities?
In the expression (\ref{eq:IPrr}) taken for $\lambda=2$, the only {\sl a priori} non $C^\infty$ piece
is in $Q_{I}^{[1]}(\lambda_0)[\ln |\lambda_0| \pm i\pi Y(-\lambda_0)]$, with the indices 
$\sigma$ and $\eta$ being omitted for simplicity.
But here $Q_I^{[1]}(\lambda_0)$ vanishes {\sl cubically} in $\lambda_0$, as can be seen by the simple
change of variable $t=x\lambda_0$ in the integral defining $Q_I^{[1]}$:
\be
Q_I^{[1]}(\lambda_0) \equiv \int_0^{\lambda_0} dt \frac{[P(t)]^2}{2} 
= \frac{A_P^2}{2} \lambda_0^3 \int_0^1 dx (x-1)^2 (\lambda_0 x-\lambda_0')^2
\ee
With the help of the relation $s Y(-s\varepsilon)=Y(s)-Y(\varepsilon)$, true for any {\yvan $s=\pm 1$} but in practice 
used with $s=P'(0)/|P'(0)|$, we obtain, by reintroducing the indices:
\be
\label{eq:diff3i}
I_\sigma^\eta(2) \stackrel{\varepsilon\to 0}{=} -\frac{(r\varepsilon)^3[1+O(\varepsilon)]}{6{P_\sigma^\eta}'(0)}
[\ln|\varepsilon| + i \pi Y(\varepsilon)] +\mathrm{C}^\infty
\ee
so that one must take the {\sl third} order derivative of  $I_\sigma^\eta(2)$ with respect to $\varepsilon$ to have
a logarithmic divergence in its real part, and a discontinuity in its imaginary part.
In the case of $J_1[P](2)$, we make the same reasoning by replacing $Q_I$ with $Q_J$:
\be
\label{eq:qjq}
Q_J^{[1]}(\lambda_0) \equiv \!\! \int_0^{\lambda_0}\!\!\!\!\! dt \frac{[P(t)]^2-[P(0)]^2}{2t} 
= \frac{A_P^2}{2} \lambda_0^2 \! \int_0^1\!\frac{dx}{x} [(x-1)^2 (\lambda_0 x-\lambda_0')^2-\lambda_0'^2]
\ee
This vanishes {\sl quadratically} in $\lambda_0$, and by taking the derivative of $J_1[P](2)$ just {\sl twice} with respect to $\varepsilon$
one obtains a logarithmic divergence in the real part and a discontinuity in the imaginary part.
However, the singularity coming from the functional $J_2$ is even more severe, even after multiplication by $[P(0)]^2/2$,
see (\ref{eq:J2Prr}) and footnote~\ref{note:Li2_asymptote} on the asymptotic behavior of the function $\bar{\mathrm{Li}}_2(x)$: As a result, the
second order derivative of $J_\sigma^\eta(2)$ with respect to $\varepsilon$ diverges as the {\sl square} of the 
logarithm of $\varepsilon$ for the real part, and, on the $\lambda_0>0$ side, as the logarithm of $\varepsilon$ for the
imaginary part.

Collecting all the contributions to $\bar{\Sigma}^{(2)}(\bar{K},\varepsilon)$, we find however that the expected 
singularity in $\varepsilon=0$ in the second order derivative do not appear, due to a perfect cancellation 
of the contributions of $j_\alpha^\eta$, $J^{-\eta}_\delta(2)$ and $J_\beta^\eta(2)$.
As in equation (\ref{eq:jtat1}), there is thus a clever combination to consider,
\be
\label{eq:comb_ast}
j_\alpha^\eta+J^{-\eta}_\delta(2)-J_\beta^\eta(2)=\mathcal{J}_\alpha^\eta-\mathcal{J}_\beta^\eta + \mathrm{C}^\infty
\ee
for which we find after rather long calculations:
\bea
\label{eq:tjab}
\mathcal{J}_\sigma^\eta&\equiv&Q_{J^\eta_\sigma}^{[1]}(\lambda_0^{(\eta\sigma)})[\ln |\lambda_0^{(\eta\sigma)}| + i \pi Y(\varepsilon)] 
\nonumber\\
&+&\frac{(r\varepsilon)^2}{2} \left[-\bar{\mathrm{Li}}_2(2/\lambda_0^{(\eta\sigma)})+i\pi 
Y(\varepsilon) \ln |\lambda_0^{(\eta\sigma)}|\right]
\eea
We used, inter alia, the duality relation (\ref{eq:dualite}), which implies $\lambda_0^{(-\eta\delta)}=-\lambda_0^{(\eta\alpha)}$, ${P_\delta^{-\eta}}'(0)=-{P_\alpha^\eta}'(0)$
and $Q_{J_\delta^{-\eta}}^{[1]}(\lambda_0^{(-\eta\delta)})=Q_{J_\alpha^{\eta}}^{[1]}(\lambda_0^{(\eta\alpha)})$,
the already encountered relation $s Y(-s\varepsilon)=Y(s)-Y(\varepsilon)$, here with $s_\eta={P_\alpha^\eta}'(0)/|{P_\alpha^\eta}'(0)|$,
the relation $\ln [|\lambda_2^{(\eta\alpha)}|/|\lambda_1^{(\eta\alpha)}|]=s_\eta \ln |\lambda_0^{(\eta\alpha)}| + \mathrm{C}^\infty$.
And also the fact that $\ln[|\lambda_0^{(\eta\alpha)}/\lambda_0^{(\eta\beta)}|]$ is a $C^\infty$ function of $\varepsilon$
in $\varepsilon=0$,
and that the expression $\bar{\mathrm{Li}}_2(1/x)-\bar{\mathrm{Li}}_2(-1/x)-(\pi^2/2)\,\mathrm{sgn}(x)$, because
it is $\int_0^x (dt/t) \ln[|t-1|/|t+1|]$ according to (\ref{eq:defLi2bar}),
is a $\mathrm{C}^\infty$ function of $x\in ]-1,1[$.
We mainly used the fact that the polynomials $P_\alpha^\eta(\lambda)$ 
and $P_\beta^\eta(\lambda)$ have the same first order derivative in $\lambda=0$, which implies
that the roots $\lambda_0^{(\eta\alpha)}$ and $\lambda_0^{(\eta\beta)}$ differ only to {\bf second} order
in $\varepsilon$, see (\ref{eq:devrac}), which explains why the clever combination 
(\ref{eq:comb_ast}) is a $\mathrm{C}^2$ function of $\varepsilon$ in $\varepsilon=0$, while 
the terms $\mathcal{J}_\alpha$ and $\mathcal{J}_\beta$ they are not $\mathrm{C}^2$.

By expanding the terms (\ref{eq:tjab}), and the combination (\ref{eq:comb_ast}), up to order three in $\varepsilon$,
thanks in particular to (\ref{eq:qjq}), then including the contribution (\ref{eq:diff3i}) of the $I_\sigma^\eta(2)$,
we finally find that the {\bf third} order derivative of $\bar{\Sigma}^{(2)}(\bar{K},\varepsilon)$
has a real part that logarithmically diverges and an imaginary part that is discontinuous on the horizontal axis:
\be
\label{eq:ssd0}
\bar{\Sigma}^{(2)}(\bar{K},\varepsilon)\stackrel{\varepsilon\to 0}{=} 
\frac{3r^4 (\bar{K}^2+r^2)}{64 \bar{K}^2 (\bar{K}^2-r^2)^2}[\varepsilon^3+O(\varepsilon^4)][\ln|\varepsilon|+i\pi Y(\varepsilon)]
+\mathrm{C}^\infty
\ee
Here the prefactor diverges in the tangent point of the parabolas $\Delta_\sigma^\eta=0$ with the horizontal axis,
for $\sigma\in \{\alpha,\beta,\gamma,\delta\}$ and $\eta=\pm$, which may be expected.

\subsection{On the oblique lines : The effect of a root equal to two}

To be complete let us study the singularities on the oblique line corresponding to $\eta=+$ or $\eta=-$ in
equation (\ref{eq:exlam2}), in a point of the plane $(\bar{K},\varepsilon)$ where one (and in practice each)
of the polynomials $P_\sigma^\eta(\lambda)$, with fixed $\sigma\in\{\beta,\gamma,\delta\}$ and $\eta$, has
a root equal to two, the other root being different from zero and two.
In the neighborhood of such a point, the two roots of $P_\sigma^\eta(\lambda)$ are of course real;
we shall note $\lambda_0^{(\eta\sigma)}$ the one that is arbitrarily close to two, 
\be
\label{eq:aplin2}
\lambda_0^{(\eta\sigma)}-2\stackrel{d_{\eta}\to 0}{=}-\frac{d_\eta}{{P_\sigma^\eta}'(2)}+O(d_{\eta}^2)
\ee
where the numerator, equal to $P_\sigma^\eta(2)$, is an algebraic distance to the singularity line: 
\be
\label{eq:def_deta}
d_{\eta} \equiv r\varepsilon+4\eta\bar{K}+4=P_\sigma^\eta(2),\ \ \forall \sigma\in\{\beta,\gamma,\delta\}
\ee
The other root $\lambda_0'^{(\eta\sigma)}$ remains outside a neighborhood of zero and two.

As seen in the previous subsection \ref{subsec:local},
singularities can originate here only from the terms $I_\sigma^\eta(2)$ and $J_\sigma^\eta(2)$.
For each term, we find at the considered point that the the third order derivative in the direction normal
to the oblique line has a real part that diverges logarithmically and an imaginary part that is discontinuous.

We first determine the part with singular derivatives of the functional $I[P]$, by omitting the indices
$\sigma\in\{\beta,\gamma,\delta\}$ and $\eta$ for simplicity. In equation (\ref{eq:IPrr}) written for 
$\lambda=2$, one has to keep the first term, as well as the third term for the root $\lambda_0$ close to two.
To isolate the contribution of this root to $u[P(2)]$, let us use the factorization (\ref{eq:decomp})
and the relation $Y(s x)=Y(-s)+s Y(x)$, where $s=\pm 1$ and $x$ is any real, taken here to be the linear approximation
(\ref{eq:aplin2}) of $\lambda_0-2$. As $P'(\lambda_0)$, $P'(2)$ and
$A_P(2-\lambda_0')$ have the same sign when $\lambda_0$ is sufficiently close to two, and as
$\ln |\lambda_0/d|$ is locally a $C^\infty$ function of $d$, we finally obtain in the neighborhood of $d=0$:
\be
\label{eq:Ideux}
I[P](2)=[Q_I^{[1]}(2)-Q_I^{[1]}(\lambda_0)] [\ln |d| + i\pi Y(d)] +\mathrm{C}^\infty
\ee
The prefactor vanishes indeed cubically in $d$, as can be seen thanks to (\ref{eq:decomp})
and to the change of variable $x=t-\lambda_0$ in the integral over $t$ that defines $Q_I$:
\be
Q_I^{[1]}(2)-Q_I^{[1]}(\lambda_0)=\frac{A_P^2}{2} \int_0^{2-\lambda_0} \!\!\! dx\, x^2 (x+\lambda_0-\lambda_0')^2
\stackrel{d\to 0}{=} \frac{d^3}{6 P'(2)} + O(d^4)
\ee

We perform the same analysis for the functional $J[P]$ which, let us recall it, was split in two contributions
according to (\ref{eq:j1pj2}). The result for  $J_1[P]$ can be deduced directly from (\ref{eq:Ideux}) by substituting the
polynomial $Q_I$ with the polynomial $Q_J$. Contrarily to (\ref{eq:Ideux}), the prefactor
\bea
Q_J^{[1]}(2)-Q_J^{[1]}(\lambda_0)=\frac{A_P^2}{2} \int_{\lambda_0}^2 \frac{dt}{t} [(t-\lambda_0)^2(t-\lambda_0')^2-\lambda_0^2\lambda_0'^2]\nonumber\\
=\frac{A_P^2\lambda_0'^2}{2}  \left[-4 u+6u^2+\frac{16(1-\lambda_0')}{3\lambda_0'^2} u^3 -\frac{4 u^4}{3\lambda_0'^2}\right],
\eea
where $u\equiv 1-\lambda_0/2$, vanishes only linearly in $d$, which leads to a singularity in the first order derivative.
However, there is a partial cancellation with the contribution of $J_2[P]$: By transforming in (\ref{eq:J2Prr}) the 
first term and the third term written for the root $\lambda_0$ which is nearest to two,
with the techniques having led to (\ref{eq:Ideux}), and by using the fact that $\bar{\mathrm{Li}}_2(2/\lambda_0)$
is a smooth function of $\lambda_0$ except in $2/\lambda_0=1$, where we have the expansion
\be
\bar{\mathrm{Li}}(2/\lambda_0)=\bar{\mathrm{Li}}_2(1/(1-u))\stackrel{u\to 0}{=} -\left[u+\frac{1}{2}u^2+\frac{1}{3} u^3+O(u^4)\right]
\ln|u| +\mathrm{C}^\infty,
\ee
we find with a rather long calculation that 
\be
J_\sigma^\eta(2)\stackrel{d_{\eta}\to 0}{=} \frac{d_{\eta}^3}{12 {P_\sigma^\eta}'(2)} [1+O(d_\eta)][\ln |d_\eta|+i\pi Y(d_\eta)], \ \ \forall
\sigma\in\{\beta,\gamma,\delta\}
\ee
where we have restored the indices.
This equivalent is precisely half of the one obtained for $I_\sigma^\eta(2)$, for which we give the following simple interpretation:
In the integral (\ref{eq:defJ}) defining $J[P](\lambda)$, here with $\lambda=2$, only the contribution of a neighborhood of
the upper bound $t=2$ can lead to singularities, since it is from there that the root $\lambda_0$ which is closest to two
enters (or exits) the interval of integration when the distance from the singularity line is varied. 
Also the term $[P(0)]^2u[P(0)]/t$ in the integrand of (\ref{eq:defJ}) can be ignored, and $t$ can be approximated
by two in the denominator of $[P(t)]^2u[P(t)]/t$. Then one indeed recovers exactly half of the integrand of $I[P](2)$.   

By collecting all the contributions thanks to the compact notation (\ref{eq:sig2scalli}), we do not find any particular cancellation
between them, so that the {\bf third} order derivative of $\bar{\Sigma}^{(2)}(\bar{K},\varepsilon)$ in the direction
normal to the oblique singularity lines has a logarithmically divergent real part and a discontinuous imaginary part:
\be
\label{eq:sig2obli}
\bar{\Sigma}^{(2)}(\bar{K},\varepsilon) \stackrel{d_\eta\to 0}{=}
\frac{3\eta r [d_\eta^3 + O(d_\eta^4)] [\ln |d_\eta| + i\pi Y(d_\eta)]}{128 \bar{K} (2+\eta\bar{K}) [(2+\eta\bar{K})^2-r^2]} 
+ \mathrm{C}^\infty
\ee
where $d_\eta$, an algebraic distance from these lines, is given by (\ref{eq:def_deta}).
Note that the denominator of the prefactor in (\ref{eq:sig2obli}) vanishes, as expected,
at the points where the considered oblique line is tangent to the parabolas $\Delta_{\sigma}^\eta=0$
of same index $\eta$, for $\sigma=\beta, \gamma$ and $\delta$. In turn, the fact that the denominator 
vanishes in $\bar{K}=0$ corresponds to the crossing points of the two oblique lines, where the distances $d_\pm$
coincide and the contributions of index $\eta=\pm$ interfere; summing them up leads to a finite prefactor.

\section{Some physical applications}
\label{sec:qap}

\subsection{Some results on the complex energy recovered}
\label{subsec:drslecr}

The analytic properties of the resolvent of the Hamiltonian $\hat{G}(z)$, more precisely of its 
matrix elements, 
forbid the resolvent to have a pole in the complex plane, out of those on the real axis associated to the discrete spectrum of $\hat{H}$.
Nevertheless, in the thermodynamic limit, $\hat{G}(z)$ has a branch cut at the location of the continuous spectrum of 
$\hat{H}$, so that the analytic continuation of $\hat{G}(z)$ from the upper half-plane $\Im z>0$ to the lower half-plane $\Im z<0$,
indicated by the exponent ${\rm p.a.}$ in what follows, can have complex poles \cite{CCT}.
 
This discussion extends to the Green's function $\omega\mapsto
\mathcal{G}(\KK,\omega)$, which is a matrix element of the
resolvent in the state $|\psi_\KK^0\rangle$ of the impurity of momentum
$\hbar\KK$ in the presence of the unperturbed Fermi sea, see equation (\ref{eq:Gkov2}).
If $\KK=\mathbf{0}$, we expect that the Green's function has one (and only one) pole
on the real axis, in $\omega_0(\mathbf{0})$, which corresponds to the only discrete eigenstate 
of $\hat{H}$, its ground state, since we have supposed here that the monomeronic branch is the 
minimal energy branch, see the introduction.
If $\KK\neq \mathbf{0}$, $\omega\mapsto\mathcal{G}(\KK,\omega)$ should not have anymore a real pole, 
since no energy argument prevents the impurity from emitting pairs of particle-hole excitations in the fermionic gas,
see the introduction; on the other hand, its analytic continuation $\mathcal{G}^{\rm p.a.}(\KK,\omega)$ to the
lower complex half-plane $\Im \omega <0$ should have a pole in $\omega=\omega_0(\KK)$ that continuously emerges from
the real pole $\omega_0(\mathbf{0})$ and that, by virtue of (\ref{eq:Gkov1}), is a solution of the implicit equation  
\be
\label{eq:impli_autoc}
\Delta E(\KK)\equiv \hbar\omega_0(\KK) = E_\KK +\Sigma^{\rm p.a.} (\KK,\omega_0(\KK))
\ee
where $\Delta E(\KK)$ is called complex energy of the impurity \cite{lettre} and $\Sigma^{\rm p.a.}(\KK,\omega)$ is
the analytic continuation of the self-energy.
Whether $\KK$ is zero or not, only the existence of a pole at the (real or not) angular frequency $\omega_0(\KK)$
allows us to state that the impurity, through the coupling with the Fermi sea, gives birth to a well defined quasi-particle,
here a monomeron. This indeed seems to be the case even in the strongly interacting regime \cite{Svistunov} 
provided that the mass of the impurity remains finite  \cite{polzg2}.
The imaginary part of the pole, 
\be
\Im \omega_0(\KK) \equiv -\frac{\Gamma_0(\KK)}{2} <0 \ \mathrm{if}\  \KK\neq \mathbf{0},
\ee
gives the rate $\Gamma_0$ at which the system leaves exponentially with time its initial state 
$|\psi_\KK^0\rangle$:
It is here a rate of emission of particle-hole pairs.
Let us recall that the probability amplitude in $|\psi_\KK^0\rangle$ also contains, in general, a power law decreasing
term, which is of little practical importance in the weak coupling regime $g\to 0^-$ \cite{CCT}.   

The results of the previous sections allow us to calculate explicitly the complex energy of the impurity
up to second order in the coupling constant $g$. One has just to replace in (\ref{eq:devsig}) the self-energy
with its approximation of order at most two, evaluated in the non-perturbed angular frequency $\omega_0^{(0)}(\KK)=E_\KK$
since $\Sigma^{(1)}$ does not depend on the angular frequency: 
\be
\Delta E(\KK) = E_\KK + \rho g + \frac{(\rho g)^2}{\epsilon_\mathrm{F}} \bar{\Sigma}^{(2)}(\bar{K},\varepsilon=0)+O(g^3)
\label{eq:enercompentdsig}
\ee
given the rescalings (\ref{eq:adim1}) and (\ref{eq:sig2adim}). We then take the limit
$\varepsilon\to 0$ in each term of the expression (\ref{eq:scalli}). 
Each polynomial $P_\sigma^\pm(\lambda)$ has a root that tends to zero and brings, see subsection \ref{subsec:horizon},
a zero contribution to the functionals $I[P]$ and $J_1[P]$, contrarily to the other root. In addition, the contribution
of the piece $J_2[P]$ vanishes due to the factor $[P(0)]^2\propto \varepsilon^2$ in (\ref{eq:j1pj2}).
From the relations $\mathrm{sgn}(y)[Y(2+\frac{y}{x})-Y(\frac{y}{x})]=Y(y)-Y(2x+y)$
and $|y|=y[Y(y)-Y(-y)]$, that hold for all pairs of non-zero real numbers $x$ and $y$ and that are used here for the coefficients
of the quadratic terms and the linear terms of the polynomial $P_\sigma^\eta(\lambda)$, we finally obtain
\be
\bar{\Sigma}^{(2)}(\bar{K},0)=C(\bar{K})+\sum_{s=0,1,r} D_s(\bar{K}) u(s-\bar{K})+D_s(-\bar{K}) u(s+\bar{K}),
\label{eq:sig2jolie}
\ee
Due to an unexpected cancellation of the contributions of the {\sl partie finie} $i_\alpha^\pm$ and
$j_\alpha^\pm$, the result does not involve as {\yvan contributing points $s\pm \bar{K}$} the (half-)coefficients $-r\pm\bar{K}$ of the linear terms
of the polynomials $P_\delta^\pm(\lambda)$.
The function $u(X)$ is here the one of equation (\ref{eq:defu}), and we have introduced the auxiliary functions
\bea
\label{eq:lescoefs}
C(\bar{K}) = \frac{3r(11+\bar{K}^2)}{20(1-r^2)}, \ \ D_0(\bar{K}) =  \frac{3\bar{K}^4}{20 r}, \ \ 
D_r(\bar{K})=-\frac{3(\bar{K}-r)^4(\bar{K}+4r)}{20 r \bar{K}(1-r^2)^2} \nonumber \\
\mbox{and}\ D_1(\bar{K}) = -\frac{3r(\bar{K}-1)^3}{20\bar{K} (1-r^2)^2}[(r^2-2)\bar{K}(\bar{K}+3)+6r^2-2]
\eea
As a whole this reproduces, in a concise form, the results of reference \cite{lettre}. It agrees with 
those of \cite{Bishop} 
which were limited (for $\KK\neq \mathbf{0}$)
to the imaginary part of the energy and to $r=1$.  
In particular, $\Re\bar{\Sigma}^{(2)}(\bar{K},0)$ vanishes indeed for large $\bar{K}$ as in \cite{lettre}, which implies
a sum rule implicitly used in reference \cite{lettre},
\be
\label{eq:rdslp}
\sum_{s=0,r,1} D_s(\bar{K}) + D_s(-\bar{K})=0 \ \forall \bar{K}
\ee
As a consequence, for $0<\bar{K}<\min(1,r)$, the imaginary part of the sum 
over $s$ in equation (\ref{eq:sig2jolie}) reduces to $-\pi D_0(\KK)$ and the
rate of emission of particle-hole pairs simplifies to 
\be
\Gamma_0^{(2)}(\KK)=\frac{(\rho g)^2}{\hbar\Ef} \frac{3\pi\bar{K}^4}{10 r}
\label{eq:gam0pk}
\ee
to order $g^2$.
According to figure~\ref{fig:zones}, $\bar{\Sigma}^{(2)}(\bar{K},0)$ is a $C^\infty$ function
of $\bar{K}$ over $\mathbb{R}^+$ except in $\bar{K}=1$, in $\bar{K}=r$ and, unfortunate
oversight of reference \cite{lettre}, in $\bar{K}=0$.

{\yvan
\subsection{Quasi-particle residue and Anderson orthogonality catastrophe}
The monomeron is a well defined quasi-particle if it has a non-zero quasi-particle residue $Z$. This
can be extracted from the propagator $\mathcal{G}(\KK,\omega)$ defined in equation (\ref{eq:Gkov1}) by isolating
a quasi-particle propagator from a regular part
\be
\mathcal{G}^{\rm p.a.}(\KK,\omega) = \frac{Z}{\hbar \omega - \hbar \omega_0} + \mathcal{G}^{\rm reg}(\KK,\omega)
\ee
where the pole $\omega_0$ is solution of the equation
\be
\label{eq:pole}
\hbar \omega_0 - E_\KK - \Sigma^{\rm p.a.}(\KK,\omega_0) = 0
\ee
Then $Z$ is simply the residue of $\mathcal{G}(\KK,\omega)$:
\be
Z = \lim_{\omega\to\omega_0} \hbar(\omega-\omega_0)\mathcal{G}^{\rm p.a.}(\KK,\omega) = \frac{1}{1-\partial_{\hbar\omega} \Sigma^{\rm p.a.}(\KK,\omega_0)}
\ee
and in the weakly attractive limit one gets the following perturbative expansion up to second order
\be
Z \stackrel{g\to 0^-}{=} 1+\partial_{\hbar\omega} \Sigma^{(2)}(\KK,E_\KK/\hbar) + O(g^3)
\ee
or using $\varepsilon$ instead of $\hbar \omega$ and taking into account equation (\ref{eq:pole}) we get 
\be
\label{eq:zed}
Z \stackrel{g\to 0^-}{=} 1+\left(\frac{\rho g}{\Ef}\right)^2 \partial_\varepsilon \bar{\Sigma}^{(2)}(\bar{K},0) + O(g^3)
\ee

Here we calculate the derivative $\partial_\varepsilon \bar{\Sigma}^{(2)}(\bar{K},\varepsilon)$ for an infinite impurity-to-fermion mass ratio $r=M/m$.
In this limit it is not difficult to see that the trinomials (\ref{eq:pa},\ref{eq:pb},\ref{eq:pc},\ref{eq:pd}) to leading order in $r$ drop off the dependence on 
$\eta=\pm 1$, and one obtains the limit
\be
p_\sigma(\lambda) = \lim_{r\to+\infty} \frac{P_\sigma^\eta(\lambda)}{r}  \ \ \forall \sigma\in \{\alpha,\beta,\gamma,\delta\}
\label{e	q:ppdefinition}
\ee
Therefore also the functionals $I_\sigma^\eta(\lambda)$ and $J_\sigma^\eta(\lambda)$, as well as their corresponding {\sl partie finie} in equation (\ref{eq:scalli}), drop off the dependence on $\eta$ to leading order in $r$, and equation (\ref{eq:sig2scalli}) becomes 
\be
\bar{\Sigma}^{(2)}(\bar{K},\varepsilon) \stackrel{r\to +\infty}{=} \frac{9}{16 r^2} [j_\alpha^+ -J_\beta^+(2)+J_\delta^+(2)] + O(1)
\ee
Using result (\ref{e	q:ppdefinition}), we find that the functionals $J_\sigma^\eta(\lambda)$ show a logarithmic divergence
\be
J_\sigma^\eta(\lambda) \stackrel{r\to +\infty}{=} r^2 \ln r \int_0^\lambda dt \ \frac{p_\sigma^2(t) - p_\sigma^2(0)}{2 t} + O(r^2)
\ee
while their corresponding {\sl partie finie} contribute only to subleading order, i.e. $\mathrm{Pf}\,[J_\sigma^\eta](+\infty)\stackrel{r\to +\infty}{=} O(r^2)$.
This leads to the following result
\be
\bar{\Sigma}^{(2)}(\bar{K},\varepsilon) \stackrel{r\to +\infty}{=} \ln r \frac{9}{16} \int_0^2 dt \ 
\frac{p_\delta^2(t) - p_\delta^2(0)-[p_\beta^2(t)-p_\beta^2(0)]}{2 t}  + O(1)
\ee
which, after the straightforward integration in $t$, becomes
\be
\bar{\Sigma}^{(2)}(\bar{K},\varepsilon) \stackrel{r\to +\infty}{=} -\frac{9}{4}\varepsilon \ln r + O(1)
\ee
Then the quasi-particle residue (\ref{eq:zed}) presents a logarithmic divergence 
\be
\lim _{g\to0^-} \frac{Z-1}{(\rho g/\Ef)^2} \stackrel{r\to +\infty}{=} -\frac{9}{4} \ln r + O(1)
\ee
in agreement with the result of reference~\cite{lettre}. As discussed in~\cite{lettre} this logarithmic divergence is a signature
of the Anderson orthogonality catastrophe stating that in the limit of $r\to+\infty$ the monomeron is not a well defined quasi-particle. 
}

\subsection{A non-perturbative regularisation of the divergence of the second order derivative of 
$\Delta E^{(2)}(\KK)$ at the Fermi surface for $M=m$ ($r=1$)}
\label{subsec:regul}

Overall, the fact of being able to analytically calculate the self-energy up to second order in $g$ has as the most striking
consequence the prediction of singularities in the third order derivative of $\Sigma^{(2)}(\KK,\omega)$.
In order to make more accessible an experimental signature, it is convenient to try to reduce the order 
of the derivatives in which these singularities appear, by identifying the singular point with highest multiplicity
in the plane $(\bar{K},\varepsilon)$. The discussion of section  \ref{sec:les_singus} {\yvan identified} in this plane
singularity lines (\ref{eq:exlam0}) and (\ref{eq:exlam2}),  
on which the polynomials $P_\sigma^\eta(\lambda)$ have roots equal to zero or two,
and singularity parabolas (\ref{eq:doubleab},\ref{eq:doublecd}) on which these polynomials have double roots.
In the half-plane $\bar{K}>0$, the singularity lines cross at $(\bar{K},\varepsilon)=(1,0)$. This point is also on one 
of the parabolas only when the impurity has the same mass of a fermion, from which the magic point
considered in this section:
\be
(\bar{K},\varepsilon) \to (1,0) \ \ \mbox{for}\ \ r=1
\ee
In particular, the associated singularities are exactly at the Fermi surface and, as suggested in reference \cite{lettre}
and as we shall see, they appear in the derivatives of order {\bf two} only.

Here the most accessible observable in a cold atom experiment seems to be the complex energy 
$\Delta E(\KK)$ of the quasi-particle, simply by radio frequency spectroscopy between an internal state of the impurity
not coupled to the fermions, and a coupled internal state. The shift and the broadening of the line due to the
presence of the fermions gives access to the real part and the imaginary part of $\Delta E(\KK)$, with an uncertainty 
which has already reached respectively $5\cdot 10^{-3} \Ef$ and $10^{-4} \Ef$ \cite{Zaccanti}.   
We assume that $\Delta E(\KK)$ can be measured with a sufficiently good precision in the neighborhood of $K=\kf$,
such that it is possible to numerically take the second order derivative with respect to $K$. 
Perturbation theory, whose results have been already published in \cite{lettre} and recovered
in subsection \ref{subsec:drslecr}, leads, for a weakly interacting limit taken at fixed $\bar{K}$ different from one, to
\bea
\lim_{g\to 0^-} \frac{d^2}{d\bar{K}^2} \frac{\Delta E(\KK)-E_\KK}{(\rho g)^2/\Ef} &\stackrel{\bar{K}\to 1}{=}& -\frac{9}{4}\ln |\bar{K}-1|
-\frac{27}{20}(2+\ln 2) \nonumber \\ &&+ \frac{9 i\pi}{4} \left[Y(\bar{K}-1)-\frac{4}{5}\right] +o(1)
\label{eq:resperturblog}
\eea
of real part that diverges logarithmically and discontinuous imaginary part. Perturbation theory is however much more vague
when we take the limit at $\bar{K}=1$,  
\be
\lim_{g\to 0^-} \left[\frac{d^2}{d\bar{K}^2} \frac{\Delta E(\KK)-E_\KK}{(\rho g)^2/\Ef}\right]_{\bar{K}=1} = +\infty
\label{eq:flou}
\ee
since it does not specify how the divergence is produced.

\subsubsection{A self-consistent heuristic approach}

How can we go beyond result (\ref{eq:flou}) by using the ingredients already available in the present work?
We must perform a non-perturbative treatment, as e.g.\ a self-consistent approximation.
The simplest thing consists in replacing the self-energy $\Sigma(\KK,\omega)$,
which appears in the implicit equation (\ref{eq:impli_autoc})  on the complex energy,
with its expansion up to order two included in $g$, $\rho g+\Sigma^{(2)}(\KK,\omega)$. A simple improvement 
of this minimalist prescription is to include the last contribution to $\Sigma^{(3)}(\KK,\omega)$ in 
equation (\ref{eq:sig3}) coming from a mean-field shift on $\omega$, given that at fixed wave vector
and angular frequency, 
\be
\Sigma^{(2)}(\KK,\omega-\rho g/\hbar)\stackrel{g\to 0^-}{=}\Sigma^{(2)}(\KK,\omega) -\frac{\rho g}{\hbar} \partial_\omega \Sigma^{(2)}(\KK,\omega)+ O(g^2)
\ee
Physically, this global shift takes into account the fact that the mean-field shift $\rho g$ experienced by the impurity 
is exactly the same in all subspaces at zero, one, two, $\ldots$ pairs of particle-hole excitations, once 
the limit of zero-range interaction has been taken.
We shall stick then to the (non-perturbative) self-consistent heuristic approximation   
\be
\label{eq:autoc1}
\Delta E(\KK)\stackrel{\mathrm{heuris}.}{=} E_\KK + \rho g + \Sigma^{(2)\mathrm{p.a.}} (\KK,\frac{\Delta E(\KK)-\rho g}{\hbar})
\ee
which can be written, in terms of a reduced unknown $\varepsilon_e$, complex effective value of the variable $\varepsilon$
(hence the index $e$), in the dimensionless compact form
\be
\label{eq:autoc2}
-{\varepsilon_e}(\bar{K})\!\stackrel{\mathrm{heuris}.}{=}\! \left(\frac{\rho g}{\Ef}\right)^{\!2}\! \bar{\Sigma}^{(2)\mathrm{p.a.}}
(\bar{K},{\varepsilon_e}(\bar{K}))\ \ \mbox{where}\ \ 
{\varepsilon_e}(\bar{K})\!\equiv\! \frac{E_\KK + \rho g - \Delta E(\KK)}{\Ef}
\ee
which we differentiate twice with respect to $\bar{K}$ to identify the useful derivatives of 
$\bar{\Sigma}^{(2)}$:
\be
\label{eq:derivcomp}
-\frac{d^2\varepsilon_e}{d\bar{K}^2}\!\stackrel{\mathrm{heuris}.}{=}\! \left(\frac{\rho g}{\Ef}\right)^2
\left[\partial^2_{\bar{K}}+2 \frac{d\varepsilon_e}{d\bar{K}} \partial_{\bar{K}}\partial_{\varepsilon}
+\left(\frac{d\varepsilon_e}{d\bar{K}}\right)^2 \partial^2_\varepsilon 
+ \frac{d^2{\varepsilon_e}}{d\bar{K}^2}\ \partial_\varepsilon \right]\bar{\Sigma}^{(2)\mathrm{p.a.}}
\ee
taken here at the point $(\bar{K},\varepsilon={\varepsilon_e}(\bar{K}))$. Let us recall that the exponent $\mbox{p.a.}$ means {\sl analytic continuation}
to the complex values of $\omega$ from the upper half-plane to the lower half-plane.

\subsubsection{Singularities of the second order derivatives of $\bar{\Sigma}^{(2)}$ and scaling law prediction}
\label{sss:singularitiesandscaling}

In order to see how the second order derivative of ${\varepsilon_e}(\bar{K})$ behaves in the neighborhood of $\bar{K}=1$
in the limit $g\to 0^-$, it is sufficient to initially determine the singularities of the second order derivatives 
of $\bar{\Sigma}^{(2)}(\KK,\varepsilon)$ for $\varepsilon$ real.
By transposing to the case $(\bar{K},\varepsilon)\to (1,0)$ the techniques developed in section \ref{sec:les_singus},
we note that some magic cancellations, as the quasi-identity of certain roots
$\lambda_0^{(\eta\alpha)}$ and $\lambda_0^{(\eta\beta)}$ of the polynomials $P_\alpha^\eta$
and $P_\beta^\eta$, which made the second order derivatives regular, do not happen anymore, and we laboriously 
end up with the following results: 
\bea
\label{eq:dkk}
\partial_{\bar{K}}^2 \bar{\Sigma}^{(2)}(\bar{K}=1,\varepsilon) &\stackrel{\varepsilon\to 0}{=}&
-\frac{9}{4} \ln |\varepsilon|+O(1) \\
\label{eq:dee}
\partial_{\varepsilon}^2 \bar{\Sigma}^{(2)}(\bar{K}=1,\varepsilon) &\stackrel{\varepsilon\to 0}{=}&
\frac{9}{128} (\ln |\varepsilon|)^2 + O(\ln |\varepsilon|) \\
\label{eq:dke}
\partial_\varepsilon\partial_{\bar{K}}\bar{\Sigma}^{(2)}(\bar{K}=1,\varepsilon) &\stackrel{\varepsilon\to 0}{=}&
\frac{9}{8} \ln |\varepsilon| + O(1)
\eea
The unknown ${\varepsilon_e}(\bar{K})$ is of second order in $g$, as well as its first order derivative,
thus the derivatives (\ref{eq:dee}) and (\ref{eq:dke}),
that diverge only logarithmically in $g$ are {\yvan suppressed} in equation (\ref{eq:derivcomp})
by the factors $d{\varepsilon_e}/d\bar{K}=O(g^2)$ and $(d{\varepsilon_e}/d\bar{K})^2=O(g^4)$.
As for the first order derivative with respect to $\varepsilon$ in (\ref{eq:derivcomp}), which does not diverge,
it is {\yvan suppressed} by the factor $(\rho g/\Ef)^2$, as can be seen after collecting with the term of the first member 
in this same equation.
Hence the drastic simplification in the limit $g\to 0^-$, even in the neighborhood of $\bar{K}=1$:
\be
\label{eq:drastique}
-\partial_{\bar{K}}^2 {\varepsilon_e}(\bar{K}) \stackrel{\mathrm{approx.}}{=} \left(\frac{\rho g}{\Ef}\right)^2
\partial_{\bar{K}}^2 \bar{\Sigma}^{(2)\mathrm{p.a.}}(\bar{K},{\varepsilon_e}(\bar{K})) + O[(\frac{\rho g}{\Ef})^4\ln |\frac{\rho g}{\Ef}|]
\ee
Our heuristic self-consistent approach thus predicts that the first member of the equation (\ref{eq:flou}), evaluated in
$\bar{K}=1$, diverges logarithmically when $g\to 0^-$:
\be
\label{eq:cestlog}
\left[\frac{d^2}{d\bar{K}^2} \frac{\Delta E(\KK)-E_\KK}{(\rho g)^2/\Ef}\right]_{\bar{K}=1} \stackrel{\mathrm{heuris}.}{=} 
-\frac{9}{4} \ln \left[\left(\frac{\rho g}{\Ef}\right)^2\right]+O(1)
\ee

It is actually possible to find this result, to make it more precise and to extend it to $\bar{K}\neq 1$,
by performing a clever calculation
of the second order derivative of $\bar{\Sigma}^{(2)}(\bar{K},\varepsilon)$ with respect to $\bar{K}$. Let us start from
the identities (\ref{eq:sig2opera}) and (\ref{eq:avec_coupure}), and obtain the second order derivatives of the integral quantities 
$I_\sigma^\eta(\lambda)$, with $\lambda=2$ or $\Lambda$, $\eta=\pm$ and $\sigma\in\{\alpha,\beta,\gamma,\delta\}$,
by taking the derivative of their defining expressions (\ref{eq:defI})
and (\ref{eq:defJ}) with respect to $\bar{K}$ under the integral sign 
[the same trick is valid for the derivative with respect to $\varepsilon$
and leads directly to (\ref{eq:dee}) and (\ref{eq:dke})].
As can be verified on the equations (\ref{eq:pa},\ref{eq:pb},\ref{eq:pc},\ref{eq:pd}),
$\partial_{\bar{K}} P_\sigma^\eta(t)= 2\eta t$ with $\varepsilon$ and $t$ fixed, thus
\be
\label{eq:deriv2astuce}
\partial_{\bar{K}}^2 I_\sigma^\eta(\lambda)=\int_0^\lambda dt\, 4t^2 u[P_\sigma^\eta(t)] \ \ \mbox{and}\ \ 
\partial_{\bar{K}}^2 J_\sigma^\eta(\lambda)=\int_0^\lambda dt\, 4t\, u[P_\sigma^\eta(t)]
\ee
where the function $u$ is the one of equation (\ref{eq:defu}) and $u^{[2]}$ in (\ref{eq:defI},\ref{eq:defJ})
one of the primitives of order two. To see which one of these terms have a finite limit 
when $(\bar{K},\varepsilon)\to (1,0)$, and can only contribute to $\partial_{\bar{K}}^2 \bar{\Sigma}^{(2)}$
as a slowly varying background, it is sufficient to replace the trinomials $P_\sigma^\eta$ with
their value for $\bar{K}=1$ and $\varepsilon=0$, see table~\ref{tab:limpoly}. 

\begin{table}[htb]
\begin{center}
\begin{tabular}{|c||c|c|}
\hline
$\sigma$ & $\eta=+$ & $\eta=-$ \\
\hline\hline
$\alpha$ & $2t^2+4t$ & $2t^2$ \\
\hline
$\beta$  & $4t$ & $0$ \\
\hline
$\gamma$ & $t^2+2t$ & $t^2-2t$ \\
\hline
$\delta$ & $2t^2$  & $2t^2-4t$  \\
\hline
\end{tabular}
\end{center}
\caption{The trinomials $P_\sigma^\eta(t)$ for the multi critical point $(\bar{K},\varepsilon)=(1,0)$.}
\label{tab:limpoly}
\end{table}

We than see that only the polynomial $P_\beta^-(t)$ leads to a divergence. As $r=1$, it manifests itself only {\sl via}
the functional $J[P]$, which is particularly neat on equation (\ref{eq:scalli}), so that
\be
\partial_{\bar{K}}^2 \bar{\Sigma}^{(2)}(\bar{K},\varepsilon) \stackrel{(\bar{K},\varepsilon)\to (1,0)}{=}
-\frac{9}{32} \partial_{\bar{K}}^2 J_\beta^-(2) + \frac{9}{40}(-17+14\ln 2 +2i\pi) +o(1)
\label{eq:d2sd}
\ee
where the additive constant was obtained by specializing to $\varepsilon=0$  and by comparing with (\ref{eq:resperturblog}).
To calculate the integral giving $\partial_{\bar{K}}^2 J_\beta^-(2)$ in (\ref{eq:deriv2astuce}),
it remains to use equation (\ref{eq:IPrr}) with $Q_I(t)=4t$ and to simplify the imaginary part with the help of
the relation $-\mathrm{sgn}(x)[Y(2+\frac{y}{x})-Y(\frac{y}{x})]=Y(y)-Y(2x+y)$ valid for any real number $x$ and $y$:
\be
\label{eq:d2jbmexpli}
\frac{1}{8} \partial_{\bar{K}}^2 J_\beta^-(2)=u(\varepsilon)+[1-(\frac{\varepsilon/4}{\bar{K}-1})^2]
\{u[\varepsilon-4(\bar{K}-1)]-u(\varepsilon)\}-\frac{1}{2}-\frac{\varepsilon/4}{\bar{K}-1}
\ee
To continue and to draw the consequences in (\ref{eq:drastique}), we must extend this result to the case 
$\Im(-\varepsilon)>0$ then analytic continue it to the case $\Im(-\varepsilon)<0$, which is what we are going to do.
Let us first note a remarkable property: The terms of the second member 
of  (\ref{eq:d2jbmexpli}) are positively homogeneous functions of $(\bar{K}-1,\varepsilon)$ of degree zero [invariant
 by global multiplication of $\bar{K}-1$ and $\varepsilon$ by any real number $\tau>0$], {\bf except} the first term.
This first term thus fixes the global value of (\ref{eq:d2jbmexpli});
as $\varepsilon$ has to be taken here of order $|{\varepsilon_e}(\bar{K}=1)|\approx (\rho g/\Ef)^2$,
it immediately leads to the logarithmic behavior (\ref{eq:cestlog}).
The other terms of (\ref{eq:d2jbmexpli})
give the dependence in $\bar{K}-1$, not described by (\ref{eq:cestlog}), which is produced on a characteristic 
scale $|\varepsilon|\approx (\rho g/\Ef)^2$. We obtain thus, within our self-consistent heuristic approach (\ref{eq:autoc1}),
the following scaling law for the second order derivative of the complex energy of the impurity in the neighborhood
of the Fermi surface ($\bar{K}\to 1$) in the weakly interacting limit ($g\to 0^-$):
\be
\label{eq:loiech}
\frac{d^2}{d\bar{K}^2} \frac{\Delta E(\KK)-E_\KK}{(\rho g)^2/\Ef}\stackrel{\mathrm{heuris}.}{=} 
-\frac{9}{4} \ln \left[\left(\frac{\rho g}{\Ef}\right)^2\right]+
F\left(\frac{4(\bar{K}-1)}{(\rho g/\Ef)^2}\right)+o(1)
\ee
where the scaling function $F(x)$ remains to be specified. A simple but remarkable consequence of this scaling law
is that the {\bf third} order derivative of the complex energy of the impurity does not tend uniformly to zero in the
limit of weak interaction: 
\be
\label{eq:ntpuvz}
\frac{d^3}{d\bar{K}^3} \frac{\Delta E(\KK)}{\Ef} \stackrel{g\to 0^-}{\centernot{\to}} 0
\ \ \mbox{in a neighborhood of}\ \ \bar{K}=1
\ee

\subsubsection{Analytic continuation to a complex energy variable and numerical emergence of the scaling law}
\label{subsubsec:paauveceendllde}

In order to see the scaling law (\ref{eq:loiech}) emerge when the strength of the interaction is reduced,
we have implemented the self-consistent heuristic program of equation (\ref{eq:autoc2}).
This led us to overcome a practical obstacle, that is to determine the analytic continuation
of $\bar{\Sigma}^{(2)}(\bar{K},\varepsilon)$ to complex values of $\varepsilon$.
Let us give the main steps that we followed to realize it. (i) Results of section \ref{sec:cedsigdlcg}
can be generalized directly to the case $\Im (-\varepsilon )>0$, since equation (\ref{eq:sig2}) has
the energy $\hbar \omega +i 0^+$ in the denominator. This corresponds to the complex upper half-plane    
for the energy variable $z$ of the resolvent of the Hamiltonian $\hat{G}(z)$, so that one can extend
$\hbar \omega$ to a positive imaginary part in equation (\ref{eq:adim1}), 
and correspondingly $\varepsilon$ to a negative imaginary part, without crossing the branch cut of the resolvent,
thus without the need for any analytic continuation.
(ii) In this favorable case $\Im (-\varepsilon )>0$, the roots of the polynomials $P_\sigma^\eta(\lambda)$ are all complex,
so we must use the form (\ref{eq:J2Pri}) of the functional $J_2[P_\sigma^\eta]$,
and the form (\ref{eq:IPri}) of the functional $I[P_\sigma^\eta](\lambda)$,
in which we must care to replace $u[P_\sigma^\eta(\lambda)]$ with 
$\ln[-P_\sigma^\eta(\lambda)]$. This follows from the remark below equation (\ref{eq:un}) and from the fact that
$-P_\sigma^\eta(\lambda)$ tends to the real axis from the upper complex half-plane when $\Im(-\varepsilon)\to 0^+$.
Alternatively this follows from the result of the integration of (\ref{eq:moychapK}) for $\epsilon$ positive non-infinitesimal, which leads
formally to $u(x\pm y-i\epsilon)=\ln(-x\mp y+i\epsilon)$ thus to $u(z)=\ln(-z)$. 
In the aforementioned expressions, let us recall it, $\ln$ and $\mathrm{Li}_2$ are the usual branch of the complex logarithm and dilogarithm function,
of branch cut $\mathbb{R}^-$ and $[1,+\infty[$.
(iii) To verify the two previous assertions (i) and (ii),
we can take the limit $\Im (-\varepsilon)\to 0^+$ in those 
generalizations of (\ref{eq:IPri}) and (\ref{eq:J2Pri}), in the case where the roots $\lambda_0^{c}$
of the polynomial $P_\sigma^\eta(\lambda)$
have real limits $\lambda_0^r$. We then have to recover exactly expressions (\ref{eq:IPrr}) and (\ref{eq:J2Prr}).
We have scrupulously verified that this is indeed the case, by using the relation $\Pi_{[0,\lambda]}(y)-\Pi_{[0,\lambda]}(x)=
Y[(\lambda-x)(\lambda-y)]-Y[xy]$ satisfied for the rectangular function $\Pi_{[0,\lambda]}$ for any real $x<y$,
as well as the property 
\be
\frac{d}{d \Im (-\varepsilon)} [P_\sigma^\eta(\lambda_0^c)]=0 \Longrightarrow
\frac{d\lambda_0^c}{d \Im (-\varepsilon)} \stackrel{\Im(-\varepsilon)\to 0^+}{\to} 
\frac{1}{P'(\lambda_0^r)}
\ee
which allows us to know if the roots $\lambda_0^c$, thus the arguments of $\ln$ and of $\mathrm{Li}_2$,
reach the real axis from the upper or lower complex half-plane. This leads to: 
\bea
\ln(\lambda-\lambda_0^c)\stackrel{\Im(-\varepsilon)\to 0^+}{\to} \ln |\lambda-\lambda_0^r| -i\pi \frac{P'(\lambda_0^r)}
{|P'(\lambda_0^r)|} Y(\lambda_0-\lambda) \\
\mathrm{Li}_2(\frac{\lambda}{\lambda_0^c}) \!\!\stackrel{\Im(-\varepsilon)\to 0^+}{\to}\!\! 
\bar{\mathrm{Li}}_2(\frac{\lambda}{\lambda_0^r})
+i\pi  \frac{P'(\lambda_0^r)}{|P'(\lambda_0^r)|} \ln \left|\frac{\lambda_0^r}{\lambda}\right|[Y(\lambda-\lambda^r_0)-Y(-\lambda^r_0)]
\eea
knowing that $\mathrm{sgn}\,(\lambda)Y(\lambda/\lambda_0^r-1)=Y(\lambda-\lambda^r_0)-Y(-\lambda^r_0)$
and $\mathrm{Li}_2(x\pm i 0^+)=\pm i\pi (\ln|x|) Y(x-1)$ for any real $x$.
In these expressions, $\lambda$ is any real number and the polynomial $P$ is the limit of the polynomial $P_\sigma^\eta$
for $\varepsilon$ real. (iv) To finally analytically continue the functionals $I[P]$ and $J[P]$, thus the self-energy
$\bar{\Sigma}^{(2)}(\KK,\varepsilon)$ from the half-plane $\Im(-\varepsilon)>0$ to the half-plane $\Im(-\varepsilon)<0$,
it is sufficient to know if the argument $Z$ of each function $\ln$ and $\mathrm{Li}_2$ moves from the upper half-plane
to the lower half-plane or vice versa. In the first case, we move the branch cut of $\ln$ from the real negative half-axis
to the purely imaginary negative half-axis, and the one of $\mathrm{Li}_2$ from $[1,+\infty[$ to $1+i\mathbb{R}^-$,  
that is we rotate them by $\pi/2$ and $-\pi/2$ respectively:
\be
\label{eq:feb}
\ln^{\mathrm{p.a.}\downarrow}\!\!Z\!=\!\ln_{3\pi/2}Z\ \mbox{and}\ \mathrm{Li}^{\mathrm{p.a.}\downarrow}_2(Z)\!=\!\mathrm{Li}_2(Z)
+\ln_\pi\!Z [\ln_\pi(1-Z)-\ln_{\pi/2}(1-Z)]
\ee
where the arrow $\downarrow$ recalls the movement of $Z$ in the complex plane, and
\be
\ln_\theta Z\equiv\ln |Z| + i \arg_\theta Z
\ee
is defined with the branch $\theta-2\pi < \arg_\theta Z\leq \theta$ of the argument of the complex number $Z$.
In the opposite case where $Z$ moves from the lower half-plane to the upper half-plane, we rotate the branch cut
of $\ln$ by an angle $-\pi/2$, so as to displace it to the purely imaginary positive half-axis, and we rotate
the one of $\mathrm{Li}_2$ by an angle $\pi/2$, so as to displace it to $1+i \mathbb{R}^+$:
\be
\label{eq:feh}
\ln^{\mathrm{p.a.}\uparrow}\!\!Z\!=\!\ln_{\pi/2} Z\ \mbox{and}\ \mathrm{Li}^{\mathrm{p.a.}\uparrow}_2(Z)\!=\!\mathrm{Li}_2(Z)
+\ln_\pi\!Z [\ln_\pi(1-Z)-\ln_{3\pi/2}(1-Z)]
\ee

\begin{figure}[htb]
\includegraphics[width=8cm,clip=]{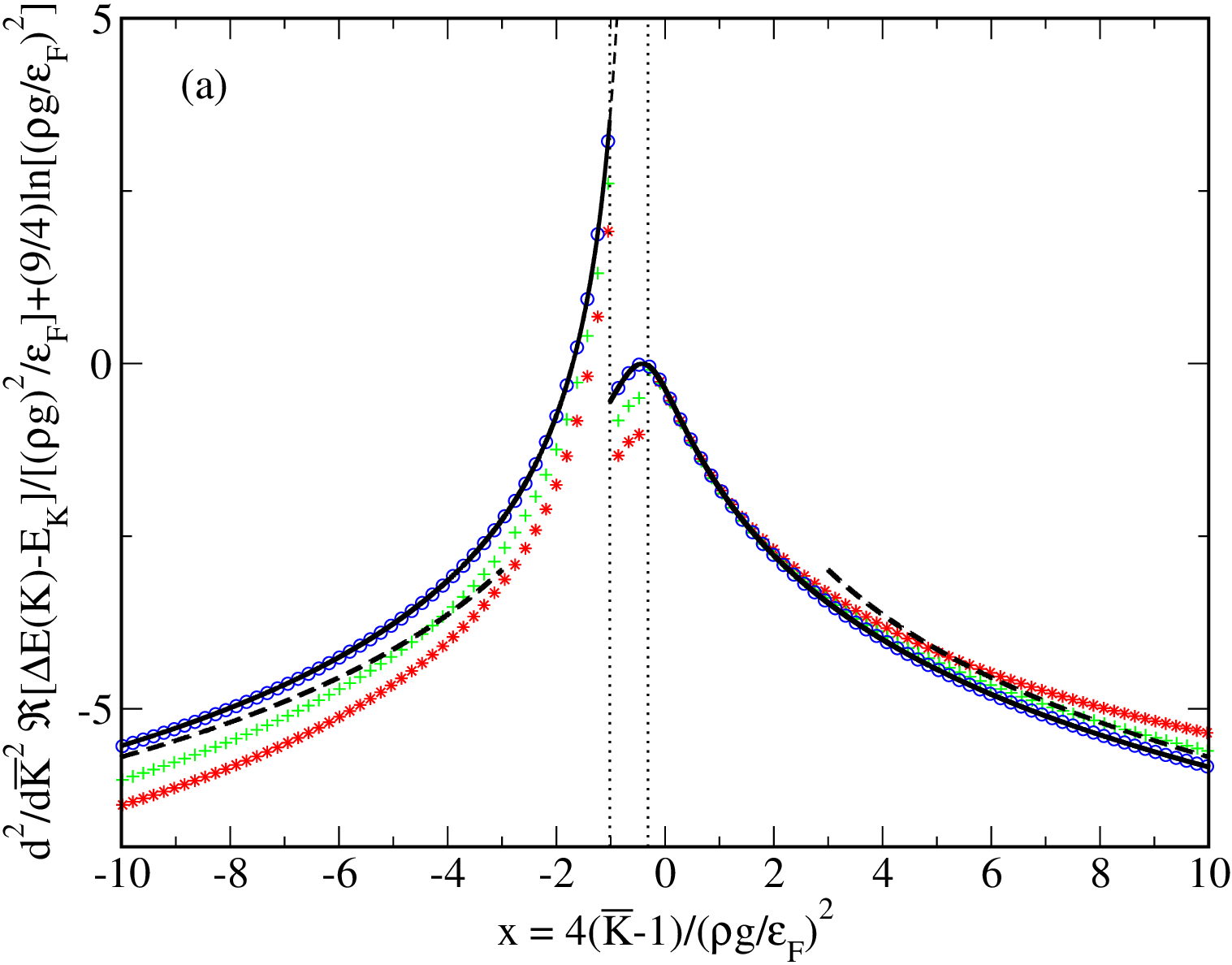}
\includegraphics[width=8cm,clip=]{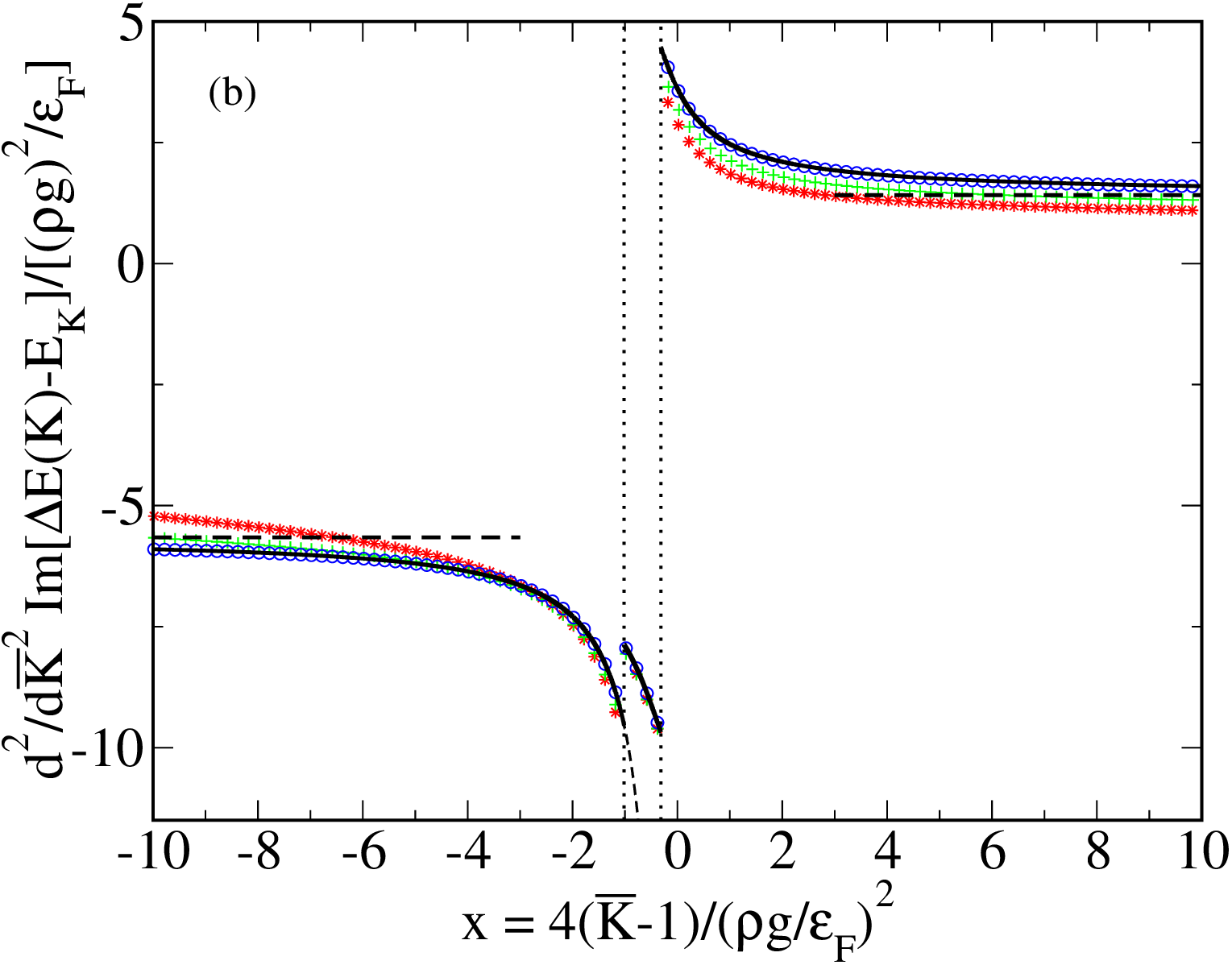}
\caption{Second order derivative of the complex energy of the impurity [(a): real part {\yvan shifted by $(9/4)\ln[(\rho g/\Ef)^2]$}, (b): imaginary part]
obtained by numerical resolution 
of the self-consistent equation (\ref{eq:autoc2}), for a mass ratio $r=M/m=1$ and for values of 
$\rho g/\Ef$ equal to $-0.15$ (red stars), $-0.1$  (green plus signs) and $-0.01$ (blue circles), the last
value having practically reached the limit $g\to 0^-$. 
In the chosen units system, we can see emerge a scaling law of the form (\ref{eq:loiech}).
Black solid line: Analytic prediction in the limit $g\to 0^-$;
the location of the corresponding discontinuities in the second order derivative are 
identified by the vertical dotted lines of abscissas $x'_{\rm jump}$ and $x_{\rm jump}$ from left to right, see
equation (\ref{eq:xsaut}).
Black thin dashed line: Analytic prediction corresponding to another possible writing of the function 
$\partial_{\bar{K}}^2 J_\beta^-(2)$ {\sl before} its analytic continuation; it differs from the black solid line only in between the two vertical
dotted lines.
Black thick dashed line: Perturbation theory (\ref{eq:resperturblog}) deduced from reference \cite{lettre}, limited
to its validity domain, that is the tails.}
\label{fig:loiech}
\end{figure}

We have numerically implemented this procedure of analytic continuation, by iteratively solving the self-consistent
equation (\ref{eq:autoc2}) and by calculating the second order derivative of $\varepsilon_e(\bar{K})$ using the 
middle-point method. We show
the result on figure  \ref{fig:loiech}, for three values of $\rho g/\Ef$ corresponding to subsequently weaker interactions.
The choice of the origin and the units on the axis presuppose a scaling law of the form (\ref{eq:loiech}),
towards which the numerics seem indeed to converge. Let us remark however, that even in this self-consistent
approach, the second order derivative of $\varepsilon_e(\bar{K})$ presents, as a function of $\bar{K}$, discontinuities
which affect the real part as well as the imaginary part, and that do not disappear when $\rho g/\Ef\to 0^-$.
Unfortunately, we shall see that the position and the number of these discontinuities do not have any physical meaning
since they depend on the form under which we write the different functions before analytically continue them.

To finish, let us show how to analytically obtain the limits of the results of figure \ref{fig:loiech}
when $g\to 0^-$, that is how to obtain an explicit expression of the corresponding scaling function 
$F(x)$, where $x=4(\bar{K}-1)/(\rho g/\Ef)^2$. We take as the starting point equations (\ref{eq:d2sd}) and (\ref{eq:d2jbmexpli}),
which is legitimate to directly extend (without analytic continuation)
to the complex values of $\varepsilon$ with $\Im(-\varepsilon)>0$, by formally consider that $u(z)=\ln(-z)$
on $\mathbb{C}$, where $\ln$ is the principal branch of the complex logarithm, to obtain
\be
\partial_{\bar{K}}^2 \bar{\Sigma}^{(2)}(\bar{K},\varepsilon)|_{\Im(-\varepsilon)>0}  \!\!
\stackrel{(\bar{K},\varepsilon)\to (1,0)}{=}\!\! -\frac{9}{4} \left\{\ln [(\frac{\rho g}{\Ef})^2]
+f_x(-\bar{\varepsilon})\right\} +C_{\mathrm{bg}} + o(1)
\ee
with $\bar{\varepsilon}\equiv \varepsilon/(\rho g/\Ef)^2$, $C_{\mathrm{bg}}=(9/20)(7\ln 2-6+i\pi)$ and
\be
\label{eq:fx1}
f_x(-\bar{\varepsilon})|_{\Im(-\bar{\varepsilon})>0}=\ln(x-\bar{\varepsilon})-\frac{\bar{\varepsilon}^2}{x^2}
\ln(1-x/\bar{\varepsilon})-\frac{\bar{\varepsilon}}{x}
\ee
where we used the fact that the function $x\mapsto \ln(x-\bar{\varepsilon})-\ln(-\bar{\varepsilon})-
\ln(1-x/\bar{\varepsilon})$, which is $C^\infty$ on $\mathbb{R}$ since the arguments of the complex logarithm 
cannot cross its branch cut $\mathbb{R}^-$, is zero in $x=0$ and of derivative zero everywhere, therefore
it is identically zero. Then, one has to analytically continue the function $f_x(-\bar{\varepsilon})$
from $\Im(-\varepsilon)>0$ to $\Im(-\varepsilon)<0$, at $\bar{K}-1$ thus $x$ fixed, by following the procedure 
exposed around equations (\ref{eq:feb}) and (\ref{eq:feh}). The argument $x-\bar{\varepsilon}$ of the first logarithm
crosses the real axis downwards; the argument $1-x/\bar{\varepsilon}$ of the second logarithm crosses the real axis
upwards if $x>0$, and downwards if $x<0$. We finally obtain
\be
\label{eq:fxpa}
f_x^{\mathrm{p.a.}\downarrow}(-\bar{\varepsilon})=\ln_{3\pi/2}(x-\bar{\varepsilon})-\frac{\bar{\varepsilon}^2}{x^2}
\ln_{\pi Y(-x) +\pi/2} (1-x/\bar{\varepsilon}) -\frac{\bar{\varepsilon}}{x}
\ee
The scaling function is then given by
\be
F(x) = -\frac{9}{4} f_x^{\mathrm{p.a.}\downarrow}(-\bar{\varepsilon}_e^{(2)}(\bar{K}=1))+ C_{\rm bg}
\ee
where $-\bar{\varepsilon}_e^{(2)}(\bar{K})=(\Delta E^{(2)}(\bar{K})-E_\KK-\rho g)/[(\rho g)^2/\Ef]=\bar{\Sigma}^{(2)}(\bar{K},0)$ 
is the value of $-\bar{\varepsilon}_e(\bar{K})$ to second order in perturbation theory, which can be deduced
from equation (\ref{eq:sig2jolie}) or from reference \cite{lettre}.

Notice on figure \ref{fig:loiech} that the numerical self-consistent results indeed converge towards 
this scaling function $F(x)$ when $\rho g/\Ef\to 0^-$.
The observed discontinuities as functions of $x$ take place when the arguments of the logarithms
in equation (\ref{eq:fxpa}) become purely imaginary and thus cross their branch cut:
\be
\label{eq:xsaut}
x_{\rm jump}=\Re [\bar{\varepsilon}_e^{(2)}(\bar{K}=1)] \ \ \mbox{and}\ \ x'_{\rm jump}=\frac{1}{\Re[1/\bar{\varepsilon}_e^{(2)}(\bar{K}=1)]}
\ee
which bring in $F(x)$ a purely imaginary jump at $x=x_{\rm jump}$ (the first logarithm has a real prefactor) and a complex jump
at $x=x'_{\rm jump}$ (the second logarithm has a complex prefactor). The values of the abscissas (\ref{eq:xsaut}) are actually 
in decreasing order, and are identified by vertical dotted lines on figure \ref{fig:loiech}.  

The positions and the numbers of these predicted discontinuities, therefore also the values of the scaling function $F(x)$,
are actually arbitrary. To see it, it is sufficient to choose for the function $f_x(-\bar{\varepsilon})$ a different but 
equivalent form {\sl before} the analytic continuation. The property set after equation (\ref{eq:fx1}) allows us to write 
\be
f_x(-\bar{\varepsilon})|_{\Im(-\bar{\varepsilon})>0}=\left(1-\frac{\bar{\varepsilon}^2}{x^2}\right) 
[\ln(x-\bar{\varepsilon})-\ln(-\bar{\varepsilon})]+\ln(-\bar{\varepsilon})-\frac{\bar{\varepsilon}}{x}
\ee
whose analytic continuation to negative values of $\Im(-\bar{\varepsilon})$
following {\sl the same} prescriptions (\ref{eq:feb}) and (\ref{eq:feh}) leads to
\be
\label{eq:fxpa2}
f_x^{\mathrm{p.a.}\downarrow}(-\bar{\varepsilon})\stackrel{\mathrm{bis}}{=}\left(1-\frac{\bar{\varepsilon}^2}{x^2}\right)
[\ln_{3\pi/2}(x-\bar{\varepsilon})-\ln_{3\pi/2}(-\bar{\varepsilon})]+\ln_{3\pi/2}(-\bar{\varepsilon})
-\frac{\bar{\varepsilon}}{x}
\ee
The corresponding scaling function $F_{\rm bis}(x)$, represented as a thin dashed line on figure~\ref{fig:loiech}, 
differs from the one deduced from (\ref{eq:fxpa}) only between the two vertical dotted lines
($x'_{\rm jump}<x<x_{\rm jump}$); it features a complex discontinuity (instead of purely imaginary)
in $x=x_{\rm jump}$, and has no longer a discontinuity in $x=x'_{\rm jump}$. 

Is it possible to make the remaining discontinuity disappear, at least on the real part of $F(x)$?
Coming from $x=-\infty$, we start in the left Riemann sheet of the function (\ref{eq:fxpa2}), that
can be followed continuously up to $x=+\infty$ by turning the angle of the branch cut from $3\pi/2$ to $2\pi$.
Similarly, starting at $x=+\infty$, the right Riemann sheet of the same function (\ref{eq:fxpa2}) can be followed 
up to $x=-\infty$, by rotating the branch cut from $3\pi/2$ to $\pi$. As 
$\ln_{2\pi}(x-\bar{\varepsilon})=2i\pi+\ln_\pi(x-\bar{\varepsilon})$ for any $x\in\mathbb{R}$ and any complex
$-\bar{\varepsilon}$ of negative imaginary part, the branches of $f_x^{\mathrm{p.a.}\downarrow}(-\bar{\varepsilon})$
in the two Riemann sheets differ by the quantity $2i\pi [1-(\bar{\varepsilon}/x)^2]$ which, neither zero nor
purely imaginary, makes a discontinuity inevitable. 

Ultimately, we do not know how to determine the true physical value of the scaling function $F(x)$,
even in the apparently innocent framework of the self-consistent approximation (\ref{eq:autoc1}),
at least for $|x|\lesssim 1$, that is for $4|\bar{K}-1|\lesssim (\rho g/\Ef)^2$.

{\yvan
\subsection{Smoothness of the complex energy $\Delta E^{(2)}(\KK)$ at finite temperature}
\label{subsec:finiteT}

Since the singularities in the derivatives of the impurity complex energy  $\Delta E^{(2)}(\KK)$ are a consequence of the 
existence of a Fermi surface it is natural to ask what happens when the temperature of the system is non zero.
In this case, one obtains
\be
\Delta E^{(2)}_T(\KK) = g^2\!\! \int_{\mathbb{R}^3}\!\! \frac{d^3q}{(2\pi)^3}\! \int_{\mathbb{R}^3}\!\! \frac{d^3k}{(2\pi)^3} \!\!
\left[\frac{2\mu}{\hbar^2 k^2}-\frac{1-\bar{n}(\kk)}{F_{\kk,\qq}(\KK,E_\KK/\hbar)}\right] \bar{n}(\qq)
\label{eq:sig2temp}
\ee
The difference with the zero-temperature result is that the Heaviside functions have been replaced by the Fermi functions
\be
\bar{n}(\kk) = \frac{1}{e^{\beta(\epsilon_\kk-\mu_{\rm F})}+1}
\label{eq:avocc}
\ee
with $\mu_{\rm F}$ being the chemical potential of the Fermi gas and $\epsilon_\kk = \hbar^2 k^2/(2m)$ the kinetic energy of a fermion
with wave vector $\kk$. In the strongly degenerate regime $k_BT\ll \mu_{\rm F}$, the chemical potential is close to the Fermi energy and
the Fermi function has a width 
\be
\delta k_{\rm typ} = \frac{T}{T_{\rm F}} \kf
\label{eq:Tscale}
\ee 
as can be seen by linearizing the dispersion relation $\epsilon_\kk$ around $k=k_{\rm F}$.
We thus expect that the divergence of the second order derivative of $\Delta E^{(2)}(\KK)$, for equal masses $m=M$, is interrupted in
$\Delta E^{(2)}_T(\KK)$ at a distance $|\bar{K}-1|\approx T/T_{\rm F}$.
To confirm this prediction in the experimentally relevant range we integrate numerically~\footnote{For a given temperature $T\neq 0$, the chemical potential $\mu_{\rm F}$ is adjusted such as to reproduce
the fermionic density at zero temperature, i.e. $\rho = \int_{\mathbb{R}^3} \frac{d^3 k}{(2\pi)^3}\bar{n}(\kk)$.} equation (\ref{eq:sig2temp}) for decreasing 
values of $T/T_{\rm F}$, see the resulting second order derivatives in figure~\ref{fig:loiechT}. 

At zero temperature, in subsection~\ref{sss:singularitiesandscaling} we have seen the emergence of a scaling law for the second order derivatives
of ${\Delta E}^{(2)}(\KK)$ of typical $K$-width given by $(\rho g/\Ef)^2 \kf$. 
We thus predict that one can observe the zero temperature
effects studied in this work even at a non-zero temperature provided that
\be
\frac{T}{T_{\rm F}} \ll \left(\frac{\rho g}{\Ef}\right)^2 
\label{eq:condTnegli}
\ee

\begin{figure}[htb]
\includegraphics[width=8cm,clip=]{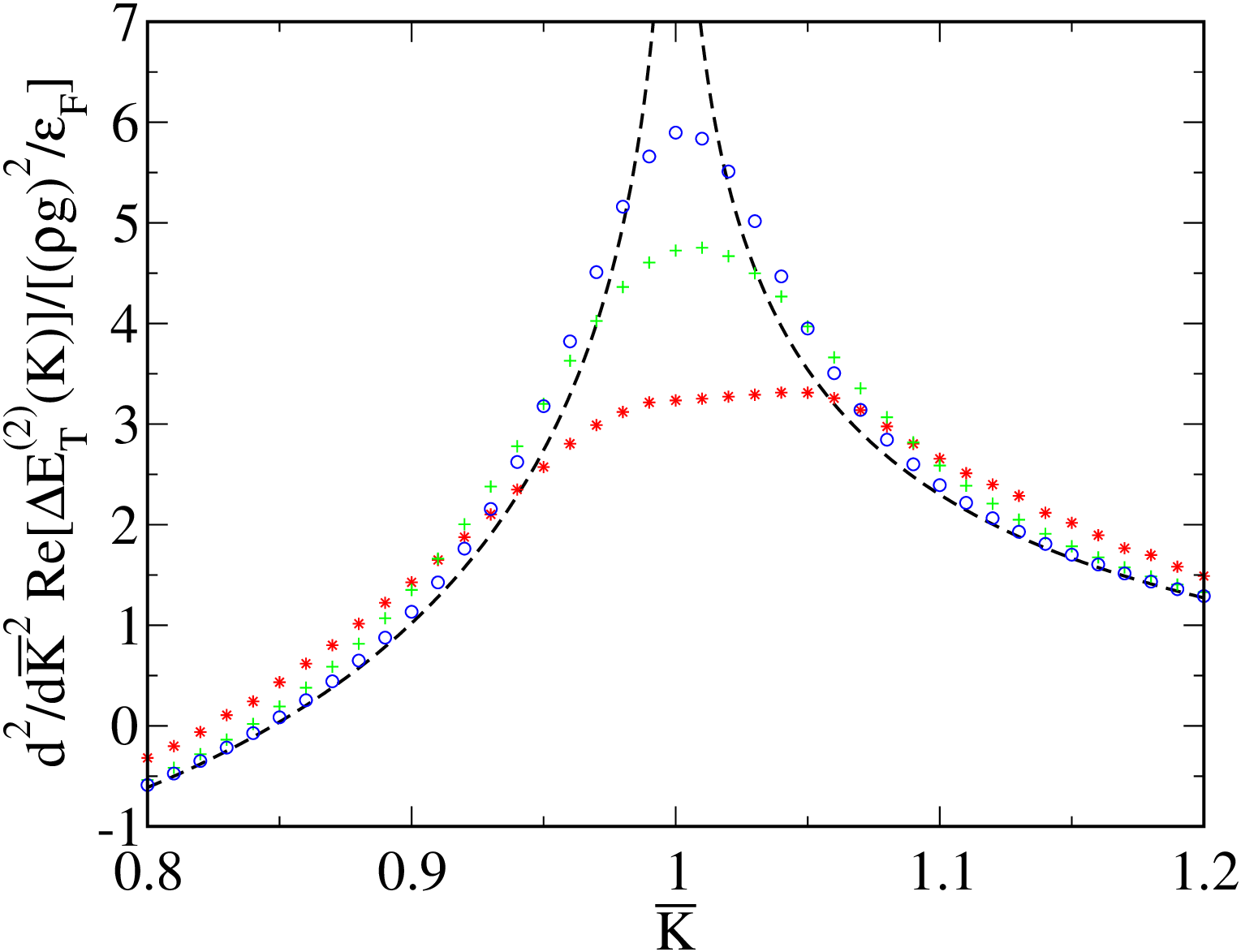}
\caption{Second order derivative of the non-zero temperature complex energy of the impurity of order two obtained by numerical integration 
of equation (\ref{eq:sig2temp}), for a mass ratio $r=M/m=1$ and for values of 
$T/T_{\rm F}$ equal to $0.1$ (red stars), $0.05$  (green plus signs), $0.025$ (blue circles).
Black thick dashed line: Zero temperature result \cite{lettre}. 
At $0<T\ll T_{\rm F}$, the  typical scale on which the divergence of the zero temperature result is interrupted is given by the difference of the 
abscissas of the crossing points between the non-zero (symbols) and zero temperature (dashed line) results. It scales linearly in $T/T_{\rm F}$.}
\label{fig:loiechT}
\end{figure}

}

\subsection{Moments of the momentum of the particle-hole pair emitted by the impurity, according to
Fermi's golden rule. Application to the damping rate and to the diffusion coefficient of the impurity momentum}
\label{subsec:lmdlidlpptepli}

In this part we suppose that the impurity is prepared at time $t=0$ in the state of well defined wave vector
$\KK$, in the presence of the unperturbed Fermi sea.
This initial state of the system is directly coupled, by the interaction potential $\hat{V}$
between the impurity and the fermions, to the excited fermionic states of one particle-hole pair,
while the impurity recoils, absorbing the corresponding momentum change. We wish to describe the evolution of 
the system at short times, in the weakly interacting regime.   

In the thermodynamic limit, the initial state is a discrete state $|\mathrm{i}\rangle$, and the final states
$|\mathrm{f}\rangle$ with one particle-hole pair belong to a continuum.
Fermi's golden rule then gives the rate of emission of one particle-hole pair by the impurity, to second
order included in the coupling constant $g$. Here, the elementary transition rate of $|\mathrm{i}\rangle$
towards $|\mathrm{f}\rangle$ simply reads $\delta \Gamma_{\mathrm{if}}=(2\pi g^2/\hbar) 
\delta(E_\mathrm{f}-E_\mathrm{i})$,
where $E_{\mathrm{i/f}}$ is the total kinetic energy of the impurity and the fermions, in the initial or
final state.   

With the calculus techniques developed in this article, 
we have easily access to all moments of the wave vector $\kk-\qq$ of the emitted particle-hole pair, as
predicted by Fermi's golden rule, that is to zeroth order in $g$:
\be
\langle (\kk-\qq)^n\rangle^{(0)}=\frac{2\pi g^2}{\hbar \Gamma_0^{(2)}(\KK)} \int_{q<\kf<k}\!\!\frac{d^3qd^3k}{(2\pi)^6}
(\kk-\qq)^n \delta(E_{\KK-\kk+\qq}+\epsilon_\kk-\epsilon_\qq-E_\KK)
\label{eq:rof}
\ee
where the integer power $\kk^n$ of a vector $\kk$ is the vector $\kk ||\kk||^{n-1}$ for odd $n$,
and the scalar $||\kk||^n$ for even $n$:
\be
\kk^n\equiv\kk ||\kk||^{n-1} \ (n\ \mbox{odd}),\ \kk^n\equiv||\kk||^n \ (n\ \mbox{even})
\label{eq:defpuisvec}
\ee
The total departure rate $\Gamma_0(\KK)$, here written to order two in $g$,
is simply the total rate of emission of a pair at this order, and it is already given by 
(\ref{eq:enercompentdsig})-(\ref{eq:sig2jolie}).
By conservation of the total momentum, this also gives  access to the moments of the momentum change
$\delta \KK=\KK_{\rm fin}-\KK$ experienced by the impurity, to order zero in $g$,
that we shall use physically in the subsection \ref{subsubsec:tacdii}:
\be
\langle (\delta \KK)^n\rangle^{(0)}=(-1)^n \langle (\kk-\qq)^n\rangle^{(0)}
\label{eq:mK}
\ee

Before proceeding to the calculation of these moments, let us give a {\bf warning}
in the limit of a zero-range interaction.
We have indeed taken such a limit in equation (\ref{eq:rof}), $b\to 0$ in the lattice model.
The momentum distribution of the impurity interacting with the fermions, however, has an asymptotic tail $C/\mathcal{K}^4$ 
at large wave numbers $\mathcal{K}$, where $C$ is called the ``contact",
so that its average kinetic energy must diverge
\cite{Tan,BraatenGen,Leggett,WernerGenSimple,WernerGenLong},
as well as the moments of $\delta \KK$ and of $\kk-\qq$ of order $n>1$.
Physically this is due to non-resonant transition processes neglected by the
Fermi's golden rule that may be easily taken into account by calculating the evolution
operator to first order in $g$. Also the expression (\ref{eq:rof}), that may be considered as 
mathematically interesting in itself and that will appear in the kinetic equation models as the
one of reference \cite{DavidHuse}, has for a zero-range interaction a clear physical meaning
only for $n=0$ and $n=1$.  	

However, expression (\ref{eq:rof}) is {\bf legitimate}
in the framework of our perturbative treatment of the lattice model interaction,
where the coupling constant $g$ tend to zero with a fixed interaction range (lattice spacing)
$b$: Contrarily to the usual experimental regime for cold atoms,
the impurity-to-fermion scattering at zero energy is supposed to be strongly non-resonant,
with a scattering length $a$  much smaller than $b$ in the absolute value, see equations 
(\ref{eq:lienga}) and (\ref{eq:g0_dev}).
The constant $C$ (the contact) is proportional to the derivative of the total energy with respect to $1/a$,
both for a zero-range interaction \cite{Tan,BraatenGen} and for the lattice model
\cite{WernerGenSimple,WernerGenLong} once $\kf b\ll 1$ and $K b\ll 1$,
and the leading correction to the energy of the ideal gas is here the mean-field one $\rho g$,
so that $C=O(\rho a^2)$ and the contribution of the term in $C/\mathcal{K}^4$ to the moments
of order $n\geq 2$ of the impurity  wave vector is $O(\rho a^2/b^{n-1})$, given that wave vectors
are restricted to the first Brillouin zone.
This contribution, even though divergent if $b\to 0$ for $a$ fixed, is actually of second order in $g$:
It is negligible in the expressions (\ref{eq:rof}) and (\ref{eq:mK}) that are of order zero in $g$.

\subsubsection{Moments of even order $n$}

We can directly reuse the calculations of subsection \ref{subsec:pisis}, with the notable simplification 
that $\hbar\omega=E_\KK$ in equation (\ref{eq:defF}), and $\varepsilon=0$ in equation (\ref{eq:adim1}).
Moreover, only the imaginary part of the result matters,
$\Im [1/F_{\kk,\qq}(\KK,E_\KK)]$ giving $\pi$ times the Dirac delta distribution that ensures
conservation of kinetic energy in Fermi's golden rule, so that
\be
\label{eq:resmompsi}
\langle (\kk-\qq)^n\rangle^{(0)}\stackrel{n\ \mathrm{even}}{=}  \frac{\kf^n (\rho g)^2}{\Ef\hbar\Gamma_0^{(2)}(\KK)}
\frac{9r}{4\bar{K}} \int_0^{+\infty} d\lambda\, \lambda^n \Im[\psi^+(\lambda)-\psi^-(\lambda)]
\ee
where $\bar{K}$ is the initial wave number of the impurity in units of $\kf$,
and $\psi^\pm(\lambda)$ is given by (\ref{eq:forme_inf}) or (\ref{eq:forme_sup}) 
if $\lambda$ is smaller or larger than $2$.
As in subsection \ref{subsec:eetddf}, we write the integral over $[2,+\infty[$
as the difference of the integrals over $[0,+\infty[$ and $[0,2]$ of the same integrand.
Since all polynomials $P_\sigma^\eta(\lambda)$ now vanish in zero, it is convenient to
introduce the reduced polynomials $p_\sigma^\eta(\lambda)\equiv P_\sigma^\eta(\lambda)/\lambda$:
\bea
\label{eq:pred1}
p_\alpha^\eta(\lambda)= (1+r)\lambda+2(r+\eta\bar{K}), &\ \ \ & p_\beta^\eta(\lambda)= (1-r)\lambda+2(r+\eta\bar{K}) \\
\label{eq:pred2}
p_\gamma^\eta(\lambda)= \lambda+2\eta\bar{K}, &\ \ \ & p_\delta^\eta(\lambda)=(1+r)\lambda+2(-r+\eta\bar{K})
\eea
all of degree one, with real root $\lambda_0^{(\eta\sigma)}$, for $\sigma\in\{\alpha,\beta,\gamma,\delta\}$
and $\eta=\pm$. The expressions (\ref{eq:forme_inf}) and (\ref{eq:forme_sup}) lead to the functions
$\Im u^{[s]}(X)=\pi Y(X) X^s/s!$, $s=2$ or $s=3$, to be taken in the values of the polynomials $P_\sigma^\eta(\lambda)$.
Since $\lambda>0$, we can replace $Y[P_\sigma^\eta(\lambda)]$ with $Y[p_\sigma^\eta(\lambda)]$.
From the powers of $P_\sigma^\eta(\lambda)$ one can pull out factors $\lambda^s$. This directly leads,
without integrating by parts, to a functional of the type $\int_0^\lambda dt \mathcal{Q}(t) Y[p(t)]$,
where $\mathcal{Q}(t)$ is any polynomial and $p(t)$ is a polynomial of degree one with real coefficients.
Writing $p(t)=at+b$, of root $\lambda_0=-b/a$, we can simplify the imaginary part of relation (\ref{eq:IPrr})
as
\be
\int_0^\lambda dt\, \mathcal{Q}(t) Y(at+b)= [\mathcal{Q}^{[1]}(\lambda)-\mathcal{Q}^{[1]}(-\textstyle{\frac{b}{a}})] Y(a\lambda+b) +
\mathcal{Q}^{[1]}(-\textstyle{\frac{b}{a}})Y(b)
\ee
In practice, $p(t)$ is one of the reduced polynomials $p_\sigma^\eta(\lambda)$, and $Q(t)$ is
the corresponding auxiliary polynomial, given by 
\bea
\mathcal{Q}_\sigma^\eta(\lambda)=\frac{\lambda^{n+1}}{48r^3}[p_\sigma^\eta(\lambda)]^2[p_\sigma^\eta(\lambda)-6r]
\ \ \forall \sigma\in\{\alpha,\beta\},  \nonumber\\
\mathcal{Q}_\gamma^\eta(\lambda)=\frac{\lambda^{n+2}}{8r^2} [p_\gamma^\eta(\lambda)]^2, \ \ \ 
\mathcal{Q}_\delta^\eta(\lambda)=-\mathcal{Q}_\alpha^{-\eta}(-\lambda)
\eea
The last identity results from the duality property $p_\delta^{\eta}(\lambda)=-p_\alpha^{-\eta}(-\lambda)$;
by simple integration over $\lambda$, it leads to the relation $\mathcal{Q}_\delta^{\eta[1]}(\lambda)=\mathcal{Q}_\alpha^{-\eta[1]}(-\lambda)$,
which allows us, later in the calculation, to eliminate the contributions in $Y(-r\pm \bar{K})$ thanks to the relation
$Y(-r\pm \bar{K})=1-Y(r\mp\bar{K})$ without any remainder.
The final result thus has the same structure as the imaginary part of equation (\ref{eq:sig2jolie}):
It is a pure combination of Heaviside functions without remainders since the quantity $C(\KK)$ in (\ref{eq:sig2jolie}) is real: 
\be
\langle (\kk-\qq)^n\rangle^{(0)}\stackrel{n\ \mathrm{even}}{=} \frac{\kf^n (\rho g)^2}{\Ef\hbar\Gamma_0^{(2)}(\KK)}
\frac{9\pi r}{4\bar{K}}\sum_{s=0,r,1}\sum_{\eta=\pm} Y(s+\eta\bar{K}) \mathcal{D}_s^\eta(\bar{K})
\label{eq:resmompair}
\ee
The prefactors $\mathcal{D}_s^\eta(\bar{K})$ of the Heaviside functions are given by
\bea
\mathcal{D}_0^\eta(\bar{K})=\eta\mathcal{Q}_\gamma^{\eta[1]}(\lambda_0^{(\eta\gamma)}), \ \
\mathcal{D}_1^\eta(\bar{K})=\sum_{\sigma=\beta,\gamma,\delta} 
\eta[\mathcal{Q}_\sigma^{\eta[1]}(2)-\mathcal{Q}_\sigma^{\eta[1]}(\lambda_0^{(\eta\sigma)})], \nonumber \\
\label{eq:pref01r}
\mathcal{D}_r^\eta(\bar{K})=\eta[\mathcal{Q}_\beta^{\eta[1]}(\lambda_0^{(\eta\beta)})
-\mathcal{Q}_\delta^{-\eta[1]}(\lambda_0^{(-\eta\delta)})] 
\eea
The primitives $\mathcal{Q}_\sigma^{\eta[1]}(\lambda)$ of the polynomials
$\mathcal{Q}_\sigma^\eta(\lambda)$ that vanish in $\lambda=0$ may be calculated,
we shall however give an explicit result only for the $s=0$ contribution:
It is extremely simple, and it fully determines the result for small values 
of $\bar{K}$, see hereafter.
One has the relation
\be
\label{eq:relpm}
\mathcal{D}_s^-(\bar{K})=-\mathcal{D}_s^+(-\bar{K}) \ \ \forall\bar{K},
\ee
due to the fact that $\psi^-(\lambda)$ can be deduced from $\psi^+(\lambda)$ in equation (\ref{eq:resmompsi}) by changing $\bar{K}$ to $-\bar{K}$.
Also, the prefactors obey a sum rule similar to equation (\ref{eq:rdslp}),
as a direct consequence of the sum rule on the auxiliary polynomials:
\be
\label{eq:rdsmom}
\sum_{\eta=\pm} \sum_{\sigma=\beta,\gamma,\delta} \eta \mathcal{Q}_\sigma^\eta(\lambda)=0\ \forall \lambda
\ \ \mbox{thus} \ \ \sum_{s=0,r,1}\sum_{\eta=\pm} \mathcal{D}_s^\eta(\bar{K}) =0 \ \forall \bar{K}
\ee
This causes the sum over $\eta$ and $s$ into equation (\ref{eq:resmompair}) to reduce to
$-\mathcal{D}_s^{-}(\bar{K})$ for $0< \bar{K}< \min(1,r)$, in which case, given the equations
(\ref{eq:enercompentdsig}), (\ref{eq:sig2jolie}), (\ref{eq:lescoefs}):
\be
\label{eq:momfkp}
\langle (\kk-\qq)^n\rangle^{(0)}\stackrel{n\ \mathrm{even}}{=} \frac{60 (2K)^n}{(n+3)(n+4)(n+5)}
\ee

\subsubsection{Moments of odd order $n$}

Odd order moments are vectors. They are invariant by rotation around the axis $\hat{\KK}\equiv \KK/K$,
that is the direction of $\KK$, so that they are parallel to that direction. The trick of averaging over $\hat{\KK}$
has now to be applied to the scalar quantity $\hat{\KK}\cdot\langle(\kk-\qq)^n\rangle^{(0)}$,
which amounts to adding a factor $w$ in the integrand of equation  (\ref{eq:moychapK})
and a factor $x/y$ in the imaginary part of (\ref{eq:deff}),
the variables $x$ and $y$ being those of (\ref{eq:deflamxy}). This leads finally to the single integral 
\be
\label{eq:resmompsit}
\langle (\kk-\qq)^n\rangle^{(0)}\stackrel{n\ \mathrm{odd}}{=}  \hat{\KK} \frac{\kf^n (\rho g)^2}
{\Ef\hbar\Gamma_0^{(2)}(\KK)}
\frac{9\pi r}{8\bar{K}^2} \int_0^{+\infty} d\lambda\, \lambda^{n-1} [\tilde{\psi}^+(\lambda)-\tilde{\psi}^-(\lambda)]
\ee
with, for $\eta=\pm$,
\be
\tilde{\psi}^\eta(\lambda)\equiv \int_{\max(1-\lambda,0)}^{1} \bar{q} d\bar{q} \int_{\max(\lambda-\bar{q},1)}^{\lambda+\bar{q}}
\bar{k}d\bar{k}
\, x Y(x+2 \eta \bar{K}\lambda)
\ee
This double integral is of the same form as (\ref{eq:defpsipm}) with $\varepsilon=0$, 
except that the function $u(X)$ of (\ref{eq:defu}) must be replaced with $\tilde{u}^\eta_\lambda(X)=(X-2\eta\bar{K}\lambda) Y(X)$
which parametrically depends on $\lambda$ and on $\eta$.
Fortunately the calculation procedure exposed below (\ref{eq:un}) applies to a generic function $u(X)$.
The explicit value of $\tilde{\psi}^\eta(\lambda)$ can then be deduced directly from expressions (\ref{eq:forme_inf}) 
and (\ref{eq:forme_sup}) by replacing $u$ with $\tilde{u}$,
and the resulting primitives of order $s$ are simply
\be
\tilde{u}_\lambda^{\eta [s]}(X) = Y^{[s+1]}(X)-2\eta\bar{K} \lambda Y^{[s]}(X),
\ \ \ \mbox{with} \ \ Y^{[s]}(X)=\frac{X^s}{s!} Y(X)
\ee
The rest follows, as in the case of even order moments, with the same reduced polynomials 
(\ref{eq:pred1},\ref{eq:pred2}), with different expressions for the auxiliary polynomials: 
\bea
{\mathcal{Q}}_\sigma^\eta(\lambda)= \frac{\lambda^{n+1}}{192 r^3} [p_\sigma^\eta(\lambda)]^2 
\left\{[p_\sigma^\eta(\lambda)]^2-4 p_\sigma^\eta(0) p_\sigma^\eta(\lambda)+48\eta r \bar{K}\right\} \ \ \forall
\sigma\in\{\alpha,\beta\}, \nonumber \\
\label{eq:auximomimp}
{\mathcal{Q}}_\gamma^\eta(\lambda)= \frac{\lambda^{n+2}}{24 r^2}[p_\gamma^\eta(\lambda)]^2 [p_\gamma^\eta(\lambda)-6\eta\bar{K}],\ \ 
{\mathcal{Q}}_\delta^\eta(\lambda)=-{\mathcal{Q}}_\alpha^{-\eta}(-\lambda)
\eea
but with the same duality relations, in particular ${\mathcal{Q}}_\delta^{\eta [1]}(\lambda)={\mathcal{Q}}_\alpha^{-\eta [1]}(-\lambda)$.
This leads to a result of the same form as  (\ref{eq:resmompair}) up to the vectorial factor $\hat{\KK}/(2\bar{K})$:
\be
\label{eq:resmomimpair}
\langle (\kk-\qq)^n\rangle^{(0)}\stackrel{n\ \mathrm{odd}}{=} \hat{\KK} \frac{\kf^n (\rho g)^2}
{\Ef\hbar\Gamma_0^{(2)}(\KK)} \frac{9\pi r}{8\bar{K}^2}
\sum_{s=0,r,1}\sum_{\eta=\pm} Y(s+\eta\bar{K}) \mathcal{D}_s^\eta(\bar{K})
\ee
Here the prefactors $\mathcal{D}_s^\eta(\bar{K})$ of the Heaviside functions are still given by the expressions (\ref{eq:pref01r}),
now written for the auxiliary polynomials (\ref{eq:auximomimp}).
As for even $n$, we have the relation (\ref{eq:relpm}), since $\tilde{\psi}^-(\lambda)$ can be deduced from $\tilde{\psi}^+(\lambda)$
in equation (\ref{eq:resmompsit}) by changing $\bar{K}$ to $-\bar{K}$, and we also have the chain of sum rules (\ref{eq:rdsmom}),
as shown by an explicit calculation on the auxiliary polynomials (\ref{eq:auximomimp}).
For $0<\bar{K}<\min(1,r)$, we thus have the equivalent of the result (\ref{eq:momfkp}), with a different factor in the denominator:
\be
\label{eq:momfki}
\langle (\kk-\qq)^n\rangle^{(0)}\stackrel{n\ \mathrm{odd}}{=} \frac{60 (2K)^n \hat{\KK}}{(n+3)(n+4)(n+6)}
\ee

\subsubsection{Damping rate and diffusion coefficient of the impurity momentum}
\label{subsubsec:tacdii}

By conservation of the total momentum, see equation (\ref{eq:mK}), and by applying Fermi's golden rule,
we can deduce from the moments (\ref{eq:rof}) of the momentum of the emitted particle-hole pair, the initial
rate of variation of the moments of the impurity momentum $\PP$ away from its initial value $\hbar \KK$:
\be
\frac{d}{dt} \langle (\PP-\hbar\KK)^n\rangle(t=0)\stackrel{g\to 0^-}{=}
(-\hbar)^n \Gamma_0^{(2)}(\KK) \langle(\kk-\qq)^n\rangle^{(0)}+O(g^3)
\ee
where the integer power of a vector is taken in the sense of (\ref{eq:defpuisvec}).
Two values of $n$ deserve further developments. The case $n=1$ corresponds
to the initial damping rate of the average momentum of the impurity, 
\be
\frac{d}{dt} \langle \PP\rangle(t=0)\equiv -\Gamma_P(\KK) \langle \PP\rangle(t=0)
\ee
It has been studied in reference \cite{StringariFermi} with the Fermi liquid theory,
which is exact for any coupling constant $g$ but only holds for an arbitrary weak $\bar{K}$.
Here, we obtain a complementary prediction, which is exact for any $\bar{K}$ but holds for an arbitrary  
weak coupling constant: To order two in $g$,
\be
\Gamma_P^{(2)}(\KK) = \Gamma_0^{(2)}(\KK)\hat{\KK}\cdot \langle (\kk-\qq)\rangle^{(0)}/K
\ee
where we recall that $\hat{\KK}\equiv \KK/K$.
As its explicit expression is of a reasonable length, we give in the usual reduced form
$\bar{\Gamma}^{(2)}_P(\bar{K})\equiv \hbar \Gamma_P^{(2)}(\KK)\Ef/(\rho g)^2$
and with the presentation adopted in reference \cite{lettre}: We distinguish
(i) the region of small wave numbers $0<\bar{K}<\min(1,r)$, where
\be
\label{eq:region1}
\bar{\Gamma}^{(2)}_P(\bar{K})=\frac{9\pi \bar{K}^4}{35 r},
\ee
(ii) the region of intermediate wave numbers for an impurity lighter than a fermion, $r<\bar{K}<1$, where
\bea
\label{eq:region2}
\bar{\Gamma}^{(2)}_P(\bar{K})=\frac{3\pi r}{35(r^2-1)^3}[(3r^4-9r^2+10)\bar{K}^4-14(r^2+3)\bar{K}^2\nonumber \\
+35 r(r^2+3)\bar{K}-35 r^2 (r^2+3) +14 r^3 (r^2+3)\bar{K}^{-1} -r^5 (r^2+3) \bar{K}^{-3}],
\eea
(iii) the intermediate region for an impurity on the contrary heavier than a fermion, $1<\bar{K}<r$, where
\bea
\label{eq:region3}
\bar{\Gamma}^{(2)}_P(\bar{K})=\frac{3\pi r}{35(r^2-1)^3}[-(1+3r^{-2})\bar{K}^4+14(3+r^2) \bar{K}^2 -35 (1+3r^2) \bar{K} 
\nonumber \\
+35 r^2(3+r^2)-14 r^2 (1+3 r^2) \bar{K}^{-1}+(3-9r^2+10r^4)\bar{K}^{-3}]
\eea
and (iv) the region of large wave numbers, $\max(1,r)<\bar{K}$, where
\be
\label{eq:region4}
\bar{\Gamma}^{(2)}_P(\bar{K})=\frac{3\pi r}{35 (1+r)^3}[35\bar{K}+14 r^2\bar{K}^{-1} -(r^4+3r^3+9r^2+9r+3)\bar{K}^{-3}]
\ee
The prediction of reference \cite{StringariFermi} for $\Gamma_P(\KK)$,  
based on the Fermi liquid theory, is indeed an equivalent of our result at $\bar{K}\to 0$
when we specialize it to the limit of weak interaction, which amounts to replacing the effective mass
$m^*_\downarrow$ of the impurity with its bare mass $M$ and the effective coupling constant of the
monomeron with its value to order one in $g$, $\gamma^{(1)}=m \kf a/(\pi \mu)$.

The case $n=2$ corresponds to the initial momentum diffusion coefficient of the impurity, that is
to the derivative of the variance of its momentum:
\be
\frac{d}{dt} [\langle \PP^2\rangle-\langle\PP\rangle^2](t=0)\equiv 2 D_P(\KK)
\ee
We obtain here its value to leading order, that is to second order in $g$:
\be
2 D_P^{(2)}(\KK) = \hbar^2 \Gamma_0^{(2)}(\KK) \langle (\kk-\qq)^2\rangle^{(0)}
\ee
For conciseness, we give the explicit value only for $0<\bar{K}<\min(1,r)$, readily deduced from equations
(\ref{eq:gam0pk}) and (\ref{eq:momfkp}):
\be
\label{eq:coefdifftotalexpli}
2 D_P^{(2)}(\KK) =\frac{(\rho g)^2}{\hbar\Ef} (\hbar \kf)^2 \frac{12\pi\bar{K}^6}{35 r}
\ee
This diffusion coefficient is actually the trace of the momentum diffusion tensor,
which is anisotropic for $\KK\neq\mathbf{0}$ and can be calculated to order $g^2$ with the techniques of this article.
From $\Gamma_P$ and $D_P$ we can deduce the damping rate of the impurity kinetic energy.
It is worth recalling the warning that follows equation (\ref{eq:rof}): Contrarily to $\Gamma_P(\KK)$,
the meaning of the coefficient $D_P(\KK)$ remains to be clarified for a zero-range interaction, if we 
describe the evolution of
the impurity coupled to the fermions beyond
the kinetic equation model of reference \cite{DavidHuse}.

\section{Conclusion}
\label{sec:conclusion}

The problem considered here of one impurity interacting with an ideal Fermi gas of spin-polarized fermions
(at zero temperature) belongs to the general class of polaronic problems. It shows with beauty and simplicity 
the very general fact that a particle, through the effect of a coupling to a system of continuous spectrum (in the
thermodynamic limit), gives birth to a quasi-particle, with an energy at rest, a mass, a dispersion relation, etc, 
different from those of the bare particle.
The problem shows a renewed interest thanks to cold atoms experiments, which involve a mixture of two species
or two spin-states of the same fermionic species.  

This problem was mainly studied at zero total momentum, that is for an immobile impurity before coupling 
to the Fermi sea. The main physical motivation was indeed  to determine the energy at rest and the effective mass 
of the quasi-particle, which allows one to estimate the equation of state of the Fermi ``liquid" consisting of a non-zero
density of impurities, and thus the critical ratio of chemical potential between the impurities and the fermions below which
this Fermi liquid is more favorable than the paired superfluid phase \cite{Chevy,LoboStringari}.
In this case, the variational approach of references \cite{Chevy,Combescot_varia,Combescot_deux} remains quantitatively 
correct in the strongly interacting regime, which allows one to go beyond the perturbative study of reference \cite{Bishop}.

The case of a moving impurity is more subtle. The quasi-particle acquires a finite lifetime in its considered momentum subspace 
since it can slow down by emitting particle-hole pairs in the Fermi sea \cite{BishopNucl,StringariFermi}, which is {\yvan not} reproduced
by the variational approach in a neighborhood of zero momentum \cite{lettre}.
So we turned to the weakly interacting regime in a perturbative calculation up to second order included in the coupling constant $g$.
Going against some stereotypes, we think that this regime is rather interesting:
Non-systematic predictions may lead to very different scaling laws, see for example \cite{WernerGenSimple}, while the strongly interacting regime
(the unitary limit) is often subject to rather technical debates on the pure numbers that are the values of the observables, and whose 
ultimate determination needs the heavy machinery of quantum Monte Carlo simulations.   

Let us now review some salient results of this long article.

Perhaps unexpectedly, the self-energy function $\Sigma(\KK,\omega)$ of the impurity,
when limited to second order included in $g$, can be expressed in a fully explicit way,
that is the sextuple integral over the momenta of the virtually created particle-hole pair defining 
the second order contribution $\Sigma^{(2)}(\KK,\omega)$ is exactly solvable, see equations (\ref{eq:sig2scalli}) and
(\ref{eq:scalli}). This allows us to show that $\Sigma^{(2)}(\KK,\omega)$ is not a smooth function of $\KK$ and $\omega$,  
which is a consequence of the discontinuity of the wave vectors distribution of the fermions $n(\kk)$ at the Fermi surface $k=\kf$.
We find in general that the first singularities appear in the third order derivatives of $\Sigma^{(2)}(\KK,\omega)$, and that the 
region of these singularities in the plane {\yvan ($K,\varepsilon= (E_\KK-\hbar\omega)/\Ef)$} has a well defined mathematical origin, see figure \ref{fig:zones}:
Either the polynomials appearing in the explicit expression (\ref{eq:sig2scalli},\ref{eq:scalli}) have a double root, in which case 
the singularities are located on some parabolas and are given by equations (\ref{eq:ssp1}), (\ref{eq:ssp2}), (\ref{eq:ssp3});     
or these polynomials have a root equal to zero or two, in which case the singularities are located on some straight lines
and are given by equation (\ref{eq:ssd0}) or by equation (\ref{eq:sig2obli}).
However, it is possible to have already some singularities in the second order derivatives of $\Sigma^{(2)}(\KK,\omega)$,
in particular in $K=\kf$ for an impurity and fermions of the same mass $m$, see equations (\ref{eq:dkk}), (\ref{eq:dee}), (\ref{eq:dke}).

Then we concentrated on a physical quantity directly accessible to cold atom experiments, {\sl via} radio-frequency
spectroscopy: This is the complex energy $\Delta E(\KK)$ of the impurity with wave vector $\KK$. 
In principle, it can be obtained from the self-energy $\Sigma(\KK,\omega)$ after analytic continuation to $\omega$ in
the lower complex half-plane, by solution of the implicit equation (\ref{eq:impli_autoc}).
To second order in $g$, the singularities in the derivatives of the self-energy directly result into singularities
of the derivatives of the complex energy with respect to $K$, which were already present in reference \cite{lettre};
at this order, one finds for equal masses that $\partial_K^2 \Delta E^{(2)}(\KK)$ tends logarithmically to $+\infty$
in $K=\kf$, which is a limitation of the perturbative approach. In order to determine the real behavior of $\partial_K^2 \Delta E(\KK)$
in the neighborhood of $K=\kf$, for $g$ close to zero but finite, in a way taking advantage of our analytical
expression for $\Sigma^{(2)}(\KK,\omega)$, we have performed a self-consistent approximation of the implicit equation for $\Delta E(\KK)$:
We have included the mean-field shift to all order in $g$ but we have replaced the self-energy by its expansion up to order $g^2$ only,
see the equations (\ref{eq:autoc1}) and (\ref{eq:autoc2}) for equal masses.
The $g\to 0^-$ resulting scaling law (\ref{eq:loiech}) leads to $\partial_K^2 [\Delta E(\KK) -\hbar^2 K^2/(2m)]\propto -g^2 \ln g^2$
in $K=\kf$, simply because the $-\ln|K-\kf|$ divergence of $\partial_K^2 \Delta E^{(2)}(\KK)$ in the perturbative theory
is interrupted in the self-consistent theory at a distance from $K=\kf$ proportional to $g^2$.
Another interesting prediction (\ref{eq:ntpuvz}) of the scaling law (\ref{eq:loiech}) is that the third-order derivative
$\partial_K^3 \Delta E(\KK)$ around $\kf$ does not tend to zero uniformly with $K$, when $g\to 0^-$.
Also, we plotted in figure \ref{fig:loiech} the function $F(x)$ that appears in the scaling law (\ref{eq:loiech}), although
it is not fully determined by our self-consistent approximation, due to an ambiguity in the analytic continuation
of $\Sigma^{(2)}(\KK,\omega)$ to values of $\omega$ with a negative imaginary part, see the end of subsection \ref{subsubsec:paauveceendllde}.
Our pioneering study is thus not final. {\yvan To be complete, we give in (\ref{eq:condTnegli}) a condition for the non-zero temperature effects
to be negligible in an experiment.}

We concluded by a physical application a bit out of the mainstream of this paper: Taking advantage of the techniques
of integral calculus that we developed to obtain (\ref{eq:scalli}), we could calculate all the moments (\ref{eq:momfkp})
and (\ref{eq:momfki}) of the momentum of the particle-hole pair emitted by the moving impurity in the Fermi sea,
according to Fermi's golden rule, that is to zeroth order in $g$.
This gives access to the damping rate of the mean momentum of the impurity, for an arbitrary initial
momentum but to second order in $g$, see equations (\ref{eq:region1},\ref{eq:region2},\ref{eq:region3},\ref{eq:region4}).
This is complementary to the Fermi liquid theory, which is non perturbative in $g$ but restricted to arbitrarily low
momenta \cite{StringariFermi}. This also gives access to the momentum diffusion coefficient of the impurity to second order
in $g$, see its fully explicit expression at low momenta on equation (\ref{eq:coefdifftotalexpli}).
These moments may of course be observed experimentally with cold atoms, provided that one is able to prepare a quasi-monochromatic
impurity wave packet, e.g.\ by accelerating a Bose-Einstein condensate; it remains then to put this wave packet in contact
with a polarized Fermi gas and to measure its momentum distribution by time-of-flight \cite{Ferlaino,Equipe_melange}.
Finally, we note that the second moment, not to mention the higher order ones, raises a paradox, that we solve in the beginning of
subsection \ref{subsec:lmdlidlpptepli} in the weakly interacting regime (but not in the regime of arbitrary interactions
of reference \cite{DavidHuse}): The impurity interacts with the fermions, so that its wave vector distribution should
have an asymptotic $1/K^4$ tail \cite{Tan,BraatenGen,Leggett,WernerGenSimple,WernerGenLong}, the variance of $\KK$ should
diverge and the notion of momentum diffusion should be meaningless.

\begin{acknowledgments}
Christian Trefzger acknowledges financial support from a European Union Marie Curie post-doctoral fellowship 
INTERPOL and from the ERC ``Consolidating Grant Thermodynamix".
We thank Carlos Lobo for discussions on the unphysical gap occurring in variational ansatz with a finite number of particle-hole
excitations, and Meera Parish for useful advice in polaronic terminology.
\end{acknowledgments}

\end{document}